\documentclass[final,5p,times,twocolumn]{elsarticle} % set two columns
\usepackage{amssymb}
\usepackage{amsmath}
\usepackage[caption=false,font=footnotesize]{subfig}
\usepackage{booktabs}				% Enables fancy typesetting of tables
\usepackage{array}					% Used for table modification
\usepackage{multirow}				% Enables table headings to bridge multiple rows
\usepackage{siunitx}				% For aligning numbers in tables by the decimal
\usepackage{flafter}				% Ensures figures and tables appear after they are mentioned in the text
\usepackage{comment}				% For making block comments
\usepackage[usenames,dvipsnames]{xcolor}	% Allows text coloring
\usepackage{flafter}				% Ensures figures and tables appear after they are mentioned in the text
\usepackage{lipsum} 				% Enables creation of random dummy text
\usepackage{setspace}               % Enables making more vertical space in an equation with multiple lines
\usepackage{marginnote}             % Enables writing notes on the margin

\usepackage{multicol}% http://ctan.org/pkg/multicols

\usepackage{amssymb}
\usepackage{soul}
\usepackage{url}
\usepackage{epstopdf}
\usepackage{rotating}
\usepackage{placeins}
\usepackage{pifont}
\usepackage{threeparttable}
\usepackage{makecell}

\biboptions{numbers,sort&compress}

\usepackage{hyperref}	% Makes pdf bookmarks, links for Tables and Figures
\usepackage{breakurl}
\usepackage{cleveref}	% Allows referencing of multiple equations together (Must be after hyperref package)

\usepackage{amsmath, colortbl}

\journal{arXiv}

\begin{document}

\begin{frontmatter}



%%%%%%%%%%%%%%%%%%%%%%%%%%%%%%%%%%%%%%%%%%%%%%%%%%%%%%%%%%%%%%%%%%%%%%%%%%%%%%%%%%
% TITLE
%%%%%%%%%%%%%%%%%%%%%%%%%%%%%%%%%%%%%%%%%%%%%%%%%%%%%%%%%%%%%%%%%%%%%%%%%%%%%%%%%%
\title{Policy-relevance of a Model Inter-comparison: Switzerland in the European Energy Transition}


%% Example of authorship
\author[RRE]{Ambra Van Liedekerke}
\author[RRE]{Blazhe Gjorgiev\corref{cor1}} \ead{gblazhe@ethz.ch}
\author[ESC]{Jonas Savelsberg}
\author[UNIGE]{Xin Wen}
\author[EPFL,SLF]{Jérôme Dujardin}
%\author[EPFL]{Albin Cintas}
\author[ZHAW]{Ali Darudi}
\author[UNIGE]{Jan-Philipp Sasse}
\author[UNIGE]{Evelina Trutnevyte}
\author[EPFL,SLF]{Michael Lehning}
\author[RRE]{Giovanni Sansavini\corref{cor1}} \ead{sansavig@ethz.ch}

\affiliation[RRE]{organization={Reliability and Risk Engineering Laboratory, Institute of Energy and Process Engineering, Department of Mechanical and Process Engineering, ETH Zurich},
            city={Zurich},
            country={Switzerland}}

\affiliation[ESC]{organization={Energy Science Center, ETH Zurich},
            city={Zurich},
            country={Switzerland}}

\affiliation[UNIGE]{organization={Renewable Energy Systems, Institute for Environmental Sciences, Section of Earth and Environmental Sciences, University of Geneva},
            city={Geneva},
            country={Switzerland}}

\affiliation[EPFL]{organization={School of Architecture, Civil and Environmental Engineering, Swiss Federal Institute of Technology in Lausanne (EPFL)},
            city={Lausanne},
            country={Switzerland}}

\affiliation[SLF]{organization={WSL Institute for Snow and Avalanche Research, SLF Davos},
            city={Davos},
            country={Switzerland}}

\affiliation[ZHAW]{organization={ZHAW},
            city={Zurich},
            country={Switzerland}}

\cortext[cor1]{Corresponding author.}

%%%%%%%%%%%%%%%%%%%%%%%%%%%%%%%%%%%%%%%%%%%%%%%%%%%%%%%%%%%%%%%%%%%%%%%%%%%%%%%%%%%%%%%%%%%%%%%%
% ABSTRACT
\begin{abstract} % 150-200 words ideally
The energy transition is reshaping electricity systems, bringing new challenges, and emphasizing the need for strategic planning.  Energy policies play a crucial role in guiding this transition. However, assessing their impacts often requires robust modeling involving multiple models and going beyond a single country’s scope, analyzing international interactions. In this study, we examine three Swiss energy policies, analyzing their impacts on both the national energy system and the cross-border electricity flows.  We use a model inter-comparison approach with four electricity system models to explore scenarios involving Swiss renewable generation targets, the Swiss market integration, and the Swiss winter import limitations, in the context of various European electricity developments. The results indicate that a renewable generation target leads to a reduction in net imports and electricity prices. Additionally, reduced market integration impacts both Swiss and European energy transitions by limiting trade benefits, underutilizing Variable Renewable Energy Sources (VRES), and increasing electricity supply costs.  Lastly, we observe that limiting Swiss winter imports adversely affects electricity trading, driving up both supply costs and electricity prices.

\end{abstract}

%%%%%%%%%%%%%%%%%%%%%%%%%%%%%%%%%%%%%%%%%%%%%%%%%%%%%%%%%%%%%%%%%%%%%%%%%%%%%%%%%%%%%%%%%%%%%%%%
% GRAPHICAL ABSTRACT
\begin{comment}
\begin{graphicalabstract}
\begin{figure}[htbp]
  \centering
    \includegraphics[width=1\textwidth]{./Figures/Flowchart.png}
\end{figure}
\end{graphicalabstract}
\end{comment}
%%%%%%%%%%%%%%%%%%%%%%%%%%%%%%%%%%%%%%%%%%%%%%%%%%%%%%%%%%%%%%%%%%%%%%%%%%%%%%%%%%%%%%%%%%%%%%%%

%%%%%%%%%%%%%%%%%%%%%%%%%%%%%%%%%%%%%%%%%%%%%%%%%%%%%%%%%%%%%%%%%%%%%%%%%%%%%%%%%%%%%%%%%%%%%%%%
%% Research highlights
\begin{comment}
\begin{highlights}
    \setlength\itemsep{0em}
    \item ...
    \item ...
    \item ...
    \item ...
    \item ...
\end{highlights}
\end{comment}
%%%%%%%%%%%%%%%%%%%%%%%%%%%%%%%%%%%%%%%%%%%%%%%%%%%%%%%%%%%%%%%%%%%%%%%%%%%%%%%%%%%%%%%%%%%%%%%%%

%%%%%%%%%%%%%%%%%%%%%%%%%%%%%%%%%%%%%%%%%%%%%%%%%%%%%%%%%%%%%%%%%%%%%%%%%%%%%%%%%%%%%%%%%%%%%%%%%
% KEYWORDS

\end{frontmatter}
%%%%%%%%%%%%%%%%%%%%%%%%%%%%%%%%%%%%%%%%%%%%%%%%%%%%%%%%%%%%%%%%%%%%%%%%%%%%%%%%%%%%%%%%%%%%%%%%%


%%%%%%%%%%%%%%%%%%%%%%%%%%%%%%%%%%%%%%%%%%%%%%%%%%%%%%%%%%%%%%%%%%%%%%%%%%%%%%%%%%%%%%%%%%%%%%%%%
% INTRODUCTION
%%%%%%%%%%%%%%%%%%%%%%%%%%%%%%%%%%%%%%%%%%%%%%%%%%%%%%%%%%%%%%%%%%%%%%%%%%%%%%%%%%%%%%%%%%%%%%%%%
\section{Introduction}
\label{intro}

%% Paragraph 1: General motivations
%  Why is this topic of research relevant?
% Blazhe to write
The energy transition is transforming our electricity systems~\cite{IEA2023a}. 
This is reflected in the evolving generation mix and the increasing electrification of the heating and transportation sectors~\cite{IEA2023b}. 
This brings rising challenges which require proper planning and resource allocation to ensure the reliability of the power supply in the future~\cite{Stankovski2023}. 
Therefore, the scientific community is placing significant efforts in developing energy system models to better understand these challenges and to inform decision-makers~\cite{DeCarolis2017}. 
However, modelers face hurdles in the selection of appropriate modeling approaches, the lack of data, the diversity of exogenous inputs, and their related uncertainties. 
Hence, the obtained modeling results often differ significantly across models and frameworks~\cite{Bistline2018}, limiting their applicability in obtaining policy-relevant findings~\cite{Huntington1982}. 
This motivates the research on model inter-comparisons comparing models in their natural environments~\cite{Heinisch2023}, such that a strict alignment on modeling assumptions and inputs is not required, but comparability is assured by harmonizing selected scenarios and key parameters~\cite{Wilson2021}. 
Such an approach increases the robustness of the policy-relevant results by identifying result consensus across models and scenarios~\cite{Harmsen2021, Siala2022}.

%-------------------------------------------------------------------------------------------------
%% Paragraph 2: Literature review 1
%  Focus on inter-comparison papers and how the comparison is performed
% Blazhe to write
Energy and electricity system model inter-comparisons are increasingly performed. 
% CH
A review of 19 studies between 2011 and 2018 with various scenarios for Switzerland in 2035 is conducted in~\cite{Xexakis2020}. 
The analysis highlights that most models show a dependence on fossil fuel-based generation and net electricity imports, contrary to what the general public expects. 
% First EDGE intercomparison
In a recent study, the authors compare three spatially resolved electricity system models to gain insights into the role of renewables in the Swiss energy transition in 2035~\cite{Heinisch2023, Trutnevyte2024}. 
In general, disagreements among these model results exist for valid reasons; however, they find Photovoltaic (PV) is a common denominator in the future generation mix. 
% CROSS
Similarly,~\cite{CROSS2024} compares six energy and electricity system models with different spatial and temporal resolutions focusing on Switzerland in 2050. 
Although the models mostly agree on PV as the driver of the energy transition, some models and scenarios obtain higher participation of wind in the supply mix. 
% Germany
In~\cite{Misconel2022}, the authors compare four electricity system models for Germany, focusing on investment decisions and operational behavior in 2030. 
For this purpose, they harmonize scenario data and show that the result differences mainly depend on the modeling approach. 
Three electricity system models focusing on electrification and load shifting with electric vehicle batteries and heat pumps are compared to study Germany's generation adequacy in 2030~\cite{Misconel2024}. 
The study results show discrepancies in load shifting, which are directly attributed to model features.
% EU
In~\cite{vanOuwerkerk2022}, the authors compare six electricity models with country-level aggregation for central Europe utilizing a highly simplified test case. 
They identify expansion deviations, which can be related to specific modeling differences.
Similarly,~\cite{Siala2022} compares five electricity models for Europe with high spatial aggregation. 
The paper argues that a multitude of modeling approaches enable the evaluation of result variance because of model uncertainty. 
% US
The "Energy Modeling Forum" (EMF) has conducted 37 model inter-comparisons on topics such as macroeconomic and climate change impacts, global energy modeling, electricity and fuel markets, and electric load forecasting~\cite{EMF2024}.
% Concluding sentence
In general, the literature mainly focuses on defining a comparison methodology and understanding what drives result commonalities and differences, while little or no focus is given to policy-relevant impacts.
The literature agrees that model inter-comparisons are powerful methods for policy-relevant studies~\cite{Huntington1982, Henke2023}. 
The main advantage lies in the analysis of commonalities and differences in results derived from various frameworks, models, assumptions, and case studies~\cite{Prina2022}. 
% Review of the theory
The theoretical groundwork for inter-comparisons applied to policy questions is laid out in~\cite{Gacitua2018}. 
This work reviews significant policy instruments for renewable energy integration and explores models and decision support tools developed for energy policy analysis.
Similarly,~\cite{Savvidis2019} evaluates model suitability for specific policy questions, and ~\cite{Riemer2023} investigates how different modeling frameworks can be compared to answer policy-relevant questions.
% Application to climate policy
Results from inter-comparisons are extensively used to assess the impact of CO2 targets and other climate policies on the energy system, as shown in~\cite{Clarke2009, Tavoni2015}.
% Practical examples - Europe 
In~\cite{Nikas2021}, the authors utilize eleven assessment, energy system, and sectoral models to assess the impact of the European climate goals with respect to costs, investments, and employment. 
% Practical example - America
Model inter-comparisons also address policy questions beyond climate goals. 
In~\cite{Huntington2020}, 17 modeling teams from the EMF analyze the North American energy integration and trade under various future energy market conditions and policy scenarios.
% Practical example - technology
Furthermore, studies focus on technology-specific policy questions, including policies and subsidy schemes for photovoltaics~\cite{Avril2012}, renewable generation technologies~\cite{Gacitua2018, Bistline2020}, and storage technologies~\cite{Giarola2021}.

%-------------------------------------------------------------------------------------------------
%% Paragraph 4: Research gaps
%  Clearly state the research- focus on the policy relevance of what the previous work is missing
% Paragraphs 4 and 5 can be fused together
% Ambra to write (assist Blazhe)
Most inter-comparison studies primarily address the theoretical aspects of best practices in model inter-comparison and focus less on policy-relevant questions. 
They examine how modeling features influence results and often emphasize significant harmonization efforts in input data. 
While some studies address policy-relevant questions, they typically focus on the impacts of climate policies on the electricity system at large. 
Furthermore, the specific interactions between countries and the effects of policy measures on these interactions have not been thoroughly examined. 
Overall, the model inter-comparisons often overlook the analysis of cross-country interaction and market integration.
% merge paragraphs
%-------------------------------------------------------------------------------------------------
%% Paragraph 5: What do we do
% Briefly describe how what you do is covering the above research gaps
% Provide a definition of model inter-comparison to minimize confusion (with benchmarking)
% Ambra to write
To tackle these research gaps, we employ four modeling frameworks analyzing the interactions between the electricity systems of Switzerland and its neighboring countries.
In particular, we define scenarios representing different realizations of policies concerning the Swiss renewable generation target, the Swiss market integration, and the Swiss winter import limitations.
We do so while concurrently considering different developments of the European electricity system.
To enhance the results' comparability across different models, we perform a harmonization of the scenario-relevant inputs as in~\cite{Trutnevyte2014}.

%-------------------------------------------------------------------------------------------------
%% Paragraph 6: The research questions and contributions
%  State the research questions and contributions
% Blazhe (research questions), Ambra (contributions)
Here, we aim to answer the following policy-relevant research questions: i) What is the effect of a renewable generation target on long-term planning for 2050? ii) Is the full potential of invested renewables utilized? iii) How is Switzerland affected by its (non-)integration into the European electricity market? iv) How is the electricity system affected by a limit on the winter net imports? 
The main contributions of the paper are threefold: 1) We analyze the future Swiss electricity system and its interactions with its neighboring countries under different policy conditions,  2) We answer policy-relevant questions, and 3) We ensure result robustness by conducting this study as a model inter-comparison.

%-------------------------------------------------------------------------------------------------
% Paragraph 7: Outline of the paper
% Add a sentence or two for each section. Link the sections (use the labels).
% Blazhe to write
The rest of the paper is structured as follows. Section~\ref{sec:method} describes the models used in the analyses. Section~\ref{sec:cstudy} details the case study, including the case study system, the scenarios, and the inter-comparison protocol. 
% Section~\ref{sec:results} compares the results of the models for a selection of representative scenarios as well as for the whole scenario set. 
Section~\ref{sec:results} compares the results.
Section~\ref{sec:discussion} discusses the relevance of the results in decision-making. 
Finally, Section~\ref{sec:results} provides a summary and future work.
% Finally, Section~\ref{sec:results} provides a summary.

%%%%%%%%%%%%%%%%%%%%%%%%%%%%%%%%%%%%%%%%%%%%%%%%%%%%%%%%%%%%%%%%%%%%%%%%%%%%%%%%%%%%%%%%%%%%%%%%
% METHOD
%%%%%%%%%%%%%%%%%%%%%%%%%%%%%%%%%%%%%%%%%%%%%%%%%%%%%%%%%%%%%%%%%%%%%%%%%%%%%%%%%%%%%%%%%%%%%%%%
\section{Method}
\label{sec:method}
% Optional: Add a paragraph outlining the steps/components of the method, i.e., a general overview
The model inter-comparison utilizes four electricity system models. 
Each model is developed by a different institution to answer a set of energy transition and policy-relevant questions. 
Here, we provide an overview of these models (Section~\ref{subsec:models}) and present their main similarities and differences (Section~\ref{subsec:simdiff}).

%-----------------------------------------------------------------------------------------------
% Describe the Models
% Provide 2 to max 3 paragraphs for each model
\subsection{Models}
\label{subsec:models}

\subsubsection{Nexus-e}
\label{subsec:nexuse}
% Ambra to provide a description
% Try to give a general overview of the model and its properties, point out some specifics that are relevant to the ongoing model intercomparison, and describe shortly new features that have not been described before. Of course, if you have, provide references (1-2) where the model has been described in detail.
% In addition, try to put several sentences on model inputs, only specific to your model. Try to use existing references as your data sources to avoid a lot of text and numbers. 

% Nexuse, CentIv, and DistIv
Nexus-e is an integrated energy systems modeling platform that combines multiple modules to simulate different aspects of the electricity system~\cite{Gjorgiev2022}. 
In this study, we utilize the CentIv and DistIv modules. 
The CentIv module computes the optimal dispatch and the optimal generation and transmission expansion from a centralized perspective, minimizing system investment and operation costs~\cite{Raycheva2023}. 
Conversely, the DistIv module focuses on the consumer perspective, optimizing investments and operational decisions to minimize consumers' electricity costs~\cite{Jo2022}.
Additionally, DistIv incorporates a national grid injection tariff, which provides additional earnings to consumers who feed electricity back into the grid. 
% CentIv & DistIv interaction
The CentIv and DistIv modules are soft-linked and solved iteratively, as detailed in~\cite{Gjorgiev2022}. 
This iterative approach ensures two critical aspects: first, that consumer decisions are informed by the operations of the centralized system, and second, that the centralized planning strategy considers the distributed investments made by consumers.

% Modeling specifics: 
% Technologies in general. 
Nexus-e utilizes a comprehensive techno-economic characterization of the generation and storage technologies~\cite{Nexuse_input_v2}. 
In particular, CentIv has a diverse set of centralized technologies, and DistIv focuses on rooftop PV and battery storage.
% Curtailment
Additionally, both modules can curtail excess Variable Renewable Energy Sources (VRES) generation. 
% DistIv curtails PV generation only if the power injection to the transmission grid exceeds the transformer limits.
% CentIv can curtail all VRES generation if generation exceeds demand or if power flow limits are reached.
% Grid modeling
CentIv fully models the transmission grid and computes the DC power flows, respecting line and transformer limits.
On the other hand, DistIv models each municipality as a single-node distribution system, without modeling the distribution network.
% Spatial resolution
CentIv has a high spatial resolution, modeling every extra-high voltage node in Switzerland. 
DistIv, instead, considers municipal resolution for electricity demand and renewable potential, using data from the Federal Statistical Office \cite{FSO} and Sonnendach \cite{sonnendach}, respectively.
% Neigbors' spatial resolution
Although Switzerland is modeled in high detail, Nexus-e represents the neighboring countries as a single node.
Their hourly dispatch is optimized together with the Swiss dispatch, although their installed capacities are fixed. 
% Hourly resolution
Both modules use an hourly resolution, simulating every second day, and considering a yearly time horizon.

\subsubsection{EXPANSE}
\label{subsec:expanse}
% Xin to provide a description
The EXPANSE (EXploration of PAtterns in Near-optimal energy ScEnarios) model is a single-year, bottom-up, perfect-foresight, linear optimization model~\cite{Sasse2023Energy,Heinisch2023}. 
EXPANSE addresses the least-cost capacity and aggregated transmission expansion and generation dispatch in the electricity system. 
The model incorporates constraints for technology ramping and start-up, power reserves, power flows on lines, balances for storage, and interconnection with neighboring countries. 
EXPANSE provides flexibility on the demand side through load shedding in extreme situations of grid congestion or generation deficit. 
On the generation side, EXPANSE provides flexibility through curtailment of VRES, and storage technologies such as pumped hydropower storage, grid-scale batteries, decentralized batteries with solar PV, and power-to-hydrogen technology.

In this study, EXPANSE has a spatial resolution of 2’136 Swiss municipalities, according to the definition of Swiss municipalities of 2023~\cite{swisstopo}. 
The model represents demand and generation balancing with 15 Swiss national clustered nodes and 4 nodes of neighboring countries. 
Each municipality-level node directly feeds into the nearest grid node within Switzerland. 
Each neighboring country is modeled at the national level, considering technology-specific generation, storage, and interconnection transmission dynamics. 
The temporal resolution in this study is six hours to balance result accuracy and computational efficiency due to the model's high spatial resolution and grid expansion capability.
More detailed documentation of the model and dataset used can be found in~\cite{Wen2023,Rubino2024,Sasse2019,Sasse2020}.

\subsubsection{FEM}
\label{subsec:fem}
% Ali to provide a description
The Future Electricity Market Model (FEM) is an electricity market model that extends the capabilities of Swissmod~\cite{Schlecht2014}, which primarily addresses market dispatch decisions. 
FEM focuses on medium- and long-term electricity market developments, particularly within Switzerland. 
In this study, FEM is formulated as a least-cost dispatch and investment model. 
Recognizing the significance of imports and exports, FEM also incorporates dispatch operations in 18 EU countries. 
Each country or market zone is represented by a single network node, with trade modeled using Net Transfer Capacity (NTC).

The FEM operates on an hourly resolution over an entire year, utilizing perfect foresight for both supply and demand aspects.
Given the critical role of hydrological systems in Switzerland, the model adheres to a hydro year starting in October and includes a simplified representation of Switzerland's hydro system. 
Additionally, the model integrates data on conventional power plant capacity and availability, fuel-specific generation costs, and country-specific renewable energy infeed time series. 

\subsubsection{OREES}
\label{subsec:orees}
% Jerome to provide a description
% OREES is an optimization model that computes the optimal investment in PV and wind capacities, as well as the optimal power dispatch. 
OREES is an optimization model computing the optimal investment in PV and wind capacities, as well as the optimal power dispatch. 
It ensures the power balance in the system through optimal DC power flow, considering the extra-high voltage grid in Switzerland.
At each simulation step, the optimization of the power dispatch considers the state of the hydropower infrastructure (reservoir levels, inflows), the expected Swiss load and generation from non-dispatchable renewable sources for the coming week and an annual strategy for the management of the large reservoirs. 
% The objective is to smoothen as much as possible the imports and exports required to balance the country.
The objective is to smoothen the imports and exports required to balance the country.
The optimal investment in new capacities is obtained through an evolution strategy as described in~\cite{Dujardin2021}. 
The objective function maximizes the net income for new photovoltaic and wind power installations in the modeling year, given their production profile and an exogenous market price of electricity. 
The work in~\cite{Bartlett2018} describes the optimal power flow algorithm used in OREES as well as the modeling of the storage and pumped-storage hydropower infrastructure.

% OREES utilizes curtailment, grid expansion, and load shedding as a source of flexibility when necessary. 
OREES utilizes curtailment, grid expansion, and load shedding as flexibility sources. 
First, curtailment of photovoltaic and wind power installations is allowed and directly impacts their revenues. 
This curtailment is performed by the optimal power flow and is used when grid congestions occur. 
Second, in case of optimization failure, the most loaded transmission lines are reinforced with 5\% capacity increments until the power dispatch is successful. 
The updated capacity of each line is then used for the following time steps. 
Finally, if the optimal power flow still fails while all lines are used below 80\% of their capacity, load shedding is imposed by progressively reducing the demand of the nodes with the highest positive netload. 
This strategy allows us to systematically reach a power dispatch with minimal load shedding and minimal line reinforcement.

%-----------------------------------------------------------------------------------------------
\subsection{Model similarities and differences}
\label{subsec:simdiff}
% Short discussion on model similarities: all cost minimal optimizations ;
The models used in this inter-comparison are similar at the core, computing optimal capacity expansion and dispatch over a yearly time horizon.
% Different objective function
However, different objective functions and modeling characteristics lead to a different decision logic embedded in the optimization problems.
Namely, Nexus-e, EXPANSE, and FEM minimize total investment and operational costs, while OREES maximizes the net incomes of PV and wind generators.
% Another major difference in OREES is that imports and exports are utilized only when power balancing cannot be performed endogenously. 
Another major difference is that OREES computes the optimal investment and the optimal dispatch in two separate problem setups. 
Additionally, the model uses imports and exports as a last resort only when power balancing cannot be performed endogenously.
Therefore, this model has lower curtailments as well as lower imports and exports than the others.
Conversely, Nexus-e's special trait is using the DistIv module, which confers a decentralized consumer perspective to this model. 
Additionally, DistIv models fixed grid fees and, hence, persistent self-consumption incentives for households. 
This results in considerably higher investments in decentralized PV and battery storage in Nexus-e compared to the other models.

% Temporal and geographical scales
Other important differences lie in the modeled technologies and in the temporal and geographical scales.
% Technologies
For instance, alpine PV is modeled only in Nexus-e and OREES, while demand-side management is modeled only in Nexus-e.
% Scales
The temporal scale spans from hourly resolution in Nexus-e, FEM, and OREES to a 6-hour resolution in EXPANSE. 
The geographical scale spans from a municipal level in EXPANSE and OREES to a cantonal level in FEM. 
Nexus-e has a combination of municipal and regional detail.
Additionally, the geographical scope is limited to Switzerland in OREES, to Switzerland and the neighboring countries in Nexus-e and EXPANSE, and extended to 18 EU countries in FEM.
Both the temporal and the geographical scales affect the model outcomes. 
A higher temporal resolution exposes the system to the variability of the time-dependent input parameters. 
A higher geographical resolution increases the level of detail.
Finally, the models also differ regarding the representation of the electricity grid. 
Nexus-e and OREES use a full representation of the Swiss transmission grid, EXPANSE has a reduced grid representation, and FEM does not have a power grid model but instead uses the NTC modeling approach. 
The representation of the electricity grid affects the permitted power flows in the model.
A higher level of detail corresponds to a more constrained system since power flow limits are enforced at each line.
This can lead to results with higher curtailments and increased system costs.
Endogenous modeling of grid expansion, as in Nexus-e, EXPANSE and OREES, however, allows to overcome some of the limits by investing in the grid.
A summary of the models' characteristics can be found in~\ref{App:Aappen}.

%%%%%%%%%%%%%%%%%%%%%%%%%%%%%%%%%%%%%%%%%%%%%%%%%%%%%%%%%%%%%%%%%%%%%%%%%%%%%%%%%%%%%%%%%%%%%%%%
% CASE STUDY
%%%%%%%%%%%%%%%%%%%%%%%%%%%%%%%%%%%%%%%%%%%%%%%%%%%%%%%%%%%%%%%%%%%%%%%%%%%%%%%%%%%%%%%%%%%%%%%%
\section{Case study}
\label{sec:cstudy}
% Give a general overview - introduce the test case: CH + neighboring countries...
We analyze the 2050 Swiss electricity system in a European context. 
Our models have a detailed representation of the Swiss electricity system and a simplified representation of Europe. 
We ensure that the models align on the fundamental inputs, particularly those enabling the interactions with Europe, and retain their original sets of parameters otherwise. 
This approach ensures that the models provide consistent outputs for comparison, although reflecting the diversity of methodologies and assumptions inherent in each modeling framework. 
The following sections describe the scenarios, the input parameters, and the inter-comparison protocol.

%-----------------------------------------------------------------------------------------------
% Describe scenarios and their parameterizations
% List outer input data where the models are aligned
\subsection{Scenario-based inputs}
\label{subsec:sbinputs}
% Introduce the scenarios
The model inter-comparison aims to produce robust results to support policy-relevant research questions. 
Hence, we assess the impact of different European developments on the Swiss electricity system. 
Moreover, we evaluate different Swiss policies: 1) a renewable target, 2) the integration into the European electricity market, and 3) the limit on net winter imports.
For each of the European and Swiss policy dimensions, two possible realizations are considered, as summarized in Table \ref{tbl:scenarios}.
Each scenario is then defined by a combination of realizations for those four dimensions (one for the European development and three for the energy policies), leading to a total of 16 scenarios.

%% European dimension
The European dimension is captured by alternative European developments in the electricity system.
The Global Ambition (GA) and Distributed Energy (DE) scenarios from the ENTSO-e Ten-Year Network Development Plan (TYNDP) provide information for the whole ENTSO-e's future energy system~\cite{Entsoe2022}. 
The GA and DE scenarios represent different European developments, assuming the energy transition will be based on centralized renewable and low-carbon technologies, and decentralized initiatives by prosumers, respectively.
% Inputs
The European dimension defines the model input parameters for the electricity demand, the installed capacities per technology type, the generation profiles for renewable generation in the neighboring countries, the electricity flows between these countries and their neighbors, and the NTC values for all countries.
OREES is the only model where the European dimension does not impact the results, as OREES models only Switzerland endogenously.

% The climate year
All input parameters are defined for the 1995 climate year.
This is one of the three climate years considered in the TYNDP 2022 scenarios, representing the scenario with intermediate severity of energy drought periods (Dunkelflaute)~\cite{Entsoe2022}.
In the ENTSO-e analysis of the severity of 30 climate years, 1995 scores as the fourth most severe.
In this study, the climate year defines the European electricity demand, the installed capacities per technology type, the generation profiles for renewable generation, and the electricity flows between the modeled European countries and their neighbors.

%% Policy dimensions
% Renewable target
The Swiss renewable generation target aims to achieve 45 TWh of yearly generation from PV, wind, and biomass. This reflects the most recent renewable targets set by the Swiss government~\cite{stromgesetz}. We simulate the realizations of the renewable target by imposing a 45 TWh (R45); alternatively, no renewable target (RNT) is enforced.
% Market integration
Additionally, this work explores the effects of full (N100) and reduced (N030) power market integration by setting NTC values to 100\% or 30\% of the TYNDP values for 2050, respectively.
The reduction to 30\% of the NTC value is an extreme and pessimistic outcome of the partial market integration~\cite{electricityCHEU}.
It is important to note that the authors deem this 30\% scenario highly unlikely, so it should not be considered for planning an efficient system. 
Nevertheless, when thinking about a robust system, it could still be considered a relevant extreme case.
% Net winter import
The net winter import limitation represents an initiative put forth by the parliament to increase generation adequacy by reducing the reliance on electricity imports in the winter months~\cite{federalactsecurityofsupply}.
% Where does the 5 TWH come from -> historical values?
In this work, the effects of the net winter imports policy are studied by imposing a strictly binding 5 TWh winter net import target (W05); alternatively, the winter net imports are not constrained (WNC).

\begin{table}[h]
\centering
\caption{Summary of the European and Swiss policy dimensions defining the inter-comparison scenarios}
\label{tbl:scenarios}
\begin{tabular}{lll}
\hline
\makecell[l]{European development\\ \textcolor{white}{space}}        & \hspace{0.2cm}\makecell[l]{Gloabal Ambition (GA), \\ Distributed Energy (DE)} \\
\makecell[l]{Renewable target\\ \textcolor{white}{space}}            & \hspace{0.2cm}\makecell[l]{No Target (RNT), \\ 45 TWh target (R45)}  \\
\makecell[l]{Market integration\\ \textcolor{white}{space}}          & \hspace{0.2cm}\makecell[l]{Full integration (N100),\\ Reduced integration (N030)} \\
\makecell[l]{Net winter import limit\\ \textcolor{white}{space}}     & \hspace{0.2cm}\makecell[l]{No Constraint (WNC), \\5 TWh constraint (W05)}  \\ \hline
\end{tabular}
\end{table}

% The combination of all possible policy realizations and European developments yields 16 scenarios in total.
% Subset of scenarios
Given the large number of scenario combinations, we select for analysis five representative scenarios, namely a reference scenario (Ref), three scenarios where a single policy is varied relative to the reference (R45, N030, W05), and a scenario incorporating all policy variations (All):
\begin{itemize}
\setlength\itemsep{-0.5em}
    \item Ref: reference scenario, with no policies active%GA\_RNT\_N100\_WNC
    \item R45: scenario with renewable generation target policy  %GA\_R45\_N100\_WNC
    \item N030: scenario with market integration reduction policy  %GA\_RNT\_N030\_WNC
    \item W05: scenario with winter net import limit policy  %GA\_RNT\_N100\_W05
    \item All: scenario with all policies active %GA\_R45\_N030\_W05
\end{itemize}
% Why we present only GA results
All the representative scenarios use the GA European development. 
This is sufficient because our findings indicate that this dimension has little influence on the models' outcomes.

%-----------------------------------------------------------------------------------------------
% Describe model inputs (link to existing work where this is outlined)
\subsection{Model-based inputs}
\label{subsec:mbinputs}
% What we align on
% In addition to the scenario-specific inputs, there are a few other fundamental inputs on which all models align.
In addition to the scenario-specific inputs, there are other fundamental inputs on which all models align.
The models align on the nuclear phase-out policy, assuming a 60-year operational lifespan for nuclear plants, and on the national electricity demand as described in~\cite{energieperspektieven2050}. 
Furthermore, all models adopt a uniform value of lost load of 10'000 Euros per MWh~\cite{voll_Kenneth}. 

% What we do not align on 
Despite the alignment of key parameters, each model retains its unique parametrization of the techno-economic characteristics of the generating technologies.
The potential of VRES, including wind, PV on buildings (rooftop and facade PV), and alpine PV, varies across models and depends on the spatial resolution of the model and the underlying study.
Input data on the network and the technical characterization of generating technologies for Nexus-e are detailed in~\cite{Nexuse_input_v2}. 
The investment and operation costs have been updated based on recent data from ~\cite{Löffler2019, Lazard2023}.%\cite{Löffler2019, Breyer2023, potencia, Lazard2023, DEA}.
The EXPANSE's input parameter dataset is presented in~\cite{Sasse2019,Sasse2020,Wen2023,Rubino2024}, FEM's in~\cite{FEM_input}, and OREES's network, PV and wind data in~\cite{Dujardin2021}~and~\cite{Bartlett2018}.

%-----------------------------------------------------------------------------------------------
% Describe the MIC
% Use the MIC protocol for inspiration: https://drive.switch.ch/index.php/f/7194124861
\subsection{Model inter-comparison protocol}
A model inter-comparison protocol is used to structure the scenarios as well as the input and output data for the case studies.
% Input
First, the input data is provided in a standardized format to all modeling teams.
This is then converted to a model-specific input data format for each model.
This input data defines the model- and scenario-based inputs, as described in the previous sections.
% Output
Eventually, model-specific results are converted to a standardized results format to facilitate the result analysis and comparison.

The results include information on installed generation capacity, dispatch, electricity imports and exports, cost of electricity supply, and electricity prices for each scenario.
Additionally, we also report the costs for subsidies or other mechanisms necessary to achieve the policy target.
% CAPEX, and OPEX, cost of electricity supply
The installed capacity and hourly dispatch are the decision variables of the electricity system models.
The cost of electricity supply is the sum of the annualized investment cost and the yearly fixed and variable operational costs.
% Electricity price
The electricity prices are the yearly average value of the hourly load-weighted-average nodal shadow prices.
They are thus computed from the energy balance dual variables. 
They, therefore, correspond to the wholesale electricity price and not to the price the end-consumers pay.
% Policies
Finally, we investigate the shadow prices of two policy constraints (R45 and W05). 
For the renewable target, this dual variable reflects the subsidy cost to achieve the target, whereas the dual variable on the Winter net import constraint reflects the cost of a potential certificate implemented to make sure the target is met. 
The dual variable on the NTC constraint does not reflect a possible decision for policymakers and, hence, is not discussed.

%%%%%%%%%%%%%%%%%%%%%%%%%%%%%%%%%%%%%%%%%%%%%%%%%%%%%%%%%%%%%%%%%%%%%%%%%%%%%%%%%%%%%%%%%%%%%%%%
% RESULTS
%%%%%%%%%%%%%%%%%%%%%%%%%%%%%%%%%%%%%%%%%%%%%%%%%%%%%%%%%%%%%%%%%%%%%%%%%%%%%%%%%%%%%%%%%%%%%%%%
\section{Results}
\label{sec:results}
% Give a general overview
% How do we answer the research questions
The four research questions outlined in Section~\ref{intro} are addressed by examining the impact of the energy policies on various factors, including the generation mix, generation curtailment, imports and exports, cost of electricity supply, and electricity prices.
Additionally, subsidies or other mechanisms required to achieve the presented policies are computed and analyzed. 
% Disclaimer on sensitivity
Results are mostly reported in absolute terms, as percent variations. 
The reason is twofold. 
First, this provides better insights into the effect of a policy measure. 
Second, cost, prices, and subsidy outputs are heavily dependent on investment cost and fuel price assumption. 
A sensitivity analysis would be necessary to discuss absolute values.
% Subset of scenarios
Given the large number of scenarios, in total 16, here we focus on the five representative scenarios described in Section~\ref{subsec:sbinputs}: Ref, R45, N030, W05, All.
This setup allows us to examine the individual and combined effects of the policies on the system evolution when compared to the reference, which does not enforce any policy. 
% Where to find the full set of results
A comprehensive set of results, including additional scenarios, is reported in the supplementary material (Figures 1-6).
% Robustness
In addition, this section addresses the robustness of the results. 
We assess the impact of policy measures on the output utilizing the outcomes from all models and scenarios.
% OREES' missing results
Results are presented for all models. 
Since OREES operates only when a renewable generation target is defined, it can provide results only for the scenarios with the 45 TWh renewable target.

%-----------------------------------------------------------------------------------------------
\subsection{Investments and generation} % - focus on R45

\subsubsection{National level}
% Show a figure of bar subplots for each model for 1) the base scenario (first row) and 2) the most restrictive scenario (second row)
% Put the alternative with 3 scenarios

\begin{figure*}[htbp!]
  \centering
  
  \begin{minipage}[b]{0.45\textwidth}
    \includegraphics[height=0.4\textheight]{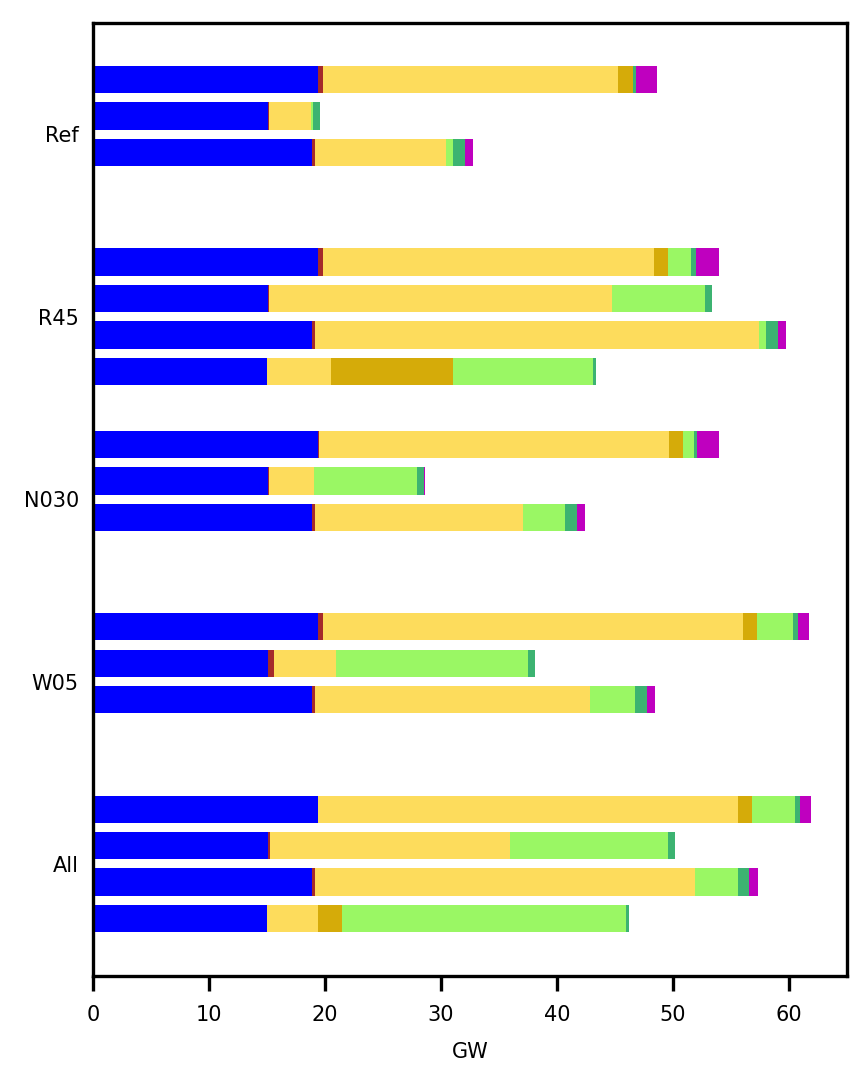}
  \end{minipage}
  \hspace{0.2cm}
  \begin{minipage}[b]{0.45\textwidth}
    \includegraphics[height=0.4\textheight]{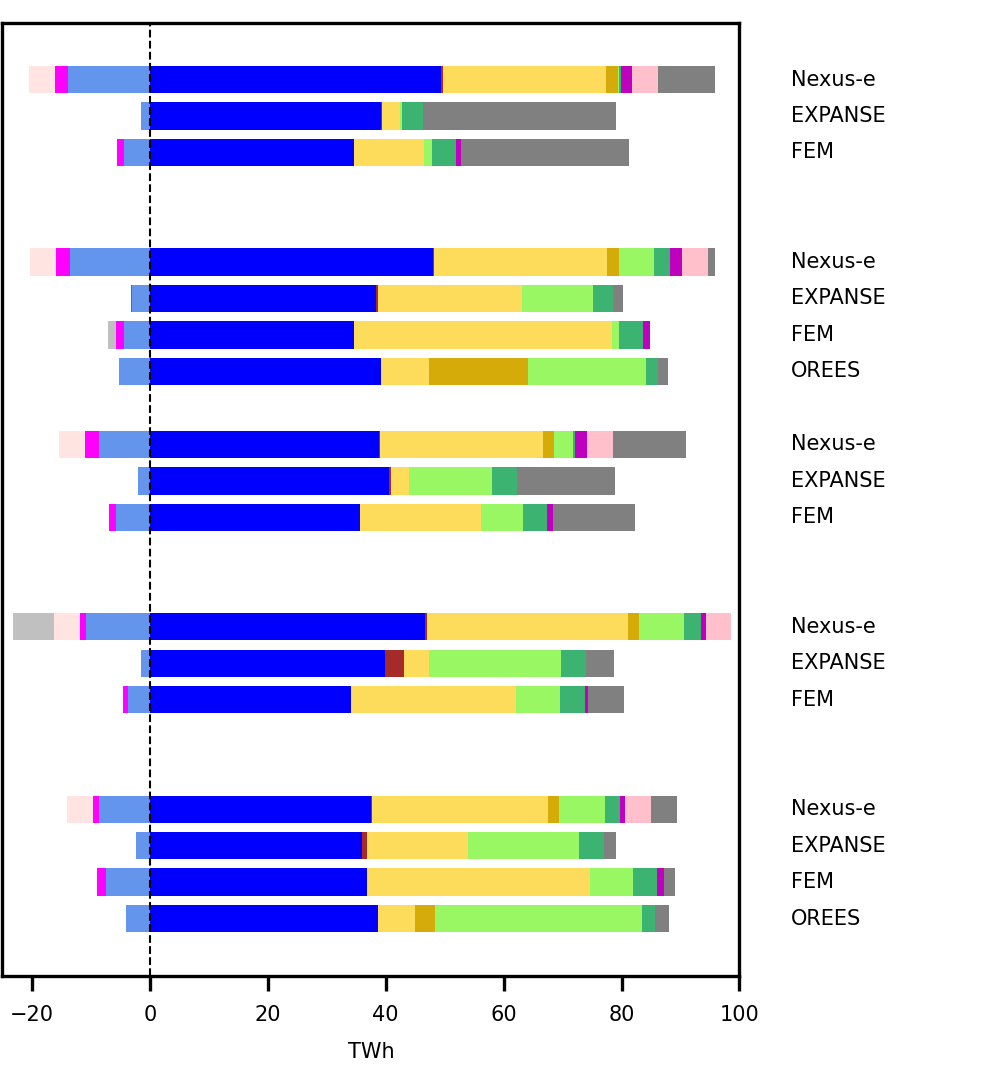}
  \end{minipage}
    
  \begin{minipage}[b]{1\textwidth}
    \hspace{-0.4cm}
    \includegraphics[width=1\linewidth]{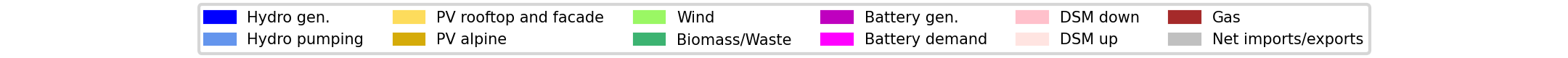}
    % \vspace{-0.5cm}
  \end{minipage}

  \caption{Installed capacity and annual generation by technology type for the representative scenarios and the four models.}
  \label{fig: generation}
\end{figure*}

Figure~\ref{fig: generation} shows the generation capacity mix (left) and the yearly generation output (right) for all models and representative scenarios in 2050.
% Introduction to the generation mix and dispatch
Overall, the policy dimension largely influences the electricity system's generation mix and operations across models.
% The influence of the models
Nevertheless, in the reference scenario without the renewable electricity target, the model results differ significantly. 
On the one side, Nexus-e has a strong uptake of installed PV capacity, due to the decentralized perspective and the persistence of self-consumption incentives assumed in DistIv.
On the other side, EXPANSE and FEM have a minor expansion of the PV capacity and largely rely on electricity imports to supply the national demand.
For those models, imports cover 43\% and 39\% of the national demand, respectively.

% R45 - target is achieved with different shares of VRES. Analyse in depth.
If the renewable target (R45) is applied, the results are more aligned since models increase their endogenous power production by investing in PV and wind generation.
There is, however, still large variability across the shares of installed VRES technologies.
For instance, Nexus-e and FEM have moderate investments in wind capacity, while EXPANSE and OREES rely on this technology to a large extent.
The models agree, however, on a minimum of 0.7 GW of installed wind capacity.
Additionally, for all models, PV plays a crucial role in increasing the national generation from VRES technologies.
Thus, all models agree that a 45 TWh renewable generation target is achievable in Switzerland.
Additionally, the differences across the models show that the optimal generation mix depends on the input assumptions, technologies modeled, and the perspective taken by the model and that there are several feasible ways to achieve the renewable target.

% N030 and W05
Investments are affected by market integration and winter import limit policies too.
Both policy measures lead to increased investments in national generation capacities.
% N030 - Increased investments due to less reliance on trades
This is due to the restricted reliance on 1) international electricity trades when N030 is imposed and 2) the need to increase internal generation in the winter months when the winter import (W05) is constrained.
% W05 - More investments in wind power - winter production
For instance, the winter import limitation (W05) causes a shift towards more installed wind power since this is characterized by a higher winter generation.
Under this policy measure, a minimum of 3.1 GW of wind capacity is being installed.

% All
In the most restrictive scenario, All (i.e., R45, N030, W05 imposed simultaneously), we observe a significant ramp-up of wind across all models.
This highlights the influence of the winter import limit and the role of wind in supplying the winter demand.
% General observation
Overall, three out of four models agree on very high shares of PV on buildings across all scenarios. 
The exception is OREES, where wind has a larger share in the generation mix, consistent with its goal to balance production and demand without additional storage, imports, or demand-side management. 
Nevertheless, in the N030 and W05 scenarios, wind is preferred over PV also by EXPANSE. 
This is because EXPANSE has less pumped-storage capacity, reducing its ability to shift excess PV generation from day to night. 
Consequently, wind, which is less intermittent on a national scale, becomes more favorable.

\subsubsection{Municipality level}
\begin{figure*}[htbp!]
  \centering
  \subfloat{\includegraphics[width=0.33\textwidth]{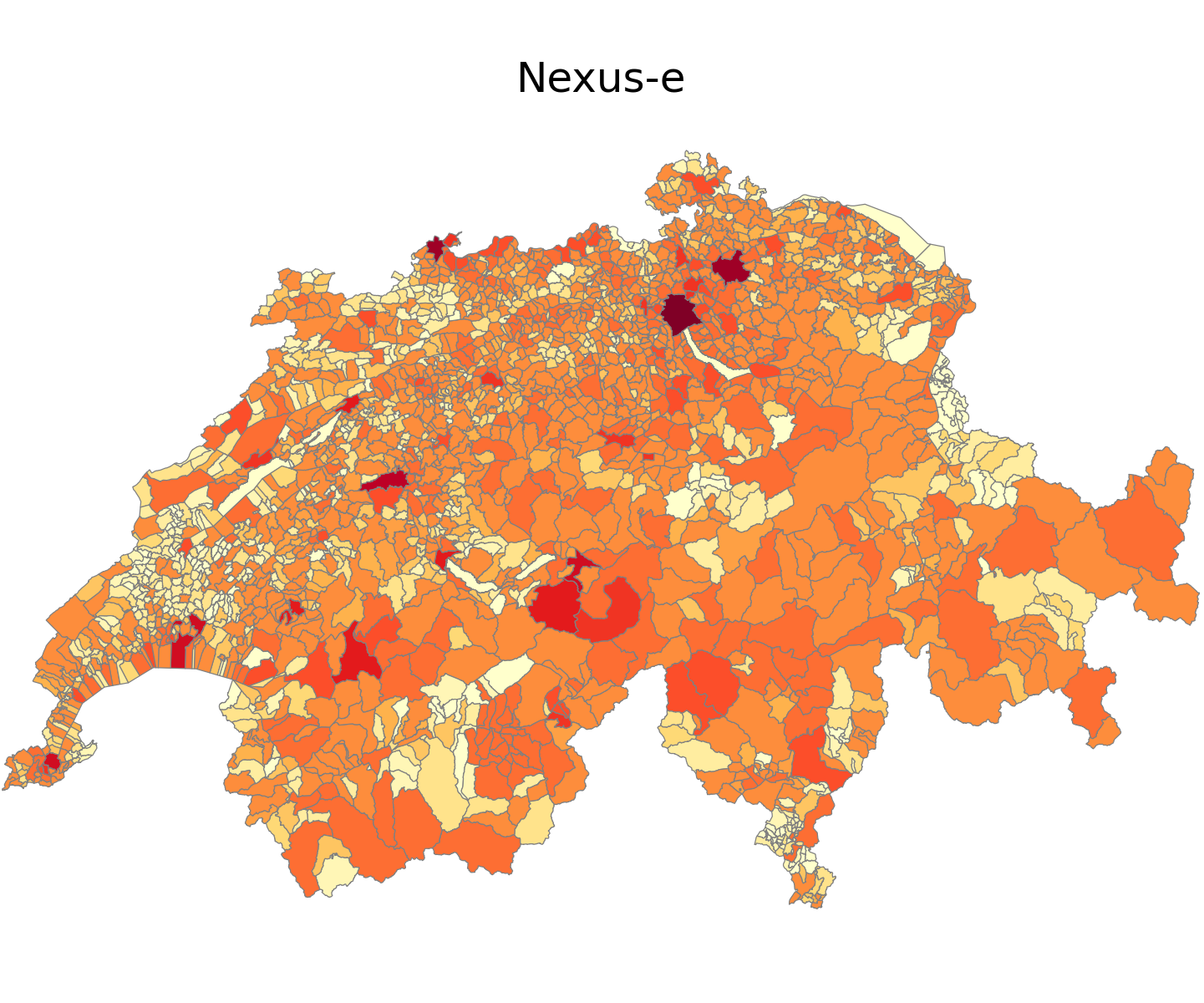}}
  % \hspace{0.15cm}
  \subfloat{\includegraphics[width=0.33\textwidth]{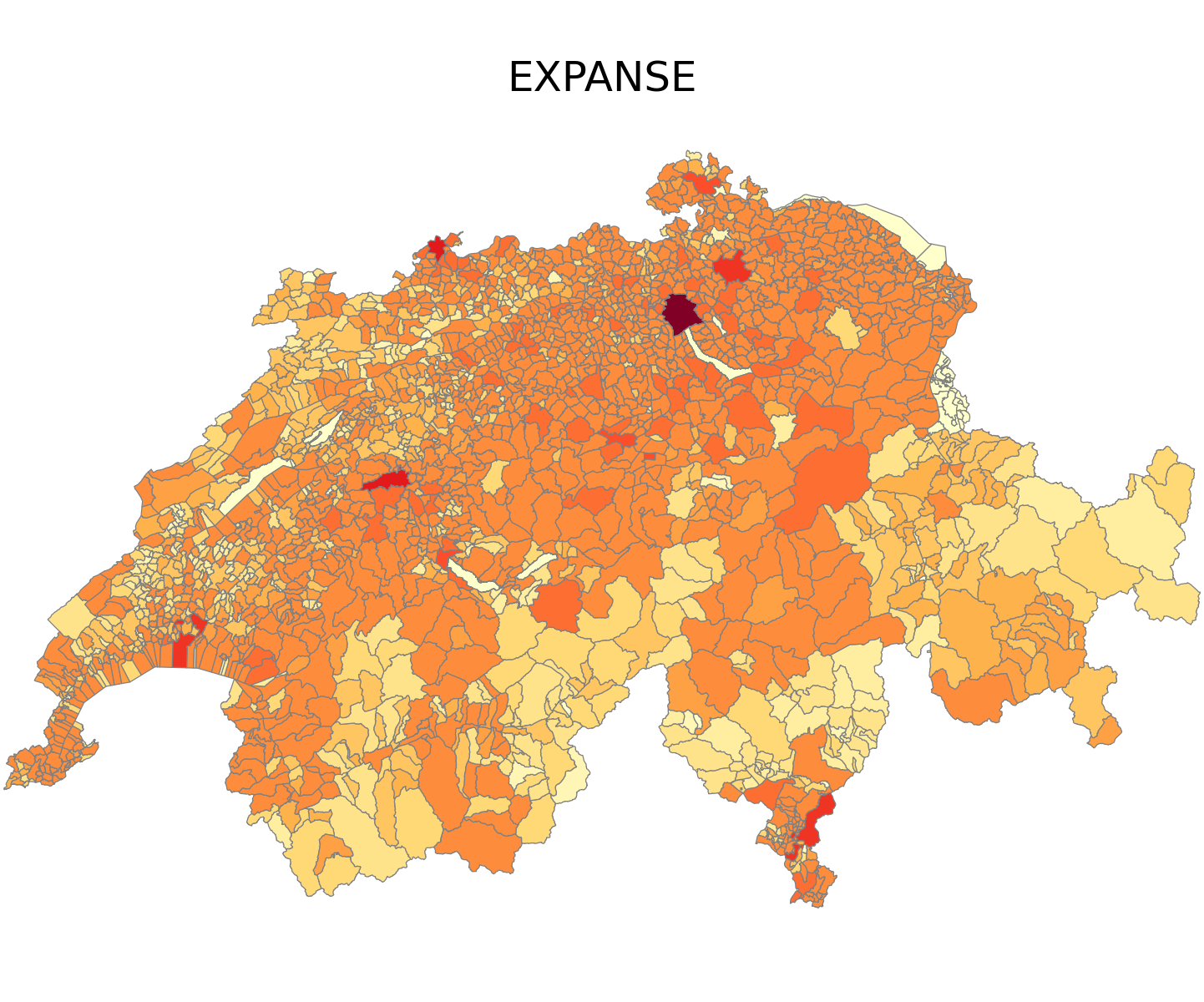}}
  % \hspace{-0.6cm}
  \subfloat{\includegraphics[width=0.33\textwidth]{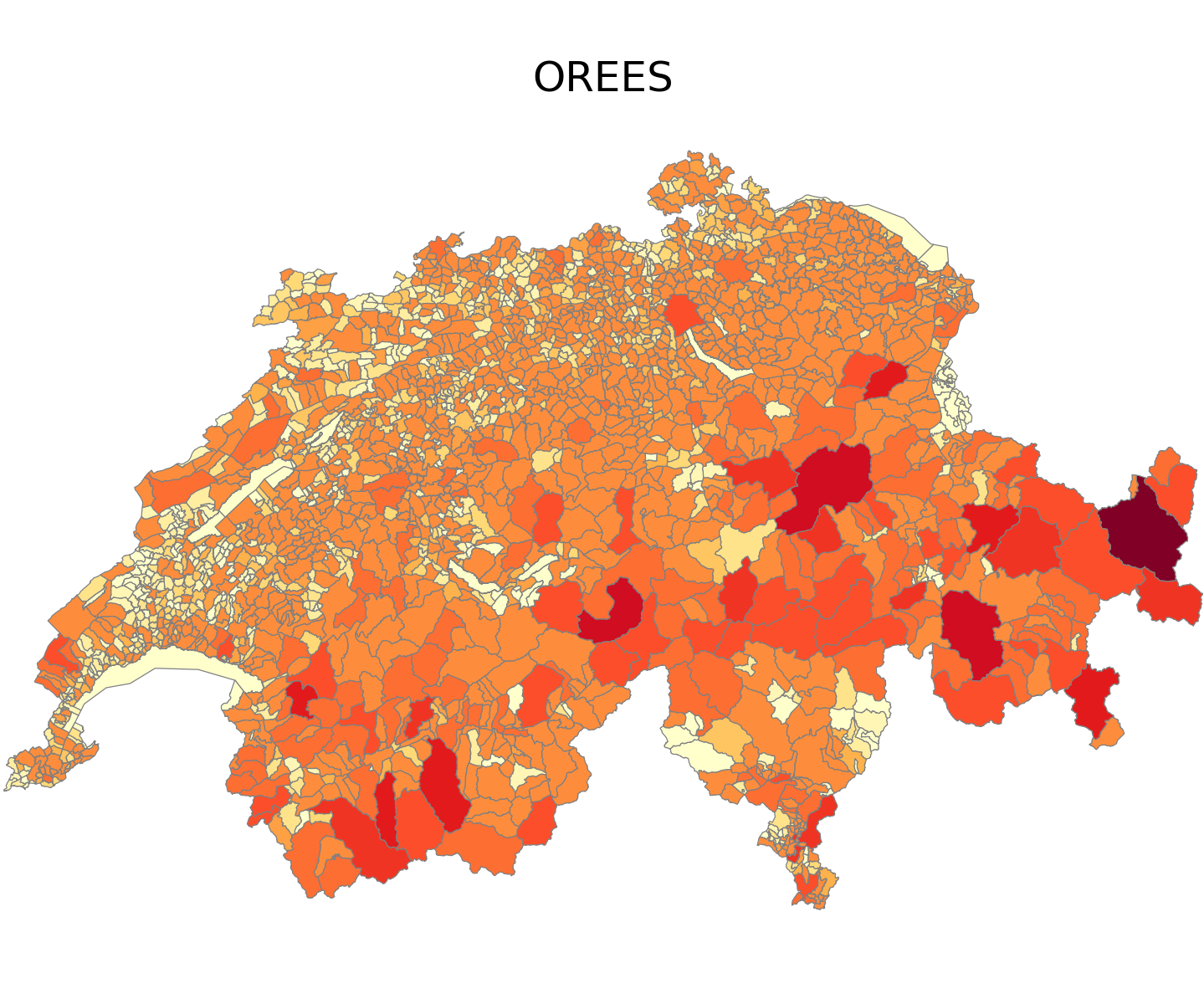}}

  \vspace{-0.3cm}
  \begin{minipage}[b]{0.1\textwidth}
  \centering
    \hspace{-1.7cm}
    \includegraphics[height=0.05\textheight, keepaspectratio]{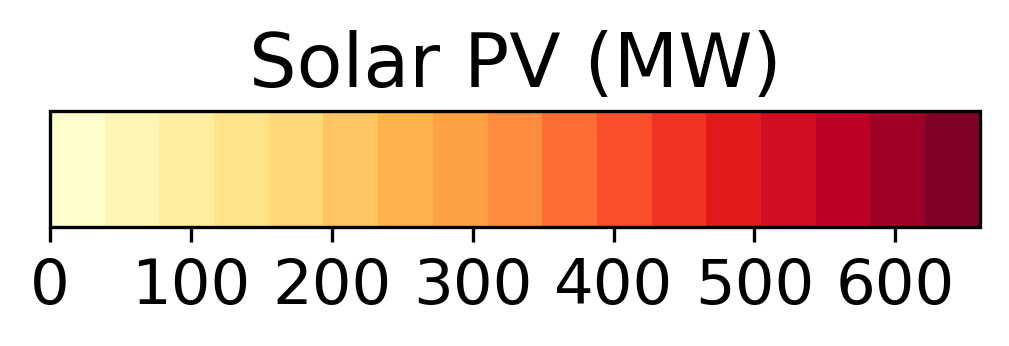}
  \end{minipage}
  
  \caption{Installed capacities of solar PV (rooftop, facade, and alpine) on municipality level for scenario R45.}
  \label{fig:municpality}

\end{figure*}

% Show a figure of four subplots for all four models in the selected scenario (to discuss)
The high spatial resolution of Nexus-e, EXPANSE, and OREES allows for a closer look at the spatial distribution of the investments in VRES technologies. 
Figure~\ref{fig:municpality} shows the PV (rooftop, facade, and alpine) investment on a municipality level for the R45 scenario across the models~\footnote{FEM results are not shown because it does not have municipality level resolution.}. 
% Comment on the differences between the model results
We observe important spatial differences in the optimal deployment of PV.
OREES  mainly in alpine PV, which has the advantage of higher generation potential, especially during winter.
Therefore, it sees high investment opportunities in the Alpine municipalities.
EXPANSE, instead, models only PV on buildings (rooftop and facade PV) and has, therefore, high investments in the urban areas where more building surface is available.
Finally, Nexus-e, suggests an intermediate solution, with high investments in urban areas, with higher population, demand, and rooftop areas, and in the Alpine region with higher potential.

%-----------------------------------------------------------------------------------------------
\subsection{Curtailment}\label{subsec:curtailment}
\begin{figure}[]
  \centering
  \begin{minipage}[b]{0.49\textwidth}
    \includegraphics[width=\textwidth]{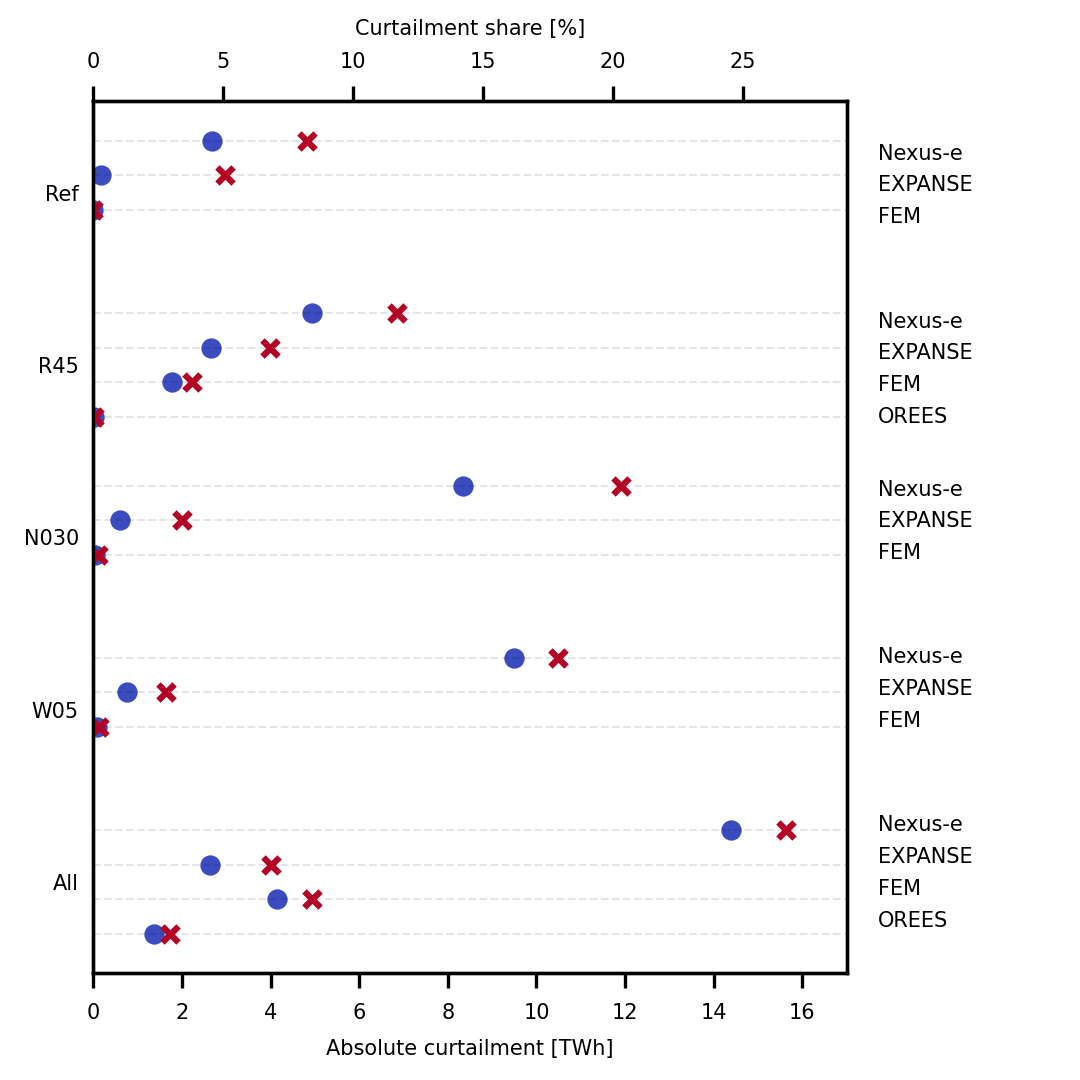}
  \end{minipage}
  
  % \vspace{-1.2cm}
  \begin{minipage}[b]{0.49\textwidth}
    \hspace{-0.6cm}
    \includegraphics[width=\textwidth]{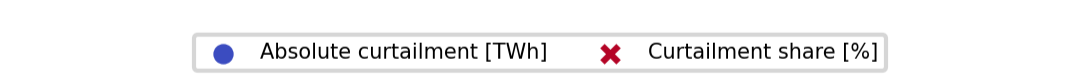}
  \end{minipage}
  % \vspace{0.5cm}

  \caption{VRES curtailment as a share of the VRES potential generation and absolute values in TWh.}
  \label{fig:curtailment}
\end{figure}

% Introduction
Due to the non-dispatchable nature of VRES, PV and wind generation often need to be curtailed to reduce their generation output if the excess electricity is not stored.
% Figure
Figure~\ref{fig:curtailment} illustrates the curtailments both as absolute values in TWh and as a share of the renewable potential generation.
% Explain why models differ so much
Results, however, differ much across models, suggesting that the efficient use of VRES strongly depends on the model perspective.
Two factors contribute to generation curtailment: power transmission limitations and the economic aspect.
On the one side, a more detailed grid representation can lead to more curtailment due to transmission grid congestions. 
Conversely, the minimum-cost objective function determines the optimal operations of the generation units, deciding whether to curtail or invest in storage for renewable generation.
The low curtailments in OREES' results suggest that curtailment is mainly driven by the models' economic perspective and not by grid congestion.
It shows that when VRES generation is prioritized over cost-optimal planning, less than 5\% of the potential generation is curtailed.
Curtailments due to distribution grid limitations are not considered here.

% R45 and N030
The policy measures with the highest impact on generation curtailment are the renewable energy target (R45) for EXPANSE and FEM and reduced market integration (N030) for Nexus-e.
% R45
The increase of installed renewable capacity in the R45 scenario leads to an increase in both the absolute curtailment and the curtailment share. 
This indicates that with higher renewable penetration, renewable generation is used less efficiently.
% N030
Reduced market integration (N030) increases generation curtailment by limiting the ability to export endogenous overproduction.
% W05
Curtailments also increase when imposing the winter constraint (W05) because of the limited trade during winter and the higher investments in VRES.
% Why nexus-e is so different?
Nexus-e has by far the highest curtailments in VRES. 
The reason is fourfold: 
1) The VRES generation profile in Nexus-e has a higher summer peak compared to EXPANSE and OREES, as shown in Figure ~\ref{fig_appen:VRES_profile} in \ref{App:Bappen},
2) The PV buildup in the DistIv module is agnostic to the market conditions and transmission grid power flows in CentIv, 
3) The optimal transmission expansion is computed from a centralized perspective and thus is agnostic to the distributed investment decisions, and 
4) This model considers the grid constraints for the full Swiss transmission grid, including the cross-border lines, leading to more constrained power flows.

%-----------------------------------------------------------------------------------------------
\subsection{Power exchange with the neighboring countries} % - focus on N030
\begin{figure*}[htbp!]
    \centering
    \includegraphics[width=1\linewidth]{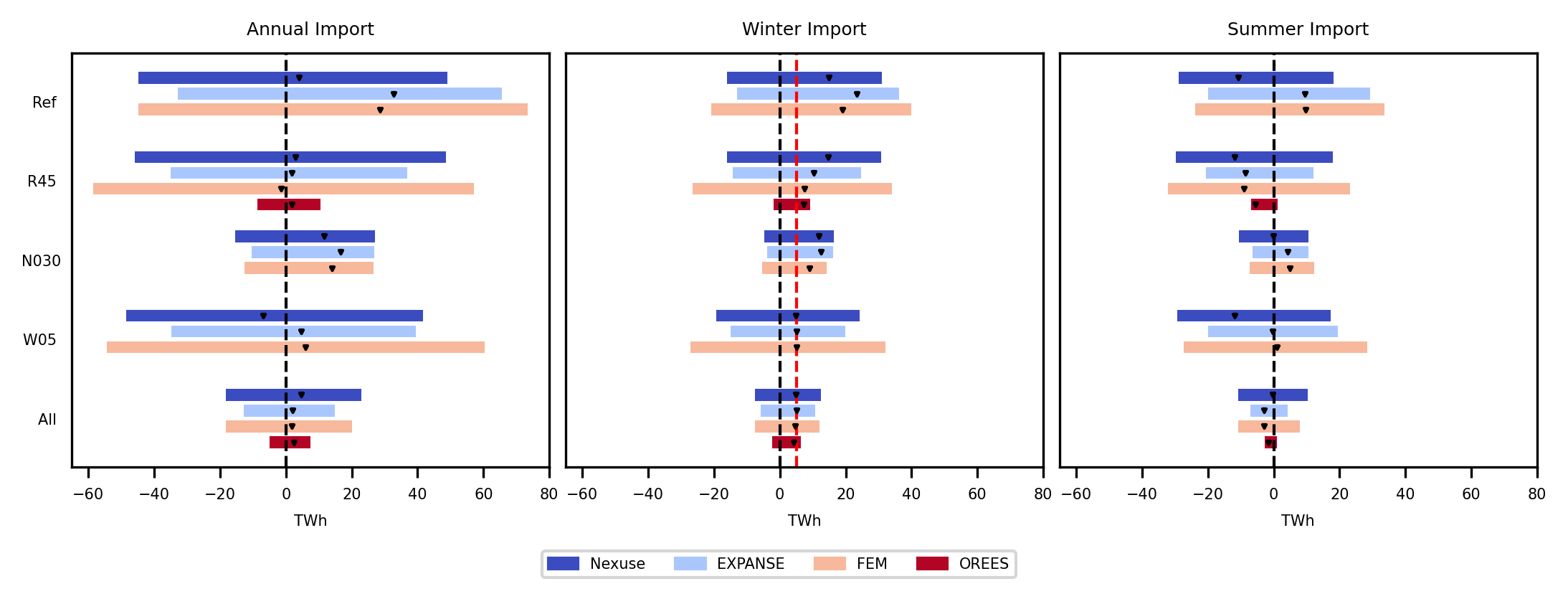}
    \caption{Imports, exports, and net imports over the full year, the winter half, and the summer half. The positive axis denotes imports; the negative axis denotes exports; the reversed triangle shows the net imports (imports - exports). The red dashed line corresponds to the 5 TWh of winter net import limit.}
    \label{fig:exchange}
\end{figure*}

% Introduction
Figure~\ref{fig:exchange} shows the annual, winter, and summer imports and exports for Switzerland. 
% The existing plots
% Ref
Annual net imports in the Ref scenario are higher for EXPANSE and FEM compared to Nexus-e due to the strong differences in endogenous generation.
% R45
The R45 policy scenario results in a decrease in the annual net imports. 
In fact, the renewable target mainly affects the summer months, when Swiss net imports are reduced by both decreasing imports and increasing exports.
This is enabled by the Swiss flexible generation and storage units, storing excess generation during the day and releasing it during the night hours. 
Thus, imports during the day are reduced due to national renewable generation, and exports are increased during the night, thanks to pumped hydro storage plants, batteries, and dams (Figure~\ref{fig_appen:typical_day} in \ref{App:Bappen}).

% N030: Reduced imports and exports. No clear effect on net imports.
The market integration reduction (N030) leads to smaller trading volumes, both for exports and imports, bringing net import values across models closer to each other. 
This suggests that the models fully utilize the existing options for trading. 
We observe an annual net import increase in the N030 compared to the Ref scenario for Nexus-e and a net import decrease for EXPANSE and FEM. 
In Nexus-e, this is due to a significant reduction in pumped storage usage, which limits summer exports. 
On the other hand, the net import decrease in EXPANSE and FEM is mainly due to the increased endogenous generation compared to the Ref scenario.

% W05: Reduction of the annual and winter imports
The winter import limit (W05), instead, has a stronger influence on the winter net imports, which are strictly limited to 5 TWh. 
This is achieved through a combination of reduced trading of Swiss flexibility (hydro pump storage), increased exports, and increased domestic generation from wind and PV.

%-----------------------------------------------------------------------------------------------
\subsection{Cost comparison} % - focus on W05
\begin{figure*}[htbp]
  % \centering
  \begin{minipage}[b]{0.31\textwidth}
    \includegraphics[height=0.31\textheight, keepaspectratio]{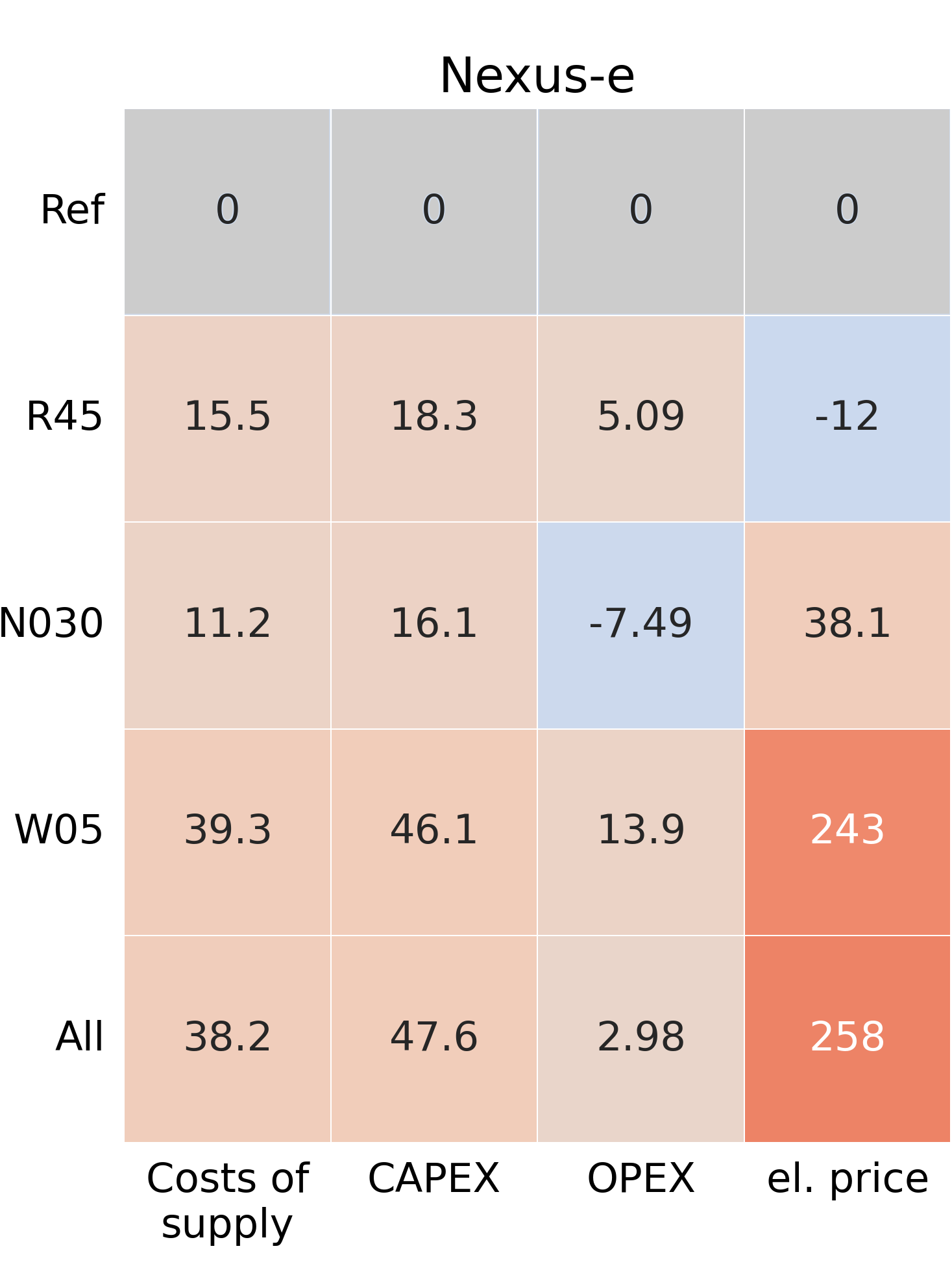}
  \end{minipage}
  \hspace{0.14cm}
  \begin{minipage}[b]{0.31\textwidth}
    \includegraphics[height=0.31\textheight, keepaspectratio]{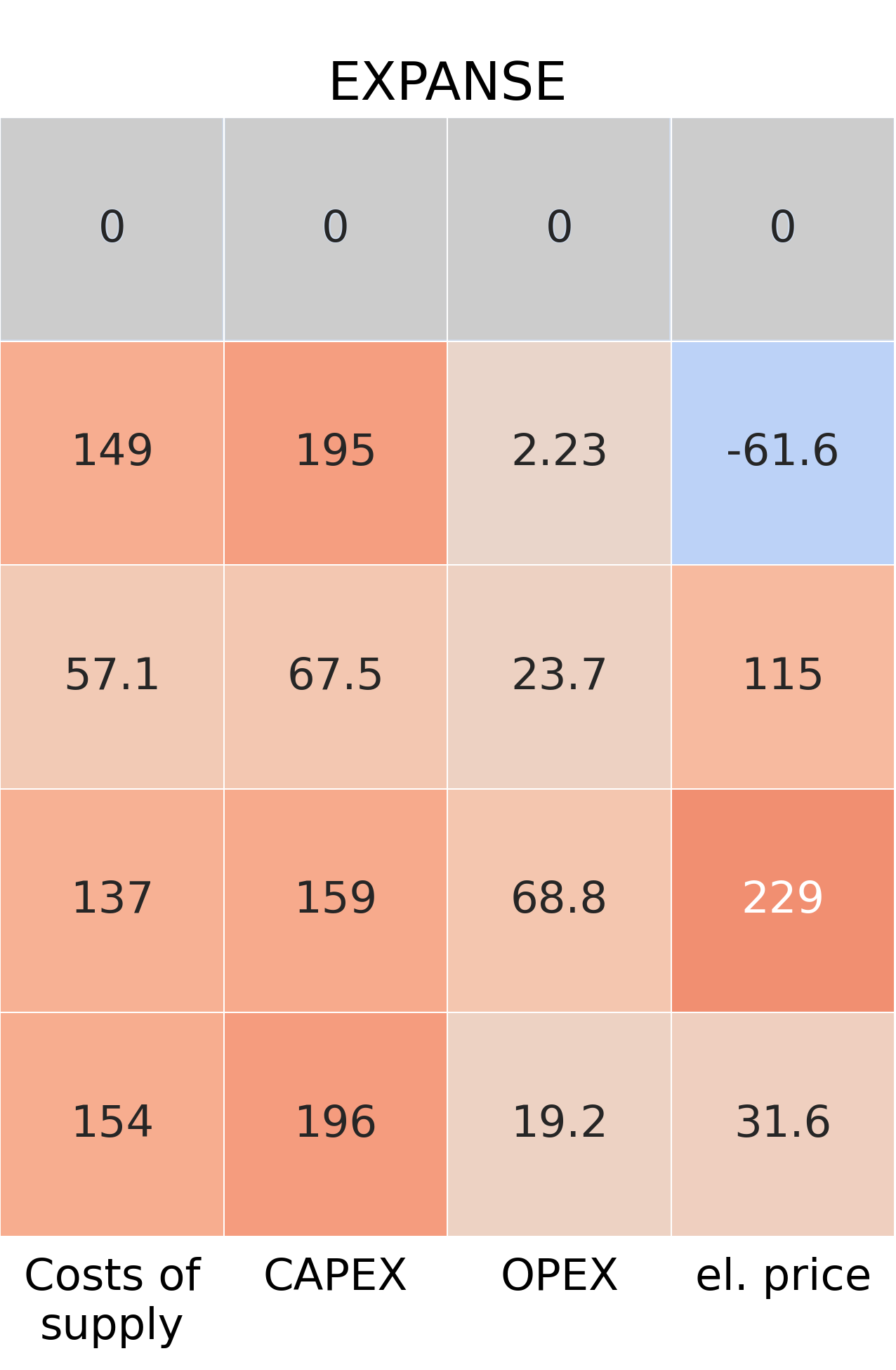}
  \end{minipage}
  \hspace{-0.6cm}
  \begin{minipage}[b]{0.31\textwidth}
    \includegraphics[height=0.31\textheight, keepaspectratio]{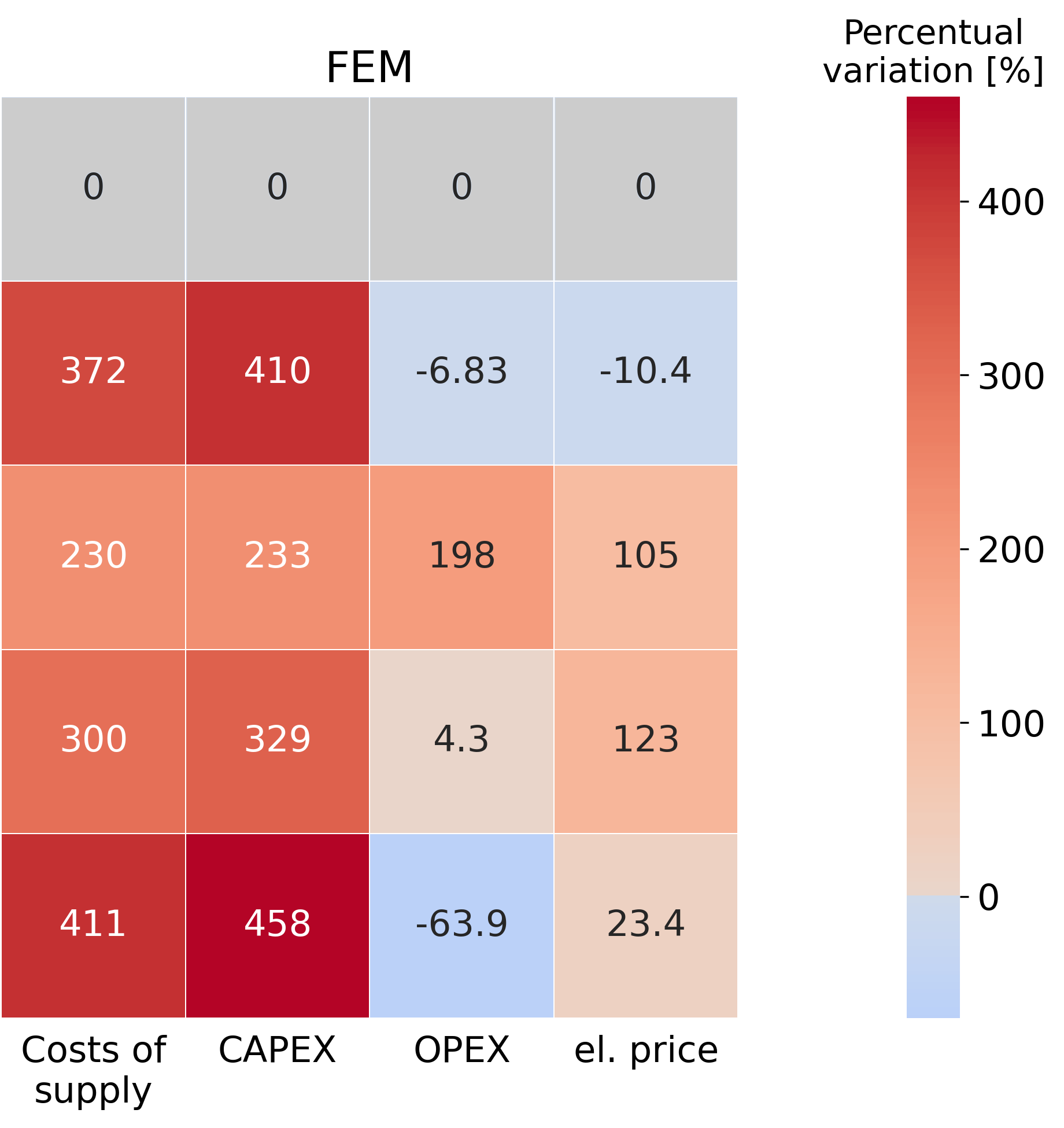}
  \end{minipage}

  \caption{Percentual variation of annualized Cost of supply, annualized investment costs (CAPEX), yearly operational costs (OPEX), and average yearly electricity price (el. price) for scenarios R45, N030, W05, and All, compared to the Ref scenario.}
  \label{fig:costs}

\end{figure*}

% Introduction to costs
Figure~\ref{fig:costs} illustrates the percent variation of the total cost of electricity supply, investment costs, operation costs, and electricity prices for the analyzed policy scenarios compared to the Ref scenario.
%
% Overview
The policy measures affect the costs and electricity prices for the different models in distinct manners.
However, common trends can be identified.
The large differences across models in the Ref scenario drive the large variations in the numeric results shown here.
In particular, we observe that FEM experiences a very high variation of CAPEX, and thus of electricity supply cost, that go up to 458 and 411\%, respectively.
These results are therefore complemented with Figure~\ref{fig:costs_all_abs_GA} in ~\ref{App:Bappen}, which shows the absolute variation of the same result parameters.
In particular, it shows that the absolute cost variations yield comparable values across the models. 
% R45
The renewable target (R45) increases the investment costs while reducing the operational costs and the electricity prices across all models.
The cheap distributed renewable generation, in fact, lowers the nodal electricity prices.
% R45 subsidies
Concurrently, to achieve the 45 TWh target, significant subsidies are needed. 
Table~\ref{tab:res_subsidy} shows the subsidy level for each model and scenario needed to achieve the renewable target of 45 TWh, i.e. the shadow price or the dual variable of the renewable constraint~\footnote{Due to the different setup of OREES, no dual variables are available.}.
Subsidy levels for FEM and EXPANSE are fairly consistent across scenarios ranging from 24 to 38 Eur per MWh of renewable generation~\footnote{This assumes a policy target that does not differentiate between different technologies. If well designed, differentiated subsidies could result in lower overall subsidy cost.}. 
In line with the high buildup of renewable generation in Nexus-e without a target that is mostly driven by households facing fixed grid fees, the subsidy level is comparably low for almost all scenarios with Nexus-e. 
However, it has to be noted that fixed grid fees and the resulting self-consumption incentives could be considered as a hidden subsidy to households.
The total subsidy cost needed to achieve the renewable target ranges from 1.1 to 1.7 billion Eur per year.

% Table 2: Renewable subsidy
\begin{table*}[h!]
\centering
\caption{Levels of renewable subsidy for the different scenarios and models [Eur/MWh].}
\footnotesize
\begin{tabular}{@{}lrrrrrr@{}}
\toprule
\textbf{EU development} & \textbf{Renewable target} & \textbf{Market integration} & \textbf{Winter net import} & \textbf{Nexus-e} & \textbf{FEM} & \textbf{EXPANSE} \\
\textbf{} & \textbf{[TWh]} & \textbf{[\%]} & \textbf{[TWh]} & \textbf{[Eur/MWh]} & \textbf{[Eur/MWh]} & \textbf{[Eur/MWh]} \\
\midrule
\textbf{GA} & 45 & 100 & NC & 18    & 29 & 37 \\ 
\textbf{GA} & 45 & 030 & NC & 14    & 28 & 38 \\ 
\textbf{GA} & 45 & 100 & 05 & 0     & 28 & 25 \\ 
\textbf{GA} & 45 & 030 & 05 & 0     & 28 & 27 \\ 
\textbf{DE} & 45 & 100 & NC & 28    & 38 & 36 \\ 
\textbf{DE} & 45 & 030 & NC & 0     & 33 & 34 \\ 
\textbf{DE} & 45 & 100 & 05 & 0     & 33 & 24 \\ 
\textbf{DE} & 45 & 030 & 05 & 0     & 33 & 27 \\ 
\bottomrule
\end{tabular}
\label{tab:res_subsidy}
\end{table*}

% N030
A reduction in market integration (N30) leads to cost and electricity price increases for Nexus-e, EXPANSE, and FEM.
The OPEX reduction in Nexus-e is caused by a concurrent increase in net imports and a decrease in endogenous generation compared to the reference scenario (Ref).

% W05 - general
The winter net import limit (W05) causes a substantial increase in costs and electricity prices across all models.
% W05 - investment
Limiting the winter net imports increases the necessary investments for endogenous generation (see Figure~\ref{fig: generation}). 
The percentual difference in CAPEX is larger for EXPANSE and FEM, compared to Nexus-e, since their CAPEX in the Ref scenario is significantly lower.
However, Nexus-e also presents a substantial increase in CAPEX costs, showing that the decentralized PV generation is not sufficient to meet such a policy target, and additional capacity for other technologies is needed to increase the winter generation.
% W05 - operation
Operational costs are increasing as well. 
This is due to the need for more operationally expensive generating technologies, such as biomass, waste, and fossil-based power generation.
% W05 - electricity price
The more expensive generation in Switzerland and the reduced availability of cheap imports from the neighboring countries also affect the electricity price, increasing it up to 243\% of the reference value.
In this study, the winter net import constraint is implemented as a binding target in all models. 
This can be imagined as a form of a certificate market where each MWh traded into and out of Switzerland in winter is subject to a certificate price. 
In the models used for this study, this certificate price can be derived from the shadow price or the dual variable of the net-winter import constraint~\footnote{OREES does not provide those dual variables.}. 
Nexus-e has fairly consistent results for the certificate cost ranging from 59 to 67 Euros per MWh, depending on the scenario (see Table \ref{tab:res_w05}). 
\begin{table*}[h!]
\centering
\caption{Certificate price corresponding to the 5 TWh winter net import limit [Eur/MWh].}
\footnotesize
\begin{tabular}{@{}lrrrrrr@{}}
\toprule
\textbf{EU development} & \textbf{Renewable target} & \textbf{Market integration} & \textbf{Winter net import} & \textbf{Nexus-e} & \textbf{FEM} & \textbf{EXPANSE} \\
\textbf{} & \textbf{[TWh]} & \textbf{[\%]} & \textbf{[TWh]} & \textbf{[Eur/MWh]} & \textbf{[Eur/MWh]} & \textbf{[Eur/MWh]} \\
\midrule
\textbf{GA} & NT & 100 & 05 & 59 & 73 & 124 \\ 
\textbf{GA} & 45 & 100 & 05 & 59 & 3 & 42 \\ 
\textbf{GA} & NT & 030 & 05 & 63 & 44 & 123 \\ 
\textbf{GA} & 45 & 030 & 05 & 63 & 0 & 42 \\ 
\textbf{DE} & NT & 100 & 05 & 60 & 86 & 109 \\
\textbf{DE} & 45 & 100 & 05 & 60 & 17 & 42 \\ 
\textbf{DE} & NT & 030 & 05 & 64 & 76 & 110 \\
\textbf{DE} & 45 & 030 & 05 & 67 & 0 & 31 \\
\bottomrule
\end{tabular}
\label{tab:res_w05}
\end{table*}
This consistency is closely linked to the high initial renewable buildup under the reference scenario (Ref) in Nexus-e. 
This results in lower impacts of the 45 TWh VRES target or the reduced import capacity scenario. 
On the other hand, results from both FEM and EXPANSE show a large variety between scenarios ranging from 0 to 124 Euros per MWh (see Table \ref{tab:res_w05}). 
In FEM, the 45 TWh target almost fully mitigates the impact of the net winter import limit of 5 TWh. 
EXPANSE yields the highest cost for import certificates but is also subject to a significant price decrease under a 45 TWh renewable target. 
The reduced import capacity has only a minor effect on certificate prices in both models.

%-----------------------------------------------------------------------------------------------
\subsection{Operations under limited winter net imports}
\begin{figure}[]
    \centering
    \includegraphics[width=0.5\textwidth]{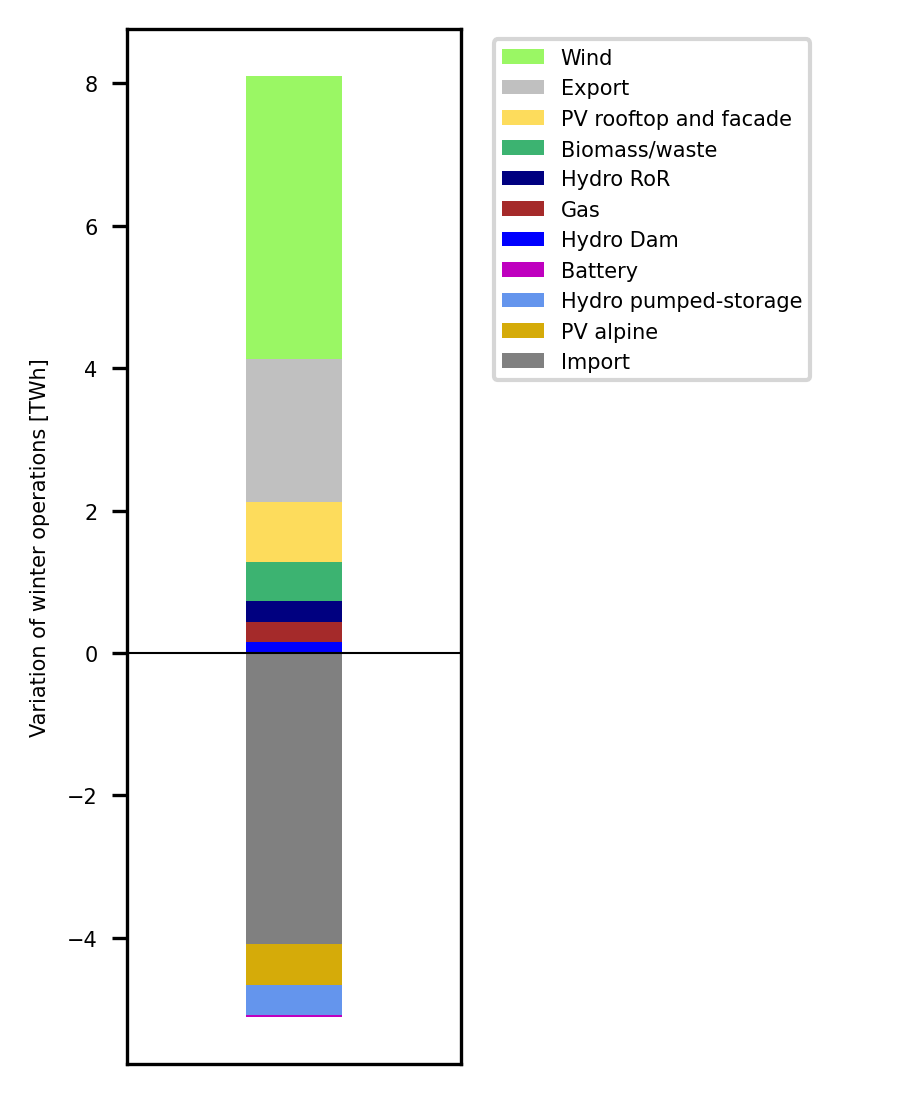}
    \caption{The variation in winter operations (in TWh) due to the winter net import limitation as an average across models and scenarios. Positive values indicate that there is an increase in operations for those technologies. Negative values show that there is a reduction in operations.}
    \label{fig:W05_operations}
\end{figure}

% Why we need to have a closer look at winter operations
The winter net import limitation has important implications for the generation mix, yearly operations, curtailments, and cost of supply as well as for the operations in the winter months.
% Introduce the figure
Figure \ref{fig:W05_operations} shows the average impact of the winter net import limit on the operations of different generation technologies in the winter months in terms of the differences in winter generation as percent variations across all models and policy scenarios.
% What does the figure tell us
Several technologies are affected by the winter net import limitation.
PV, wind, and biomass increase their generation under this policy measure, while hydro-pumped storage reduces its operations.
% Focus on imports, exports and pumped hydro
The increase in generation from PV, wind, and biomass is due to the larger installed capacities of these technologies. 
The underlying effect is that both import reductions but also export increases can be employed to meet the winter net import limitation. 
This implies that additional capacities are added not only based on whether they can provide electricity when it is needed in the Swiss system but also when it can be exported to neighboring countries. 
The reduction in hydro-pumped storage usage, instead, is a purely operational decision.
Switzerland has a high share of pump storage capacity and its flexibility is used for arbitrage on other European markets by importing at low price hours and exporting at high price hours. 
Such arbitrage is strongly penalized in winter months under the winter net import constraint because pump storage arbitrage inherently involves energy losses, meaning that any pump storage activity increases net imports.
Potentially unintended effects result from the net winter import limit on top of the high additional cost of Swiss electricity supply due to the certificate market.

%-----------------------------------------------------------------------------------------------
\subsection{Robustness of results}
\begin{figure*}[]
    \centering
    \includegraphics[width=1\linewidth]{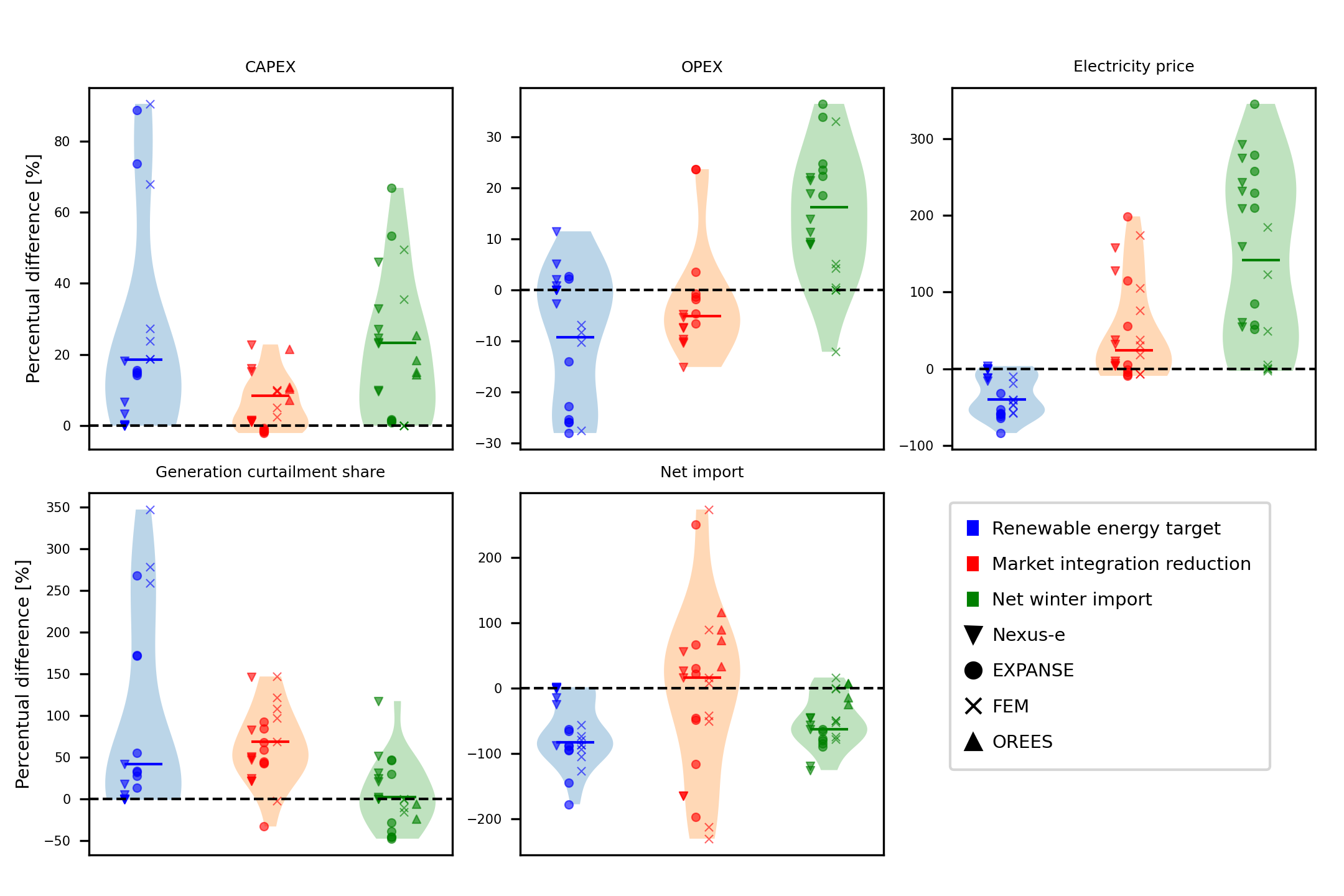}
    \vspace{-0.5cm}
    \caption{Impact of the three policy measures on CAPEX, OPEX, electricity prices, generation curtailment share, and net import as percent variation. For each policy measure, the results of a scenario with the active policy are compared to those of the same scenario without that policy. Extreme values are omitted for better readability but considered in the computation of the median value (horizontal lines).}
    \label{fig:violin}
\end{figure*}

% Why do we need to look at the robustness
% This section summarizes results for all scenarios and across all models to better understand where results agree or diverge.
% Introduction to the plot: explain how it is computed
Figure~\ref{fig:violin} shows the impact of the three policy measures on CAPEX, OPEX, electricity price, generation curtailment share, and net imports as percent variation for all scenarios and across
all models.
For each policy measure, the results of a scenario with the active policy are compared to those of the same scenario without that policy.
Extreme values are omitted for better readability but considered in the computation of the median value.
% Give a general overview of where models agree and disagree - R45 and W05 strong impact. N030 rather unsure
% R45
The renewable target (R45) consistently leads to increased CAPEX and generation curtailment share, as well as decreased OPEX, electricity prices, and net imports. % OPEX variation for Nexuse is 0 or -2%
Overall, the renewable target has a strong impact on most of the presented results.
Interestingly, for Nexus-e, only two scenarios are affected by this policy measure since, in most cases, the renewable generation already exceeds 45 TWh due to assumed incentives in the model for PV self-consumption.

% N030
The reduced market integration (N30) leads to divergent results across models and scenarios.
It has a clear impact on the generation curtailment share, which increases due to reduced trading availability.
The reduced trading also tends to reduce exports, thus increasing net imports.
In those scenarios with reduced exports and increased curtailment, internal generation is lower, thus lowering total OPEX.
This is reflected in the OPEX per unit of generated electricity, which, instead, increases when market integration is restricted, with a median value of +3.5\% (Figure \ref{fig:violin_relopex} in \ref{App:Bappen}).
Additionally, results point towards a likely increase in CAPEX and electricity prices.

% W05
The winter net import limit (W05) decreases the net imports and leads to increased CAPEX, OPEX, and Swiss electricity prices.
Interestingly, this policy measure has no clear impact on the curtailment of VRES.
This is attributed to the fact that satisfying this strict constraint requires a different generation mix, with higher shares of wind power and less PV.
Thus, although VRES capacity is high, there is relatively less overproduction in summer and, thus, less generation curtailment.

%%%%%%%%%%%%%%%%%%%%%%%%%%%%%%%%%%%%%%%%%%%%%%%%%%%%%%%%%%%%%%%%%%%%%%%%%%%%%%%%%%%%%%%%%%%%%%%%
% DISCUSSION
%%%%%%%%%%%%%%%%%%%%%%%%%%%%%%%%%%%%%%%%%%%%%%%%%%%%%%%%%%%%%%%%%%%%%%%%%%%%%%%%%%%%%%%%%%%%%%%%
\section{Policy-relevant findings}
\label{sec:discussion}
%---------------------------------------------------------------------
% \subsection{Policy-relevant findings}

% Discuss the takeaways with respect to research questions and others
% 1) RES target
% Ambra to write
\textit{Effects of a renewable generation target:}
% Big influence on all analyzed output parameters when a centralized perspective is used
The renewable generation target has a significant impact on all analyzed output when a centralized perspective is used.
% Decentralized perspective (+ injection tariff): rooftop PV is more appealing and R45 has influence only in few scenarios
When a decentralized perspective is adopted with self-consumption incentives and fixed grid fees (as in Nexus-e), the optimal investments lead to more than 45 TWh of annual potential VRES generation even when the renewable policy is not applied.
The policy is thus irrelevant for Nexus-e in most of the scenarios.
% Investments and generation mix -> several distinct optimal solutions depending on model assumptions and input data
Results show that with the renewable target, the optimal generation mix varies across models, stressing the importance of the underlying assumptions and input data while demonstrating that there are different solutions to meet the renewable target.
% Exchange with neighbors
Additionally, with the target in place, the electricity exchange with the neighbors changes towards lower net imports due to more domestic production and more exports of low-value electricity. 
Batteries and pumped storage charge during hours of excess PV generation and discharge when less renewable generation is available.
Similarly, hydro dams also shift their generation to the night hours.
% Cost increase vs electricity price decrease -> comment on dual variable and subsidy level
Investing in more renewable generation comes with a median investment cost increase of 19\% and a median electricity price decrease of 40\%. 
%
% Add a wrap-up conclusion/statement
In three out of four models, the target must be enforced through significant subsidies to achieve the expected VRES shares. 
Only one model achieves the target without it being enforced. 
This suggests that to achieve the target, one has to account for the potential self-consumption incentives of the decentralized investors.
Finally, the 45 TWh is higher than a purely economical optimal renewable buildup for Switzerland. 
Hence, significant direct or indirect subsidies would be needed. 

% 2) The utilization of RES (i.e., generation curtailment)
% Ambra to write
\textit{Curtailment of VRES:}
The higher shares of VRES installed under all three policy measures are, in most cases, accompanied by higher shares of curtailed generation.
Renewable curtailments have to be distinguished into economic curtailments and grid-based curtailments. 
The former is part of the system's optimal solution. 
In this case, the curtailed generation has zero system value, and would otherwise require additional storage beyond the system optimal level of storage. 
The latter results from insufficient grid capacity.
The results indicate that economic curtailments play a significant role when integrating high shares of VRES.
Reduced market integration constrains the exchange with the neighboring countries, thus limiting the use of VRES generation and increasing curtailments.
% OREES proves that even with high grid detail, it is possible to operate the system such that less than 5% VRES generation is curtailed.
% Putting a focus on income optimization for wind and solar producers rather than reducing overall costs of supply results in lower levels of curtailment, as in OREES' results. 
% At the same time, this curtailment results from an overall increased cost of supply due to a less cost-efficient use of other system resources compared to a cost-minimization approach. 
% When there are large penetrations of VRES, it is more a question of economics: is it worth storing?
%
% W05 on curtailment
Interestingly, limiting the winter net imports has no clear impact on the curtailment due to the shift towards more wind power in the generation mix.
% Add a wrap-up conclusion/statement
In summary, we find that VRES curtailments are part of an efficient system with high VRES shares. However, reduced market integration is an important hurdle for the Swiss and European energy transition.

% 3) Effects of NTC 
% Ambra, Blazhe, Jonas (focus on econmoc side)
\textit{Effects of reduced market integration:}
Reducing the integration in the European electricity market has diverging effects across models and scenarios.
Strikingly this policy does not reduce net imports, although it leads to reduced trading volumes.
Additionally, it has a negative impact on CAPEX, electricity prices, and curtailments. 
% Add a wrap-up conclusion/statementsuch 
Overall, such a significant reduction in trading capacity harms the Swiss and the European energy transition by constraining international trade and thus reducing the utilization of VRES.

% 4) Effects of winter constraint
\textit{Effects of winter constraint:} 
% W05 - 
The winter constraint has the highest impact on the cost of supply and electricity prices. 
More expensive generation units are required to strictly limit the net winter imports to 5 TWh.
Additionally, unintended effects play an important role in the electricity trade with the neighboring countries and the deployment of flexible generation in Switzerland.
% Add a wrap-up conclusion/statement
While the overall aim of the winter net import limit is to reduce import dependency, there might be better suited instruments - for example, a well-designed capacity mechanism or security reserve. 
Identifying a fitting mechanism for Switzerland should be subject to further research.

\section{Conclusions}
\label{sec:conclusions}

% Brief description of what is done (3 sentences)
In this work, we compare four highly resolved electricity system models for Switzerland in 2050. 
The models use a common scenario setup with TYNDP-2022 scenarios as boundary conditions for Europe. 
The models align on implementing a renewable generation target, a reduction in NTC, and a limit on the winter net imports to make the model comparison compatible with the main research questions.

% Summarize the results: Are the research questions answered? - Add one to two sentences for each research question.
% RQ 1
The renewable generation target has a major influence on the electricity system, leading to increased VRES installed capacity and, therefore, increased CAPEX costs of about 19\%.
However, it also leads to reduced operational costs (-9\%), electricity prices (-40\%), and net imports (-83\%). 
In general, we find that in three out of four models, the target must be enforced to achieve the expected VRES shares. 
This suggests that direct or indirect subsidies are needed to achieve a large buildup of renewable generation.
% RQ 2
We find that VRES curtailments are part of an efficient system with high VRES shares.
We also observe that the renewable generation target and the limited trading lead to higher generation curtailments, while the limited net imports in winter have a less decisive impact on curtailments.
% RQ 3
The study shows that limiting international trading capacities has a less clear effect on the cost of electricity supply and the electricity price, showing divergent outcomes across models and scenarios.
However, reduced trading capacity negatively affects the Swiss and the European energy transition by reducing the potential benefits of trading and hence reducing the utilization of VRES while increasing the cost of electricity supply.
% RQ 4
Finally, limiting the net winter imports leads to both a reduction of imports as well as an increase in exports to meet the target. 
Additional capacities are added to support exports to neighboring countries and thus maintain the 5TWh net import target.
Overall, limiting the net winter imports harms electricity trading leading to increased costs of electricity supply (increasing CAPEX by 23 and OPEX by 16\%) and electricity prices (141\%).

% Future work (3 sentences)
In the future, we aim to further improve the models and data assumptions. 
Therefore, we will repeat the inter-comparison with a strong focus on the parameters leading to high result variability across all models.
Consequently, the goal will be to identify pathways less prone to input parameter uncertainties.
Additionally, future work will investigate import dependency and security of supply under different scenarios. 
Although import dependency is an often-discussed topic in Switzerland, no clear quantifications have been performed to our knowledge.

%%%%%%%%%%%%%%%%%%%%%%%%%%%%%%%%%%%%%%%%%%%%%%%%%%%%%%%%%%%%%%%%%%%%%%%%%%%%%%%%%%%%%%%%%%%%%%%%
% CRediT authorship contribution statement
%%%%%%%%%%%%%%%%%%%%%%%%%%%%%%%%%%%%%%%%%%%%%%%%%%%%%%%%%%%%%%%%%%%%%%%%%%%%%%%%%%%%%%%%%%%%%%%%
\section*{CRediT authorship contribution statement}
\textbf{Ambra Van Liedekerke:} Writing – original draft, Visualization, Investigation, Formal analysis, Software, Conceptualization.
\textbf{Blazhe Gjorgiev:} Writing – original draft, Investigation, Conceptualization.
\textbf{Jonas Savelsberg:} Writing – original draft, Investigation.
\textbf{Xin Wen:} Writing – review \& editing, Software, Investigation.
\textbf{Jerome Dujardin:} Writing – review \& editing, Software, Investigation.
\textbf{Ali Darudi:} Writing – review \& editing, Software, Investigation.
\textbf{Jan-Philipp Sasse:} Software.
\textbf{Evelina Trutnevyte:}  Writing – review \& editing, Supervision, Funding acquisition.
\textbf{Michael Lehning:} Writing – review \& editing, Supervision, Funding acquisition.
\textbf{Giovanni Sansavini:} Writing – review \& editing, Supervision, Conceptualization, Funding acquisition.

%%%%%%%%%%%%%%%%%%%%%%%%%%%%%%%%%%%%%%%%%%%%%%%%%%%%%%%%%%%%%%%%%%%%%%%%%%%%%%%%%%%%%%%%%%%%%%%%
% Declaration of Competing Interest
%%%%%%%%%%%%%%%%%%%%%%%%%%%%%%%%%%%%%%%%%%%%%%%%%%%%%%%%%%%%%%%%%%%%%%%%%%%%%%%%%%%%%%%%%%%%%%%%
\section*{Declaration of Competing Interest}
The authors declare no competing interests.

%%%%%%%%%%%%%%%%%%%%%%%%%%%%%%%%%%%%%%%%%%%%%%%%%%%%%%%%%%%%%%%%%%%%%%%%%%%%%%%%%%%%%%%%%%%%%%%%
% Declaration of use of generative AI
%%%%%%%%%%%%%%%%%%%%%%%%%%%%%%%%%%%%%%%%%%%%%%%%%%%%%%%%%%%%%%%%%%%%%%%%%%%%%%%%%%%%%%%%%%%%%%%%
\section*{Declaration of generative AI and AI-assisted technologies in the writing process}
During the preparation of this work the author(s) used ChatGPT in order to improve the text. After using this tool, the author(s) reviewed and edited the content as needed and take(s) full responsibility for the content of the published article.

%%%%%%%%%%%%%%%%%%%%%%%%%%%%%%%%%%%%%%%%%%%%%%%%%%%%%%%%%%%%%%%%%%%%%%%%%%%%%%%%%%%%%%%%%%%%%%%%
% ACKNOWLEDGEMENTS
%%%%%%%%%%%%%%%%%%%%%%%%%%%%%%%%%%%%%%%%%%%%%%%%%%%%%%%%%%%%%%%%%%%%%%%%%%%%%%%%%%%%%%%%%%%%%%%%
\section*{Acknowledgements}
\label{Acknowledgements}
The research published in this report was carried out with the support of the Swiss Federal Office of Energy SFOE as part of the SWEET consortium EDGE. The authors bear sole responsibility for the conclusions and results. We thank Boubat Matthieu for executing the Nexus-e computations.

%%%%%%%%%%%%%%%%%%%%%%%%%%%%%%%%%%%%%%%%%%%%%%%%%%%%%%%%%%%%%%%%%%%%%%%%%%%%%%%%%%%%%%%%%%%%%%%%
% APPENDIX
%%%%%%%%%%%%%%%%%%%%%%%%%%%%%%%%%%%%%%%%%%%%%%%%%%%%%%%%%%%%%%%%%%%%%%%%%%%%%%%%%%%%%%%%%%%%%%%%
\appendix
%-----------------------------------------------------------------------------------------------
\section{Model overview}%Scenario description
\label{App:Aappen}

Table~\ref{tab:model_summary} summarizes the main characteristics of the models utilized in this inter-comparison, namely Nexus-e, EXPANSE, FEM, and OREES. 

\begin{table*}[]
\caption{Overview of main characteristics of the electricity systems models used in this study.}
\resizebox{\textwidth}{!}{%
\begin{tabular}{@{}lllll@{}}
\toprule
\textbf{}               & \textbf{NEXUS-E}           & \textbf{EXPANSE}                                                         & \textbf{FEM}                                        & \textbf{OREES}                        \\ \midrule
Model type              & Linear Optimization        & Linear Optimization  & Quadratic Optimization                            & Evolution strategy (optimization) with optimal power flow     \\
Objective               & Total costs minimization   & Total costs minimization                                                 & Total costs minimization                            & PV and wind generator income maximization   \\
Perfect foresight       & yes                        & yes                                                                      & yes                                                 & yes                                     \\
Developed at            & ETH                        & UniGE                                                                    & ZHAW (and Uni Basel)                                & EPFL                                  \\
Model environment       & Matlab, Python             & Python                                                                   & Python                                              & Matlab                                \\ \midrule
Spatial resolution      & Cantons, municipalities    & Municipalities                                                           & 7 Greater regions (Grossregionen)                   & 1.6 x 2.3km (PV),  1.1km grid (Wind), \\
Temporal resolution     & 1 hour                     & 6 hours                                                         & 1 hour                                              & 15 min / 1 hour                       \\
Temporal scope          & 1 year                     & 1 year                                                                   & 1 year                                              & 1 year                                \\ \midrule
Power system nodes      & 165                        & 15 in Switzerland, 19 in total  & 7 CH, 1 node per neighbor of CH and their neighbors & 169                                   \\
Grid expansion          & Yes                        & Yes                                                                      & No                                                  & No                                    \\ \midrule
Storages                & Pumped hydro storage, Batteries                  & Pumped hydro storage, Batteries, Hydrogen                                                      & Pumped hydro storage, Batteries, Hydrogen                                 & Pumped hydro storage, Batteries                             \\
Demand-side management & Yes                        & No                                                                       & No                                                 & No                                    \\
Policy                  & PV subsidies, FiT, CO2 tax & No                                                                       & No                                                  & No                                    \\ \bottomrule
\end{tabular}}
\label{tab:model_summary}
\end{table*}

%-----------------------------------------------------------------------------------------------
\section{Additional results} % If needed, rename accordingly 
\label{App:Bappen}

\begin{figure}[]
  \centering
  \begin{minipage}[b]{0.49\textwidth}
    \includegraphics[width=\textwidth]{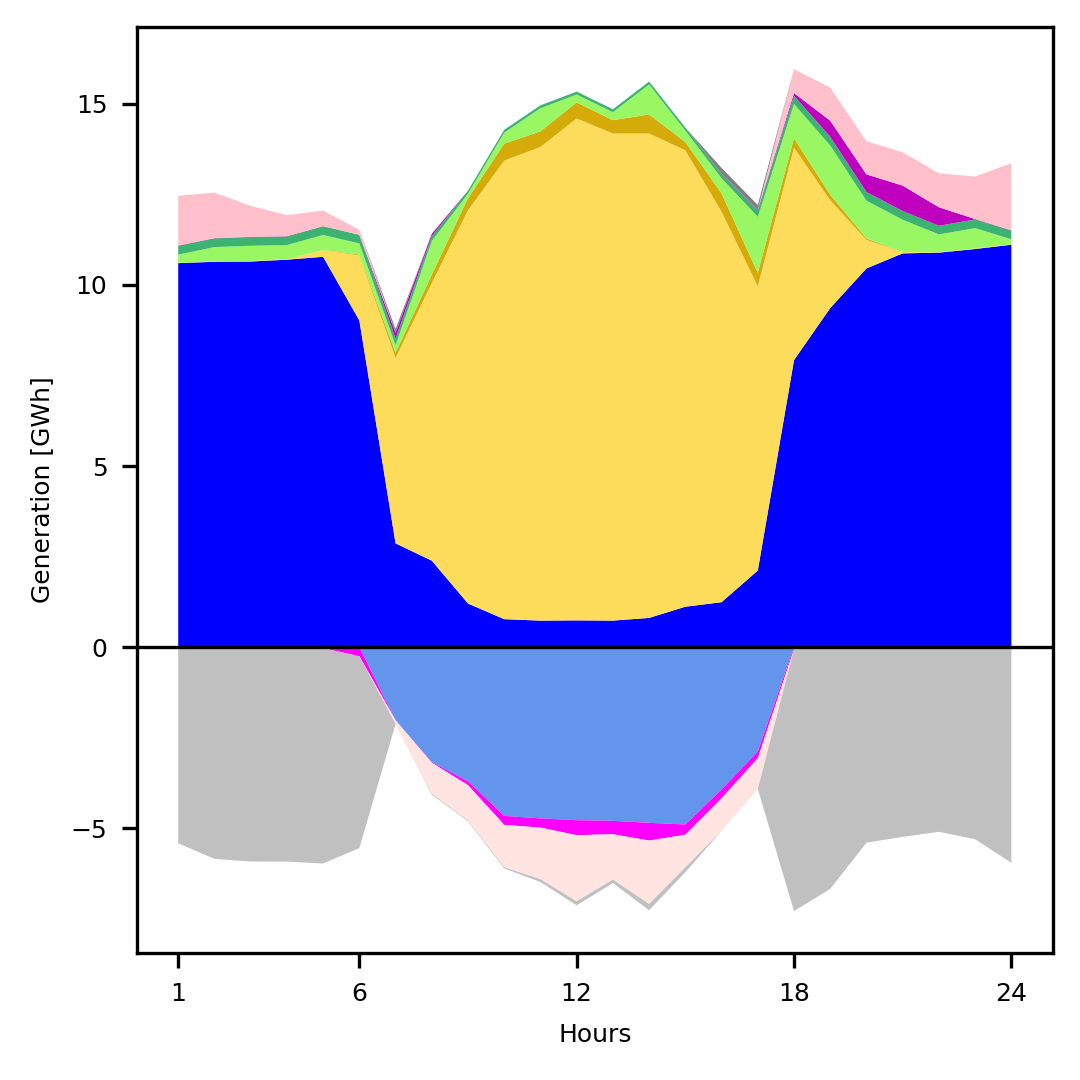}
  \end{minipage}
    
  \begin{minipage}[b]{0.49\textwidth}
    \hspace{0.4cm}
    \includegraphics[width=\linewidth]{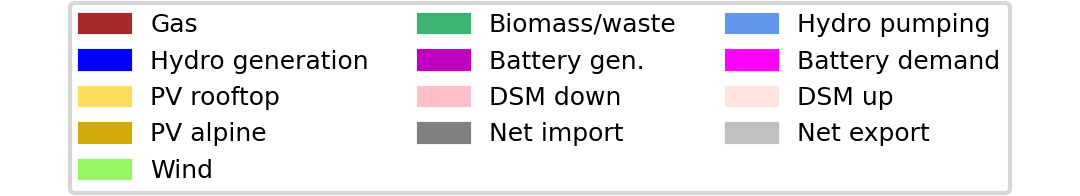}
  \end{minipage}

  \caption{Operations of a typical summer day for the R45 scenario. It illustrates how hydro and battery storage is used during the day to store excess PV generation and during the night to export electricity to the neighboring countries.}
  \label{fig_appen:typical_day}
\end{figure}

\begin{figure}[]
  \centering
  
  \begin{minipage}[b]{0.49\textwidth}
    \includegraphics[width=\textwidth]{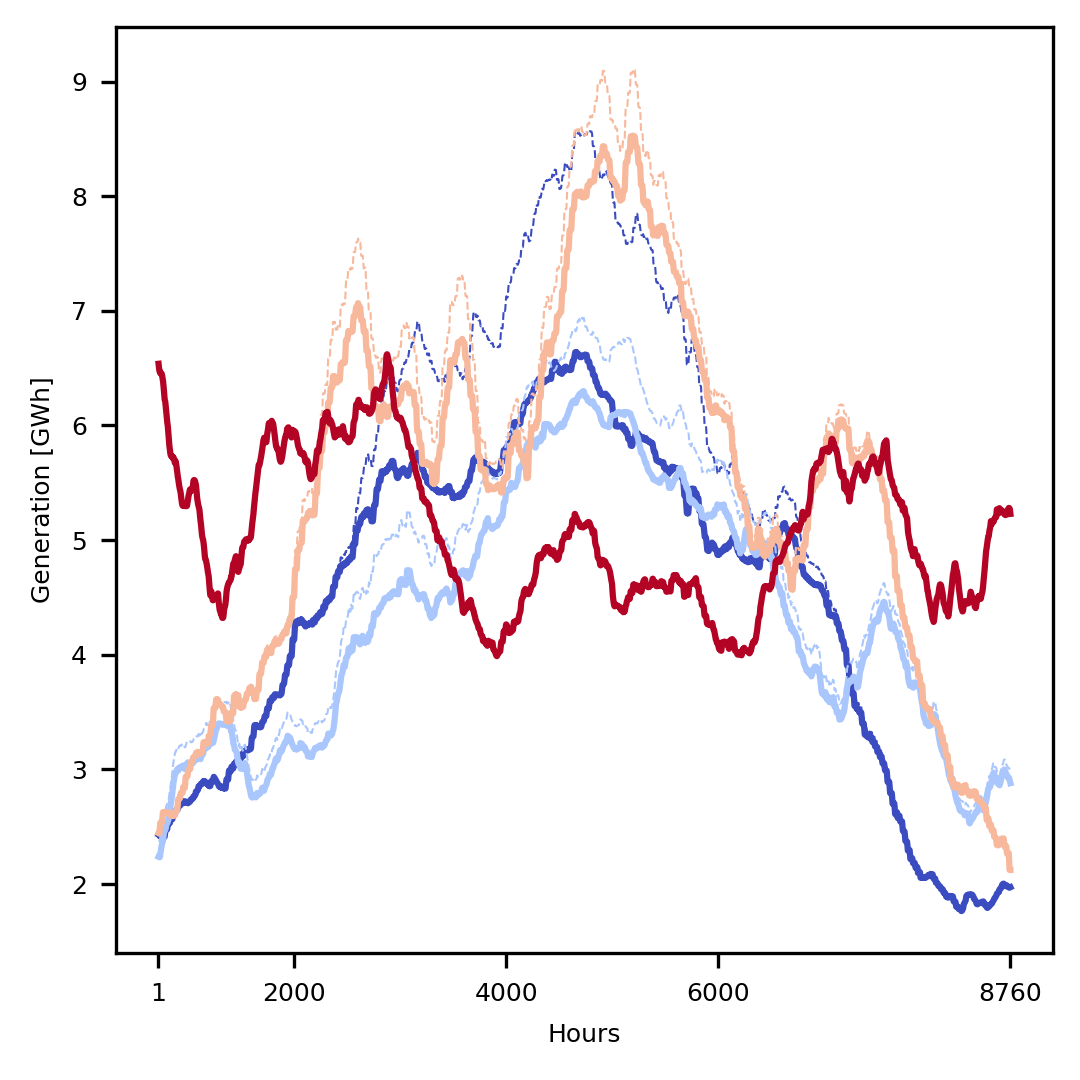}
  \end{minipage}
    
  \begin{minipage}[b]{0.40\textwidth}
    \hspace{0.3cm}
    % \centering
    \includegraphics[width=\textwidth]{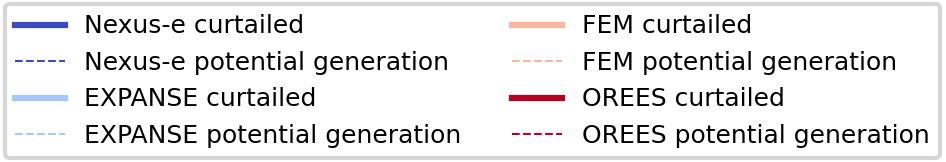}
  \end{minipage}

  \caption{The VRES generation profile for all models, before and after curtailment, referred to as \textit{potential generation} and \textit{curtailed}, respectively.}
  \label{fig_appen:VRES_profile}
\end{figure}

\begin{figure*}[]
  % \centering
  \begin{minipage}[b]{0.31\textwidth}
    \includegraphics[height=0.31\textheight, keepaspectratio]{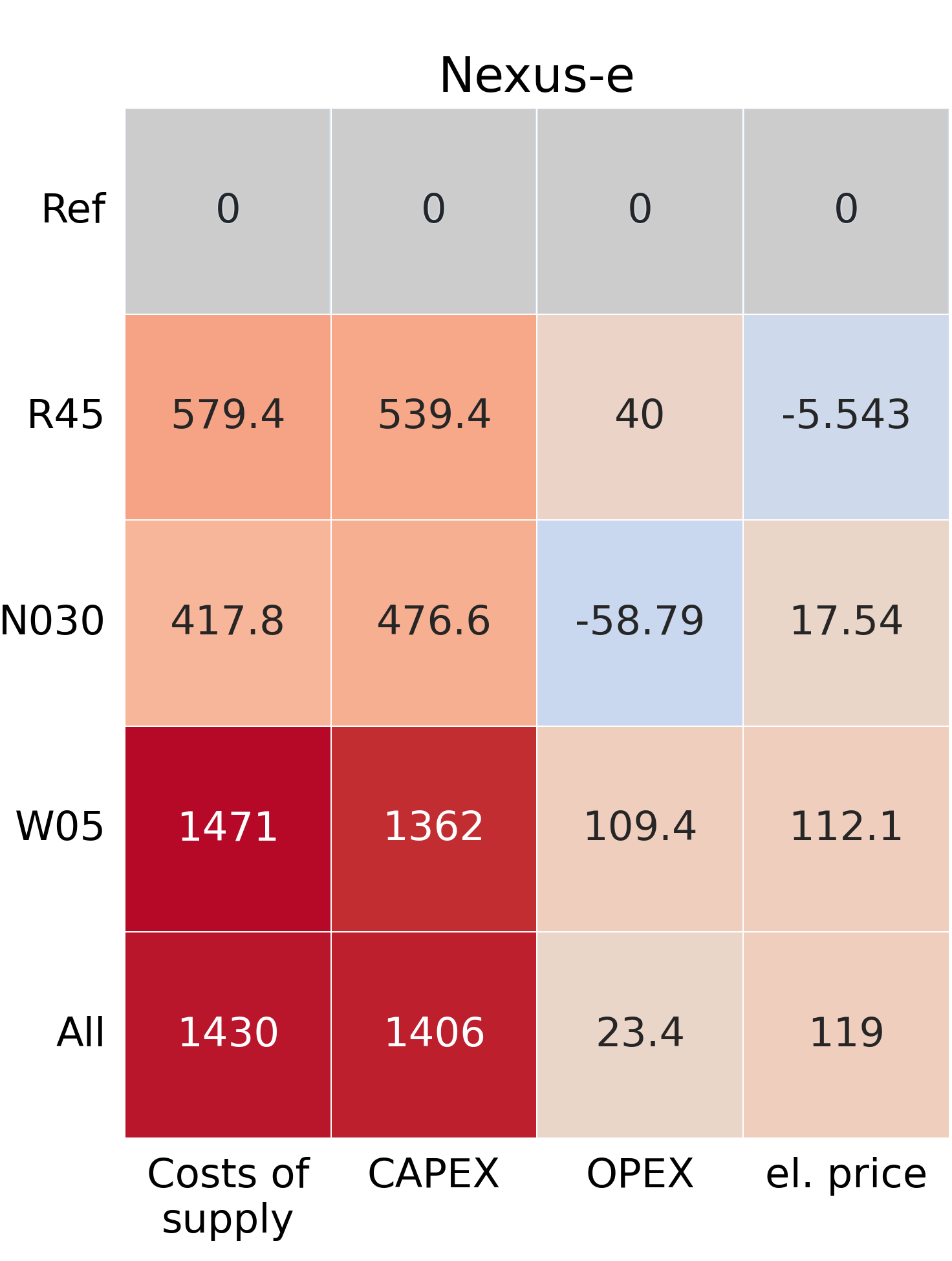}
  \end{minipage}
  \hspace{0.14cm}
  \begin{minipage}[b]{0.31\textwidth}
    \includegraphics[height=0.31\textheight, keepaspectratio]{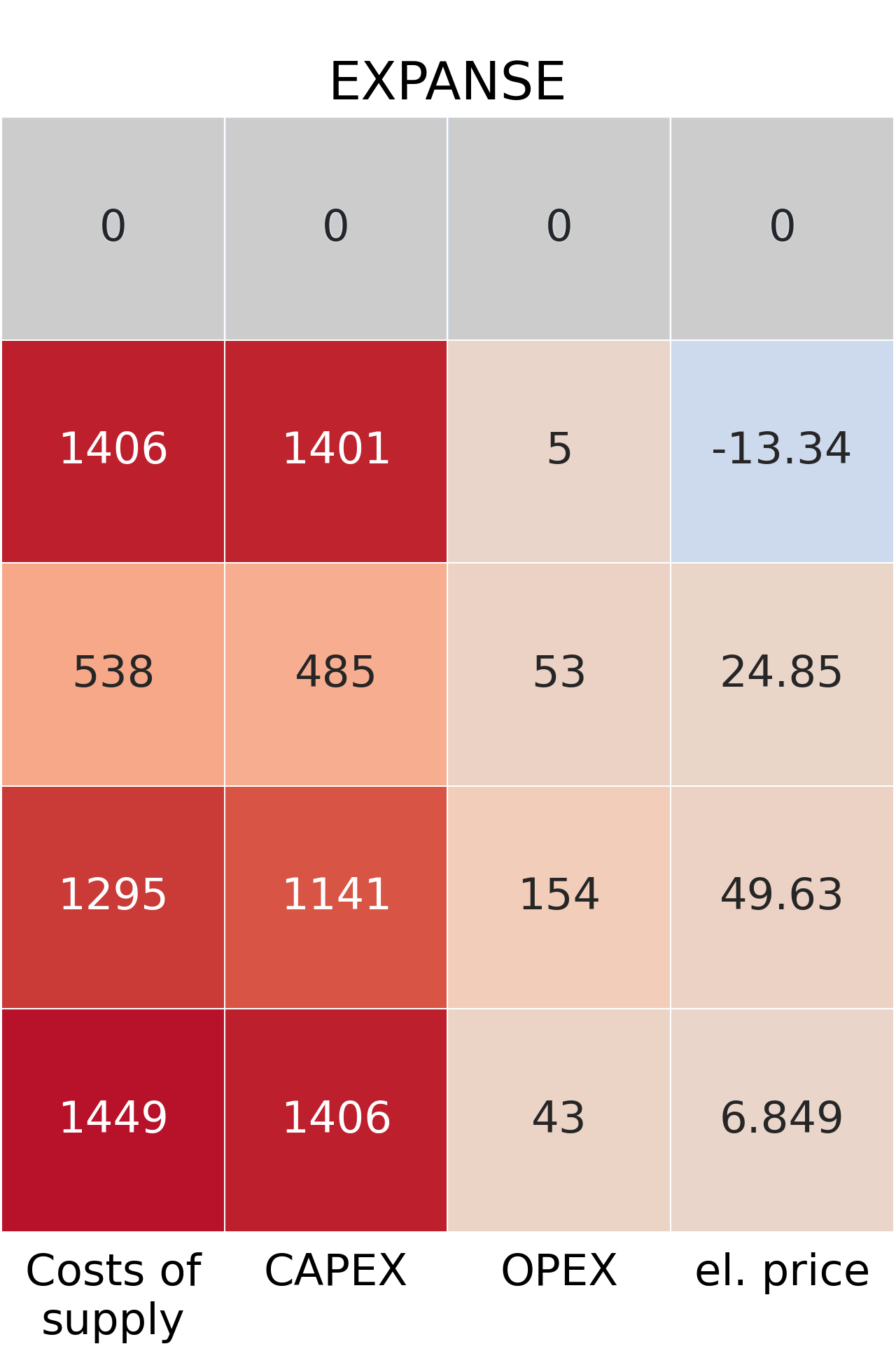}
  \end{minipage}
  \hspace{-0.6cm}
  \begin{minipage}[b]{0.31\textwidth}
    \includegraphics[height=0.31\textheight, keepaspectratio]{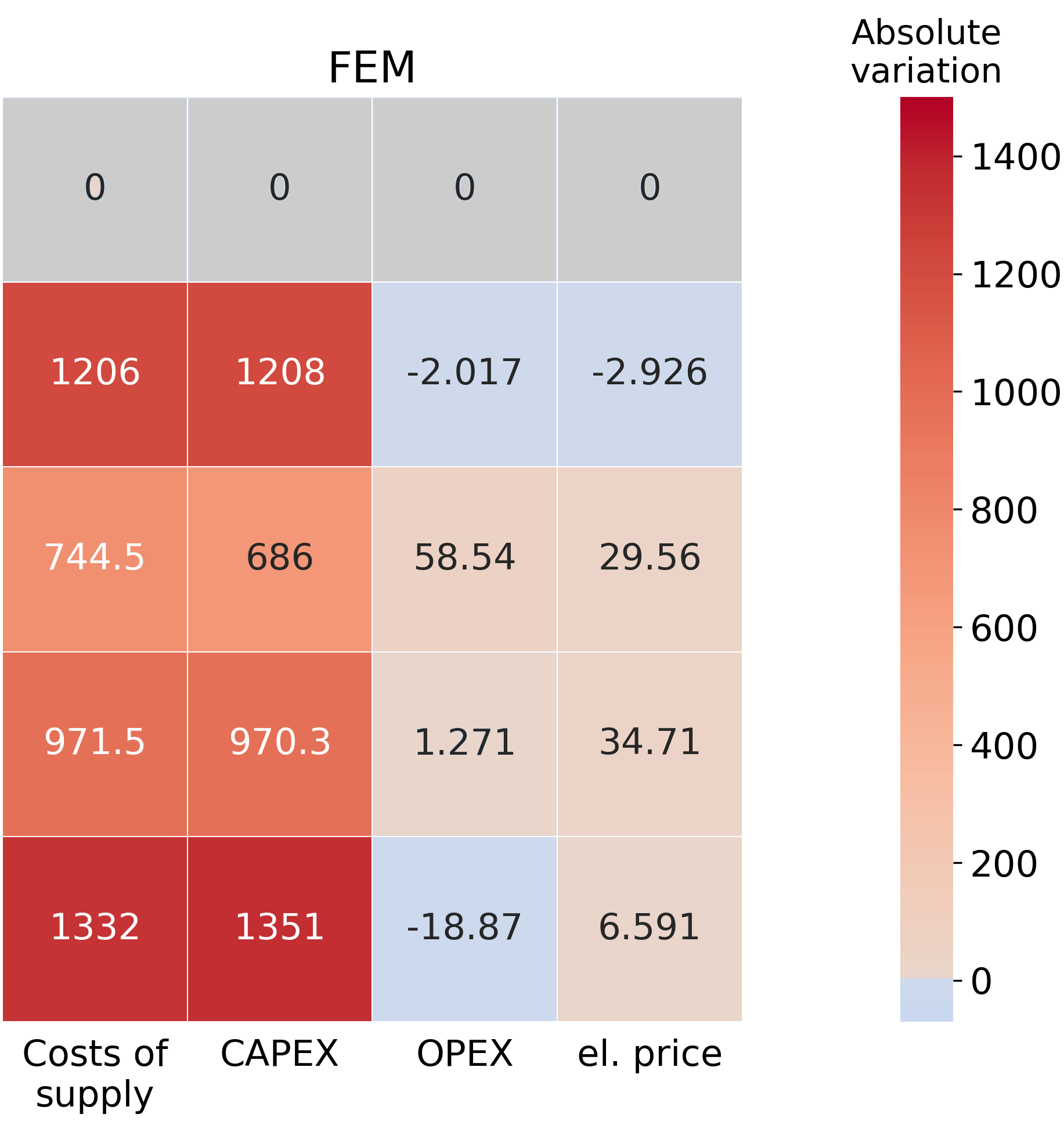}
  \end{minipage}
  
  \caption{Absolute variation of annualized Cost of supply, annualized investment costs (CAPEX), yearly operational costs (OPEX), and average yearly electricity price (el. price) for all scenarios, compared to the Ref scenario. Results for the GA European dimension. Cost values are provided in Mio Eur and electricity price is in Eur/MWh.}
  \label{fig:costs_all_abs_GA}
\end{figure*}

\begin{figure}[]
    \centering
    \begin{minipage}[b]{0.25\textwidth}
        \includegraphics[height=0.25\textheight, keepaspectratio]{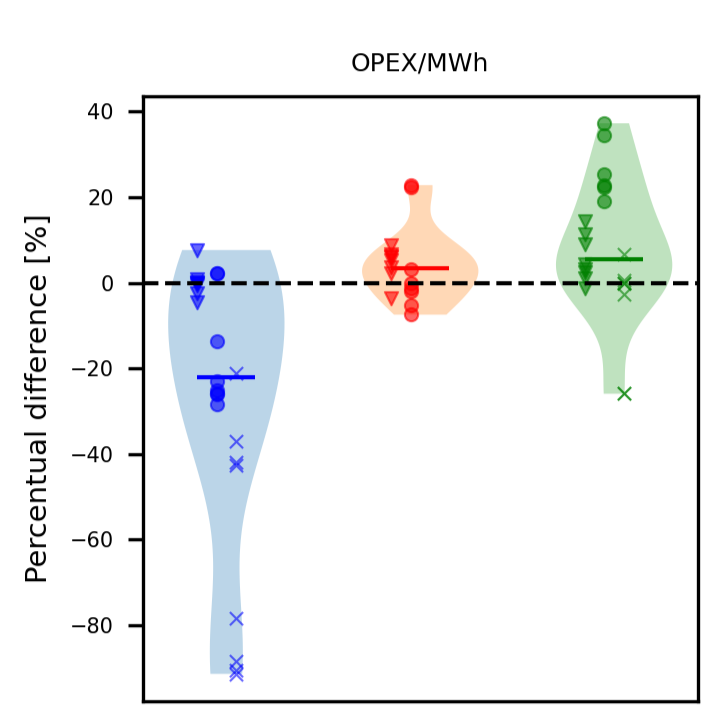}
    \end{minipage}
    \hspace{1.3cm}
    \begin{minipage}[b]{0.15\textwidth}
        \includegraphics[height=0.15\textheight, keepaspectratio]{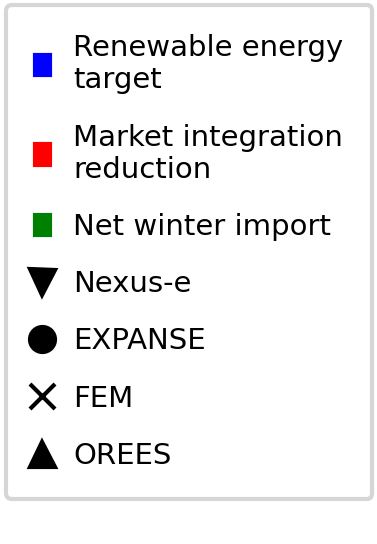}
    \end{minipage}
    \caption{Impact of the three policy measures on OPEX per unit of generated electricity, as percent variation values. For each policy measure, the results of a scenario with the active policy are compared to those of the same scenario without that policy. Extreme values are omitted for better readability but considered in the computation of the median value (horizontal lines).}
    \label{fig:violin_relopex}
\end{figure}

Figure~\ref{fig_appen:typical_day} shows the operations of a typical summer day for the R45 scenario extracted from Nexus-e. 
It shows, with an hourly resolution, how storage units store excess PV generation during the day and generate electricity during the night, enabling electricity exports to the neighboring countries.
Figure~\ref{fig_appen:VRES_profile} illustrates the VRES generation profile before and after curtailment. 
Nexus-e shows a higher generation potential during summer prior to curtailment compared to EXPANSE and OREES, explaining the higher curtailment shares observed in Nexus-e. 
FEM has a similar generation profile before curtailment but exhibits lower levels of curtailment. 
This is due in part to its exclusion of cross-border line limits, which in Nexus-e restrict exports to neighboring countries, leading to VRES curtailment, as well as other factors discussed in~\ref{subsec:curtailment}.
Figure~\ref{fig:costs_all_abs_GA} shows the absolute variation of the analyzed output parameters for the policy scenarios (R45, N030, W05, All) compared to the Ref scenario.
Figure~\ref{fig:violin_relopex} shows the impact of the three policy measures on OPEX per unit of generated electricity and highlights that a market integration reduction leads to higher generation costs per unit of generated electricity. 
Therefore, the reduction in the trading volumes causes the electricity system to operate more expensive generating units.

%%% OPERATIONS
% Typical summer operations

% VRES profile
% All costs - absolute variation - GA

%%% ROBUSTNESS

%%%%%%%%%%%%%%%%%%%%%%%%%%%%%%%%%%%%%%%%%%%%%%%%%%%%%%%%%%%%%%%%%%%%%%%%%%%%%%%%%%%%%%%%%%%%%%%%
% REFERENCES
%%%%%%%%%%%%%%%%%%%%%%%%%%%%%%%%%%%%%%%%%%%%%%%%%%%%%%%%%%%%%%%%%%%%%%%%%%%%%%%%%%%%%%%%%%%%%%%%
\bibliographystyle{elsarticle-num-names}
\bibliography{refs}

\begin{thebibliography}{54}
\expandafter\ifx\csname natexlab\endcsname\relax\def\natexlab#1{#1}\fi
\providecommand{\url}[1]{\texttt{#1}}
\providecommand{\href}[2]{#2}
\providecommand{\path}[1]{#1}
\providecommand{\DOIprefix}{doi:}
\providecommand{\ArXivprefix}{arXiv:}
\providecommand{\URLprefix}{URL: }
\providecommand{\Pubmedprefix}{pmid:}
\providecommand{\doi}[1]{\href{http://dx.doi.org/#1}{\path{#1}}}
\providecommand{\Pubmed}[1]{\href{pmid:#1}{\path{#1}}}
\providecommand{\bibinfo}[2]{#2}
\ifx\xfnm\relax \def\xfnm[#1]{\unskip,\space#1}\fi
%Type = Techreport
\bibitem[{IEA(2023{\natexlab{a}})}]{IEA2023a}
\bibinfo{author}{IEA}, \bibinfo{title}{{Electricity Market Report 2023}}, \bibinfo{type}{Technical Report}, International Energy Agency, \bibinfo{address}{Paris, France}, \bibinfo{year}{2023}{\natexlab{a}}. \URLprefix \url{https://www.iea.org/reports/electricity-market-report-2023}.
%Type = Techreport
\bibitem[{IEA(2023{\natexlab{b}})}]{IEA2023b}
\bibinfo{author}{IEA}, \bibinfo{title}{{Electricity Grids and Secure Energy Transitions}}, \bibinfo{type}{Technical Report}, International Energy Agency, \bibinfo{address}{Paris, France}, \bibinfo{year}{2023}{\natexlab{b}}. \URLprefix \url{https://www.iea.org/reports/electricity-grids-and-secure-energy-transitions}.
%Type = Article
\bibitem[{Stankovski et~al.(2023)Stankovski, Gjorgiev, Locher, and Sansavini}]{Stankovski2023}
\bibinfo{author}{A.~Stankovski}, \bibinfo{author}{B.~Gjorgiev}, \bibinfo{author}{L.~Locher}, \bibinfo{author}{G.~Sansavini},
\newblock \bibinfo{title}{Power blackouts in {E}urope: {A}nalyses, key insights, and recommendations from empirical evidence},
\newblock \bibinfo{journal}{Joule} \bibinfo{volume}{7} (\bibinfo{year}{2023}) \bibinfo{pages}{2468--2484}. \DOIprefix\doi{10.1016/j.joule.2023.09.005}.
%Type = Article
\bibitem[{DeCarolis et~al.(2017)DeCarolis, Daly, Dodds, Keppo, Li, McDowall, Pye, Strachan, Trutnevyte, Usher, Winning, Yeh, and Zeyringer}]{DeCarolis2017}
\bibinfo{author}{J.~DeCarolis}, \bibinfo{author}{H.~Daly}, \bibinfo{author}{P.~Dodds}, \bibinfo{author}{I.~Keppo}, \bibinfo{author}{F.~Li}, \bibinfo{author}{W.~McDowall}, \bibinfo{author}{S.~Pye}, \bibinfo{author}{N.~Strachan}, \bibinfo{author}{E.~Trutnevyte}, \bibinfo{author}{W.~Usher}, \bibinfo{author}{M.~Winning}, \bibinfo{author}{S.~Yeh}, \bibinfo{author}{M.~Zeyringer},
\newblock \bibinfo{title}{Formalizing best practice for energy system optimization modelling},
\newblock \bibinfo{journal}{Applied Energy} \bibinfo{volume}{194} (\bibinfo{year}{2017}) \bibinfo{pages}{184--198}. \DOIprefix\doi{10.1016/j.apenergy.2017.03.001}.
%Type = Article
\bibitem[{Bistline et~al.(2018)Bistline, Hodson, Rossmann, Creason, Murray, and Barron}]{Bistline2018}
\bibinfo{author}{J.~E. Bistline}, \bibinfo{author}{E.~Hodson}, \bibinfo{author}{C.~G. Rossmann}, \bibinfo{author}{J.~Creason}, \bibinfo{author}{B.~Murray}, \bibinfo{author}{A.~R. Barron},
\newblock \bibinfo{title}{Electric sector policy, technological change, and {U.S.} emissions reductions goals: Results from the {EMF} 32 model intercomparison project},
\newblock \bibinfo{journal}{Energy Economics} \bibinfo{volume}{73} (\bibinfo{year}{2018}) \bibinfo{pages}{307--325}. \DOIprefix\doi{https://doi.org/10.1016/j.eneco.2018.04.012}.
%Type = Article
\bibitem[{Huntington et~al.(1982)Huntington, Weyant, and Sweeney}]{Huntington1982}
\bibinfo{author}{H.~G. Huntington}, \bibinfo{author}{J.~P. Weyant}, \bibinfo{author}{J.~L. Sweeney},
\newblock \bibinfo{title}{Modeling for insights, not numbers: the experiences of the energy modeling forum},
\newblock \bibinfo{journal}{Omega} \bibinfo{volume}{10} (\bibinfo{year}{1982}) \bibinfo{pages}{449--462}. \DOIprefix\doi{https://doi.org/10.1016/0305-0483(82)90002-0}.
%Type = Article
\bibitem[{Heinisch et~al.(2023)Heinisch, Dujardin, Gabrielli, Jain, Lehning, Sansavini, Sasse, Schaffner, Schwarz, and Trutnevyte}]{Heinisch2023}
\bibinfo{author}{V.~Heinisch}, \bibinfo{author}{J.~Dujardin}, \bibinfo{author}{P.~Gabrielli}, \bibinfo{author}{P.~Jain}, \bibinfo{author}{M.~Lehning}, \bibinfo{author}{G.~Sansavini}, \bibinfo{author}{J.~P. Sasse}, \bibinfo{author}{C.~Schaffner}, \bibinfo{author}{M.~Schwarz}, \bibinfo{author}{E.~Trutnevyte},
\newblock \bibinfo{title}{Inter-comparison of spatial models for high shares of renewable electricity in {S}witzerland},
\newblock \bibinfo{journal}{Applied Energy} \bibinfo{volume}{350} (\bibinfo{year}{2023}) \bibinfo{pages}{121700}. \DOIprefix\doi{https://doi.org/10.1016/j.apenergy.2023.121700}.
%Type = Article
\bibitem[{Wilson et~al.(2021)Wilson, Guivarch, Kriegler, van Ruijven, van Vuuren, Krey, Schwanitz, and Thompson}]{Wilson2021}
\bibinfo{author}{C.~Wilson}, \bibinfo{author}{C.~Guivarch}, \bibinfo{author}{E.~Kriegler}, \bibinfo{author}{B.~van Ruijven}, \bibinfo{author}{D.~P. van Vuuren}, \bibinfo{author}{V.~Krey}, \bibinfo{author}{V.~J. Schwanitz}, \bibinfo{author}{E.~L. Thompson},
\newblock \bibinfo{title}{Evaluating process-based integrated assessment models of climate change mitigation},
\newblock \bibinfo{journal}{Climatic Change} \bibinfo{volume}{166} (\bibinfo{year}{2021}) \bibinfo{pages}{3}. \DOIprefix\doi{10.1007/s10584-021-03099-9}.
%Type = Article
\bibitem[{Harmsen et~al.(2021)Harmsen, Kriegler, van Vuuren, van~der Wijst, Luderer, Cui, Dessens, Drouet, Emmerling, Morris, Fosse, Fragkiadakis, Fragkiadakis, Fragkos, Fricko, Fujimori, Gernaat, Guivarch, Iyer, Karkatsoulis, Keppo, Keramidas, Köberle, Kolp, Krey, Krüger, Leblanc, Mittal, Paltsev, Rochedo, van Ruijven, Sands, Sano, Strefler, Arroyo, Wada, and Zakeri}]{Harmsen2021}
\bibinfo{author}{M.~Harmsen}, \bibinfo{author}{E.~Kriegler}, \bibinfo{author}{D.~P. van Vuuren}, \bibinfo{author}{K.-I. van~der Wijst}, \bibinfo{author}{G.~Luderer}, \bibinfo{author}{R.~Cui}, \bibinfo{author}{O.~Dessens}, \bibinfo{author}{L.~Drouet}, \bibinfo{author}{J.~Emmerling}, \bibinfo{author}{J.~F. Morris}, \bibinfo{author}{F.~Fosse}, \bibinfo{author}{D.~Fragkiadakis}, \bibinfo{author}{K.~Fragkiadakis}, \bibinfo{author}{P.~Fragkos}, \bibinfo{author}{O.~Fricko}, \bibinfo{author}{S.~Fujimori}, \bibinfo{author}{D.~Gernaat}, \bibinfo{author}{C.~Guivarch}, \bibinfo{author}{G.~Iyer}, \bibinfo{author}{P.~Karkatsoulis}, \bibinfo{author}{I.~Keppo}, \bibinfo{author}{K.~Keramidas}, \bibinfo{author}{A.~Köberle}, \bibinfo{author}{P.~Kolp}, \bibinfo{author}{V.~Krey}, \bibinfo{author}{C.~Krüger}, \bibinfo{author}{F.~Leblanc}, \bibinfo{author}{S.~Mittal}, \bibinfo{author}{S.~Paltsev}, \bibinfo{author}{P.~Rochedo}, \bibinfo{author}{B.~J. van Ruijven}, \bibinfo{author}{R.~D. Sands}, \bibinfo{author}{F.~Sano},
  \bibinfo{author}{J.~Strefler}, \bibinfo{author}{E.~V. Arroyo}, \bibinfo{author}{K.~Wada}, \bibinfo{author}{B.~Zakeri},
\newblock \bibinfo{title}{Integrated assessment model diagnostics: key indicators and model evolution},
\newblock \bibinfo{journal}{Environmental Research Letters} \bibinfo{volume}{16} (\bibinfo{year}{2021}) \bibinfo{pages}{054046}. \DOIprefix\doi{10.1088/1748-9326/abf964}.
%Type = Article
\bibitem[{Siala et~al.(2022)Siala, Mier, Schmidt, Torralba-Díaz, Sheykhha, and Savvidis}]{Siala2022}
\bibinfo{author}{K.~Siala}, \bibinfo{author}{M.~Mier}, \bibinfo{author}{L.~Schmidt}, \bibinfo{author}{L.~Torralba-Díaz}, \bibinfo{author}{S.~Sheykhha}, \bibinfo{author}{G.~Savvidis},
\newblock \bibinfo{title}{Which model features matter? {A}n experimental approach to evaluate power market modeling choices},
\newblock \bibinfo{journal}{Energy} \bibinfo{volume}{245} (\bibinfo{year}{2022}) \bibinfo{pages}{123301}. \DOIprefix\doi{https://doi.org/10.1016/j.energy.2022.123301}.
%Type = Article
\bibitem[{Xexakis et~al.(2020)Xexakis, Hansmann, Volken, and Trutnevyte}]{Xexakis2020}
\bibinfo{author}{G.~Xexakis}, \bibinfo{author}{R.~Hansmann}, \bibinfo{author}{S.~P. Volken}, \bibinfo{author}{E.~Trutnevyte},
\newblock \bibinfo{title}{Models on the wrong track: Model-based electricity supply scenarios in {S}witzerland are not aligned with the perspectives of energy experts and the public},
\newblock \bibinfo{journal}{Renewable and Sustainable Energy Reviews} \bibinfo{volume}{134} (\bibinfo{year}{2020}) \bibinfo{pages}{110297}. \DOIprefix\doi{https://doi.org/10.1016/j.rser.2020.110297}.
%Type = Misc
\bibitem[{Trutnevyte(2024)}]{Trutnevyte2024}
\bibinfo{author}{E.~Trutnevyte}, \bibinfo{title}{Renewable energy outlook for switzerland}, \bibinfo{year}{2024}. \URLprefix \url{https://doi.org/10.13097/archive-ouverte/unige:172640}. \DOIprefix\doi{10.13097/archive}.
%Type = Misc
\bibitem[{{SWEET-CROSS}(2024)}]{CROSS2024}
\bibinfo{author}{{SWEET-CROSS}}, \bibinfo{title}{{CROSS model result comparison}}, \bibinfo{howpublished}{\url{https://sweet-cross.ch/results/}}, \bibinfo{year}{2024}. \bibinfo{note}{Accessed: 2024-07-12}.
%Type = Article
\bibitem[{Misconel et~al.(2022)Misconel, Leisen, Mikurda, Zimmermann, Fraunholz, Fichtner, Möst, and Weber}]{Misconel2022}
\bibinfo{author}{S.~Misconel}, \bibinfo{author}{R.~Leisen}, \bibinfo{author}{J.~Mikurda}, \bibinfo{author}{F.~Zimmermann}, \bibinfo{author}{C.~Fraunholz}, \bibinfo{author}{W.~Fichtner}, \bibinfo{author}{D.~Möst}, \bibinfo{author}{C.~Weber},
\newblock \bibinfo{title}{Systematic comparison of high-resolution electricity system modeling approaches focusing on investment, dispatch and generation adequacy},
\newblock \bibinfo{journal}{Renewable and Sustainable Energy Reviews} \bibinfo{volume}{153} (\bibinfo{year}{2022}) \bibinfo{pages}{111785}. \DOIprefix\doi{https://doi.org/10.1016/j.rser.2021.111785}.
%Type = Article
\bibitem[{Misconel et~al.(2024)Misconel, Zimmermann, Mikurda, Möst, Kunze, Gnann, Kühnbach, Speth, Pelka, and Yu}]{Misconel2024}
\bibinfo{author}{S.~Misconel}, \bibinfo{author}{F.~Zimmermann}, \bibinfo{author}{J.~Mikurda}, \bibinfo{author}{D.~Möst}, \bibinfo{author}{R.~Kunze}, \bibinfo{author}{T.~Gnann}, \bibinfo{author}{M.~Kühnbach}, \bibinfo{author}{D.~Speth}, \bibinfo{author}{S.~Pelka}, \bibinfo{author}{S.~Yu},
\newblock \bibinfo{title}{Model coupling and comparison on optimal load shifting of battery electric vehicles and heat pumps focusing on generation adequacy},
\newblock \bibinfo{journal}{Energy} \bibinfo{volume}{305} (\bibinfo{year}{2024}) \bibinfo{pages}{132266}. \DOIprefix\doi{https://doi.org/10.1016/j.energy.2024.132266}.
%Type = Article
\bibitem[{{van Ouwerkerk} et~al.(2022){van Ouwerkerk}, Gils, Gardian, Kittel, Schill, Zerrahn, Murmann, Launer, Torralba-Díaz, and Bußar}]{vanOuwerkerk2022}
\bibinfo{author}{J.~{van Ouwerkerk}}, \bibinfo{author}{H.~C. Gils}, \bibinfo{author}{H.~Gardian}, \bibinfo{author}{M.~Kittel}, \bibinfo{author}{W.-P. Schill}, \bibinfo{author}{A.~Zerrahn}, \bibinfo{author}{A.~Murmann}, \bibinfo{author}{J.~Launer}, \bibinfo{author}{L.~Torralba-Díaz}, \bibinfo{author}{C.~Bußar},
\newblock \bibinfo{title}{Impacts of power sector model features on optimal capacity expansion: A comparative study},
\newblock \bibinfo{journal}{Renewable and Sustainable Energy Reviews} \bibinfo{volume}{157} (\bibinfo{year}{2022}) \bibinfo{pages}{112004}. \DOIprefix\doi{https://doi.org/10.1016/j.rser.2021.112004}.
%Type = Misc
\bibitem[{{Stanford University}(2024)}]{EMF2024}
\bibinfo{author}{{Stanford University}}, \bibinfo{title}{Energy modeling forum}, \bibinfo{year}{2024}. \URLprefix \url{https://emf.stanford.edu/}, \bibinfo{note}{(Accessed: 22.07.2024)}.
%Type = Article
\bibitem[{Henke et~al.(2023)Henke, Dekker, Lombardi, Pietzcker, Fragkos, Zakeri, Rodrigues, Sitarz, Emmerling, Fattahi, Longa, Tatarewicz, Fotiou, Lewarski, Huppmann, Kavvadias, van~der Zwaan, and Usher}]{Henke2023}
\bibinfo{author}{H.~Henke}, \bibinfo{author}{M.~Dekker}, \bibinfo{author}{F.~Lombardi}, \bibinfo{author}{R.~Pietzcker}, \bibinfo{author}{P.~Fragkos}, \bibinfo{author}{B.~Zakeri}, \bibinfo{author}{R.~Rodrigues}, \bibinfo{author}{J.~Sitarz}, \bibinfo{author}{J.~Emmerling}, \bibinfo{author}{A.~Fattahi}, \bibinfo{author}{F.~D. Longa}, \bibinfo{author}{I.~Tatarewicz}, \bibinfo{author}{T.~Fotiou}, \bibinfo{author}{M.~Lewarski}, \bibinfo{author}{D.~Huppmann}, \bibinfo{author}{K.~Kavvadias}, \bibinfo{author}{B.~van~der Zwaan}, \bibinfo{author}{W.~Usher},
\newblock \bibinfo{title}{Comparing energy system optimization models and integrated assessment models: Relevance for energy policy advice},
\newblock \bibinfo{journal}{Open Research Europe} \bibinfo{volume}{3} (\bibinfo{year}{2023}) \bibinfo{pages}{69}. \DOIprefix\doi{10.12688/openreseurope.15590.1}.
%Type = Article
\bibitem[{Prina et~al.(2022)Prina, Nastasi, Groppi, Misconel, Garcia, and Sparber}]{Prina2022}
\bibinfo{author}{M.~G. Prina}, \bibinfo{author}{B.~Nastasi}, \bibinfo{author}{D.~Groppi}, \bibinfo{author}{S.~Misconel}, \bibinfo{author}{D.~A. Garcia}, \bibinfo{author}{W.~Sparber},
\newblock \bibinfo{title}{Comparison methods of energy system frameworks, models and scenario results},
\newblock \bibinfo{journal}{Renewable and Sustainable Energy Reviews} \bibinfo{volume}{167} (\bibinfo{year}{2022}). \DOIprefix\doi{10.1016/j.rser.2022.112719}.
%Type = Article
\bibitem[{Gacitua et~al.(2018)Gacitua, Gallegos, Henriquez-Auba, Lorca, Negrete-Pincetic, Olivares, Valenzuela, and Wenzel}]{Gacitua2018}
\bibinfo{author}{L.~Gacitua}, \bibinfo{author}{P.~Gallegos}, \bibinfo{author}{R.~Henriquez-Auba}, \bibinfo{author}{Lorca}, \bibinfo{author}{M.~Negrete-Pincetic}, \bibinfo{author}{D.~Olivares}, \bibinfo{author}{A.~Valenzuela}, \bibinfo{author}{G.~Wenzel},
\newblock \bibinfo{title}{A comprehensive review on expansion planning: Models and tools for energy policy analysis},
\newblock \bibinfo{journal}{Renewable and Sustainable Energy Reviews} \bibinfo{volume}{98} (\bibinfo{year}{2018}) \bibinfo{pages}{346--360}. \DOIprefix\doi{10.1016/j.rser.2018.08.043}.
%Type = Article
\bibitem[{Savvidis et~al.(2019)Savvidis, Siala, Weissbart, Schmidt, Borggrefe, Kumar, Pittel, Madlener, and Hufendiek}]{Savvidis2019}
\bibinfo{author}{G.~Savvidis}, \bibinfo{author}{K.~Siala}, \bibinfo{author}{C.~Weissbart}, \bibinfo{author}{L.~Schmidt}, \bibinfo{author}{F.~Borggrefe}, \bibinfo{author}{S.~Kumar}, \bibinfo{author}{K.~Pittel}, \bibinfo{author}{R.~Madlener}, \bibinfo{author}{K.~Hufendiek},
\newblock \bibinfo{title}{The gap between energy policy challenges and model capabilities},
\newblock \bibinfo{journal}{Energy Policy} \bibinfo{volume}{125} (\bibinfo{year}{2019}) \bibinfo{pages}{503--520}. \DOIprefix\doi{10.1016/j.enpol.2018.10.033}.
%Type = Article
\bibitem[{Riemer et~al.(2023)Riemer, Wachsmuth, Boitier, Elia, Al-Dabbas, Şirin Alibaş, Chiodi, and Neuner}]{Riemer2023}
\bibinfo{author}{M.~Riemer}, \bibinfo{author}{J.~Wachsmuth}, \bibinfo{author}{B.~Boitier}, \bibinfo{author}{A.~Elia}, \bibinfo{author}{K.~Al-Dabbas}, \bibinfo{author}{Şirin Alibaş}, \bibinfo{author}{A.~Chiodi}, \bibinfo{author}{F.~Neuner},
\newblock \bibinfo{title}{How do system-wide net-zero scenarios compare to sector model pathways for the {EU}? {A} novel approach based on benchmark indicators and index decomposition analyses},
\newblock \bibinfo{journal}{Energy Strategy Reviews} \bibinfo{volume}{50} (\bibinfo{year}{2023}). \DOIprefix\doi{10.1016/j.esr.2023.101225}.
%Type = Article
\bibitem[{Clarke et~al.(2009)Clarke, Edmonds, Krey, Richels, Rose, and Tavoni}]{Clarke2009}
\bibinfo{author}{L.~Clarke}, \bibinfo{author}{J.~Edmonds}, \bibinfo{author}{V.~Krey}, \bibinfo{author}{R.~Richels}, \bibinfo{author}{S.~Rose}, \bibinfo{author}{M.~Tavoni},
\newblock \bibinfo{title}{International climate policy architectures: Overview of the {EMF} 22 international scenarios},
\newblock \bibinfo{journal}{Energy Economics} \bibinfo{volume}{31} (\bibinfo{year}{2009}). \DOIprefix\doi{10.1016/j.eneco.2009.10.013}.
%Type = Article
\bibitem[{Tavoni et~al.(2015)Tavoni, Kriegler, Riahi, Vuuren, Aboumahboub, Bowen, Calvin, Campiglio, Kober, Jewell, Luderer, Marangoni, Mccollum, Sluisveld, Zimmer, and Zwaan}]{Tavoni2015}
\bibinfo{author}{M.~Tavoni}, \bibinfo{author}{E.~Kriegler}, \bibinfo{author}{K.~Riahi}, \bibinfo{author}{D.~P.~V. Vuuren}, \bibinfo{author}{T.~Aboumahboub}, \bibinfo{author}{A.~Bowen}, \bibinfo{author}{K.~Calvin}, \bibinfo{author}{E.~Campiglio}, \bibinfo{author}{T.~Kober}, \bibinfo{author}{J.~Jewell}, \bibinfo{author}{G.~Luderer}, \bibinfo{author}{G.~Marangoni}, \bibinfo{author}{D.~Mccollum}, \bibinfo{author}{M.~V. Sluisveld}, \bibinfo{author}{A.~Zimmer}, \bibinfo{author}{B.~V.~D. Zwaan},
\newblock \bibinfo{title}{Post-2020 climate agreements in the major economies assessed in the light of global models},
\newblock \bibinfo{journal}{Nature Climate Change} \bibinfo{volume}{5} (\bibinfo{year}{2015}) \bibinfo{pages}{119--126}. \DOIprefix\doi{10.1038/nclimate2475}.
%Type = Article
\bibitem[{Nikas et~al.(2021)Nikas, Elia, Boitier, Koasidis, Doukas, Cassetti, Anger-Kraavi, Bui, Campagnolo, Miglio, Delpiazzo, Fougeyrollas, Gambhir, Gargiulo, Giarola, Grant, Hawkes, Herbst, Köberle, Kolpakov, Mouël, McWilliams, Mittal, Moreno, Neuner, Perdana, Peters, Plötz, Rogelj, Sognnæs, de~Ven, Vielle, Zachmann, Zagamé, and Chiodi}]{Nikas2021}
\bibinfo{author}{A.~Nikas}, \bibinfo{author}{A.~Elia}, \bibinfo{author}{B.~Boitier}, \bibinfo{author}{K.~Koasidis}, \bibinfo{author}{H.~Doukas}, \bibinfo{author}{G.~Cassetti}, \bibinfo{author}{A.~Anger-Kraavi}, \bibinfo{author}{H.~Bui}, \bibinfo{author}{L.~Campagnolo}, \bibinfo{author}{R.~D. Miglio}, \bibinfo{author}{E.~Delpiazzo}, \bibinfo{author}{A.~Fougeyrollas}, \bibinfo{author}{A.~Gambhir}, \bibinfo{author}{M.~Gargiulo}, \bibinfo{author}{S.~Giarola}, \bibinfo{author}{N.~Grant}, \bibinfo{author}{A.~Hawkes}, \bibinfo{author}{A.~Herbst}, \bibinfo{author}{A.~C. Köberle}, \bibinfo{author}{A.~Kolpakov}, \bibinfo{author}{P.~L. Mouël}, \bibinfo{author}{B.~McWilliams}, \bibinfo{author}{S.~Mittal}, \bibinfo{author}{J.~Moreno}, \bibinfo{author}{F.~Neuner}, \bibinfo{author}{S.~Perdana}, \bibinfo{author}{G.~P. Peters}, \bibinfo{author}{P.~Plötz}, \bibinfo{author}{J.~Rogelj}, \bibinfo{author}{I.~Sognnæs}, \bibinfo{author}{D.~J.~V. de~Ven}, \bibinfo{author}{M.~Vielle}, \bibinfo{author}{G.~Zachmann},
  \bibinfo{author}{P.~Zagamé}, \bibinfo{author}{A.~Chiodi},
\newblock \bibinfo{title}{Where is the {EU} headed given its current climate policy? {A} stakeholder-driven model inter-comparison},
\newblock \bibinfo{journal}{Science of the Total Environment} \bibinfo{volume}{793} (\bibinfo{year}{2021}). \DOIprefix\doi{10.1016/j.scitotenv.2021.148549}.
%Type = Article
\bibitem[{Huntington et~al.(2020)Huntington, Bhargava, Daniels, Weyant, Avraam, Bistline, Edmonds, Giarola, Hawkes, Hansen, Johnston, Molar-Cruz, Nadew, Siddiqui, Vaillancourt, and Victor}]{Huntington2020}
\bibinfo{author}{H.~G. Huntington}, \bibinfo{author}{A.~Bhargava}, \bibinfo{author}{D.~Daniels}, \bibinfo{author}{J.~P. Weyant}, \bibinfo{author}{C.~Avraam}, \bibinfo{author}{J.~Bistline}, \bibinfo{author}{J.~A. Edmonds}, \bibinfo{author}{S.~Giarola}, \bibinfo{author}{A.~Hawkes}, \bibinfo{author}{M.~Hansen}, \bibinfo{author}{P.~Johnston}, \bibinfo{author}{A.~Molar-Cruz}, \bibinfo{author}{M.~Nadew}, \bibinfo{author}{S.~Siddiqui}, \bibinfo{author}{K.~Vaillancourt}, \bibinfo{author}{N.~Victor},
\newblock \bibinfo{title}{Key findings from the core north american scenarios in the {EMF}34 intermodel comparison},
\newblock \bibinfo{journal}{Energy Policy} \bibinfo{volume}{144} (\bibinfo{year}{2020}). \DOIprefix\doi{10.1016/j.enpol.2020.111599}.
%Type = Article
\bibitem[{Avril et~al.(2012)Avril, Mansilla, Busson, and Lemaire}]{Avril2012}
\bibinfo{author}{S.~Avril}, \bibinfo{author}{C.~Mansilla}, \bibinfo{author}{M.~Busson}, \bibinfo{author}{T.~Lemaire},
\newblock \bibinfo{title}{Photovoltaic energy policy: Financial estimation and performance comparison of the public support in five representative countries},
\newblock \bibinfo{journal}{Energy Policy} \bibinfo{volume}{51} (\bibinfo{year}{2012}) \bibinfo{pages}{244--258}. \DOIprefix\doi{10.1016/j.enpol.2012.07.050}.
%Type = Article
\bibitem[{Bistline et~al.(2020)Bistline, Brown, Siddiqui, and Vaillancourt}]{Bistline2020}
\bibinfo{author}{J.~E. Bistline}, \bibinfo{author}{M.~Brown}, \bibinfo{author}{S.~A. Siddiqui}, \bibinfo{author}{K.~Vaillancourt},
\newblock \bibinfo{title}{Electric sector impacts of renewable policy coordination: A multi-model study of the north american energy system},
\newblock \bibinfo{journal}{Energy Policy} \bibinfo{volume}{145} (\bibinfo{year}{2020}). \DOIprefix\doi{10.1016/j.enpol.2020.111707}.
%Type = Article
\bibitem[{Giarola et~al.(2021)Giarola, Molar-Cruz, Vaillancourt, Bahn, Sarmiento, Hawkes, and Brown}]{Giarola2021}
\bibinfo{author}{S.~Giarola}, \bibinfo{author}{A.~Molar-Cruz}, \bibinfo{author}{K.~Vaillancourt}, \bibinfo{author}{O.~Bahn}, \bibinfo{author}{L.~Sarmiento}, \bibinfo{author}{A.~Hawkes}, \bibinfo{author}{M.~Brown},
\newblock \bibinfo{title}{The role of energy storage in the uptake of renewable energy: A model comparison approach},
\newblock \bibinfo{journal}{Energy Policy} \bibinfo{volume}{151} (\bibinfo{year}{2021}). \DOIprefix\doi{10.1016/j.enpol.2021.112159}.
%Type = Article
\bibitem[{Trutnevyte et~al.(2014)Trutnevyte, Barton, Áine O'Grady, Ogunkunle, Pudjianto, and Robertson}]{Trutnevyte2014}
\bibinfo{author}{E.~Trutnevyte}, \bibinfo{author}{J.~Barton}, \bibinfo{author}{Áine O'Grady}, \bibinfo{author}{D.~Ogunkunle}, \bibinfo{author}{D.~Pudjianto}, \bibinfo{author}{E.~Robertson},
\newblock \bibinfo{title}{Linking a storyline with multiple models: A cross-scale study of the {UK} power system transition},
\newblock \bibinfo{journal}{Technological Forecasting and Social Change} \bibinfo{volume}{89} (\bibinfo{year}{2014}) \bibinfo{pages}{26--42}. \DOIprefix\doi{10.1016/j.techfore.2014.08.018}.
%Type = Article
\bibitem[{Gjorgiev et~al.(2022)Gjorgiev, Garrison, Han, Landis, van Nieuwkoop, Raycheva, Schwarz, Yan, Demiray, Hug, Sansavini, and Schaffner}]{Gjorgiev2022}
\bibinfo{author}{B.~Gjorgiev}, \bibinfo{author}{J.~B. Garrison}, \bibinfo{author}{X.~Han}, \bibinfo{author}{F.~Landis}, \bibinfo{author}{R.~van Nieuwkoop}, \bibinfo{author}{E.~Raycheva}, \bibinfo{author}{M.~Schwarz}, \bibinfo{author}{X.~Yan}, \bibinfo{author}{T.~Demiray}, \bibinfo{author}{G.~Hug}, \bibinfo{author}{G.~Sansavini}, \bibinfo{author}{C.~Schaffner},
\newblock \bibinfo{title}{Nexus-e: A platform of interfaced high-resolution models for energy-economic assessments of future electricity systems},
\newblock \bibinfo{journal}{Applied Energy} \bibinfo{volume}{307} (\bibinfo{year}{2022}). \DOIprefix\doi{10.1016/j.apenergy.2021.118193}.
%Type = Article
\bibitem[{Raycheva et~al.(2023)Raycheva, Gjorgiev, Hug, Sansavini, and Schaffner}]{Raycheva2023}
\bibinfo{author}{E.~Raycheva}, \bibinfo{author}{B.~Gjorgiev}, \bibinfo{author}{G.~Hug}, \bibinfo{author}{G.~Sansavini}, \bibinfo{author}{C.~Schaffner},
\newblock \bibinfo{title}{Risk-informed coordinated generation and transmission system expansion planning: A net-zero scenario of switzerland in the {European} context},
\newblock \bibinfo{journal}{Energy} \bibinfo{volume}{280} (\bibinfo{year}{2023}) \bibinfo{pages}{128090}. \DOIprefix\doi{10.1016/j.energy.2023.128090}.
%Type = Article
\bibitem[{Han et~al.(2022)Han, Garrison, and Hug}]{Jo2022}
\bibinfo{author}{X.~Han}, \bibinfo{author}{J.~Garrison}, \bibinfo{author}{G.~Hug},
\newblock \bibinfo{title}{Techno-economic analysis of {PV}-battery systems in {Switzerland}},
\newblock \bibinfo{journal}{Renewable and Sustainable Energy Reviews} \bibinfo{volume}{158} (\bibinfo{year}{2022}). \DOIprefix\doi{10.1016/j.rser.2021.112028}.
%Type = Misc
\bibitem[{Garrison et~al.(2022)Garrison, Gjorgiev, Pranjal, Raycheva, Schwarz, Bardow, Turhan, Hug, Sansavini, and Schaffner}]{Nexuse_input_v2}
\bibinfo{author}{J.~Garrison}, \bibinfo{author}{B.~Gjorgiev}, \bibinfo{author}{J.~Pranjal}, \bibinfo{author}{E.~Raycheva}, \bibinfo{author}{M.~Schwarz}, \bibinfo{author}{A.~Bardow}, \bibinfo{author}{D.~Turhan}, \bibinfo{author}{G.~Hug}, \bibinfo{author}{G.~Sansavini}, \bibinfo{author}{C.~Schaffner}, \bibinfo{title}{Nexus-e: Input data and system setup\_v2}, \bibinfo{howpublished}{\url{www.nexus-e.org}}, \bibinfo{year}{2022}.
%Type = Misc
\bibitem[{{Federal Statistical Office}(2024)}]{FSO}
\bibinfo{author}{{Federal Statistical Office}}, \bibinfo{title}{Portraits of the communes}, \bibinfo{year}{2024}. \URLprefix \url{https://www.bfs.admin.ch/bfs/en/home/statistics/regional-statistics/regional-portraits-key-figures/communes.html}.
%Type = Misc
\bibitem[{son(2024)}]{sonnendach}
\bibinfo{title}{Sonnendach.ch und sonnenfassade.ch: Berechnung von potenzialen in gemeinden}, \bibinfo{howpublished}{\url{www.toitsolaire.ch}}, \bibinfo{year}{2024}.
%Type = Article
\bibitem[{Sasse and Trutnevyte(2023)}]{Sasse2023Energy}
\bibinfo{author}{J.~P. Sasse}, \bibinfo{author}{E.~Trutnevyte},
\newblock \bibinfo{title}{Cost-effective options and regional interdependencies of reaching a low-carbon {European} electricity system in 2035},
\newblock \bibinfo{journal}{Energy} \bibinfo{volume}{282} (\bibinfo{year}{2023}). \DOIprefix\doi{10.1016/j.energy.2023.128774}.
%Type = Misc
\bibitem[{{Federal Office of Topography}(2024)}]{swisstopo}
\bibinfo{author}{{Federal Office of Topography}}, \bibinfo{title}{swisstopo}, \bibinfo{year}{2024}. \URLprefix \url{https://www.swisstopo.admin.ch/en}, \bibinfo{note}{(Accessed: 31.07.2024)}.
%Type = Article
\bibitem[{Wen et~al.(2023)Wen, Heinisch, Müller, Sasse, and Trutnevyte}]{Wen2023}
\bibinfo{author}{X.~Wen}, \bibinfo{author}{V.~Heinisch}, \bibinfo{author}{J.~Müller}, \bibinfo{author}{J.~P. Sasse}, \bibinfo{author}{E.~Trutnevyte},
\newblock \bibinfo{title}{Comparison of statistical and optimization models for projecting future pv installations at a sub-national scale},
\newblock \bibinfo{journal}{Energy} \bibinfo{volume}{285} (\bibinfo{year}{2023}). \DOIprefix\doi{10.1016/j.energy.2023.129386}.
%Type = Article
\bibitem[{Rubino et~al.(tted)Rubino, Killenberger, Sasse, Wang, Zielonka, and Trutnevyte}]{Rubino2024}
\bibinfo{author}{G.~Rubino}, \bibinfo{author}{C.~Killenberger}, \bibinfo{author}{J.-P. Sasse}, \bibinfo{author}{Z.~Wang}, \bibinfo{author}{N.~Zielonka}, \bibinfo{author}{E.~Trutnevyte},
\newblock \bibinfo{title}{Spatial strategies for siting variable renewable energy sources to ensure weather resilience in switzerland},
\newblock \bibinfo{journal}{Applied Energy}  (\bibinfo{year}{2024, submitted}).
%Type = Article
\bibitem[{Sasse and Trutnevyte(2019)}]{Sasse2019}
\bibinfo{author}{J.~P. Sasse}, \bibinfo{author}{E.~Trutnevyte},
\newblock \bibinfo{title}{Distributional trade-offs between regionally equitable and cost-efficient allocation of renewable electricity generation},
\newblock \bibinfo{journal}{Applied Energy} \bibinfo{volume}{254} (\bibinfo{year}{2019}). \DOIprefix\doi{10.1016/j.apenergy.2019.113724}.
%Type = Article
\bibitem[{Sasse and Trutnevyte(2020)}]{Sasse2020}
\bibinfo{author}{J.~P. Sasse}, \bibinfo{author}{E.~Trutnevyte},
\newblock \bibinfo{title}{Regional impacts of electricity system transition in central {Europe} until 2035},
\newblock \bibinfo{journal}{Nature Communications} \bibinfo{volume}{11} (\bibinfo{year}{2020}). \DOIprefix\doi{10.1038/s41467-020-18812-y}.
%Type = Misc
\bibitem[{Schlecht and Weigt(2014)}]{Schlecht2014}
\bibinfo{author}{I.~Schlecht}, \bibinfo{author}{H.~Weigt}, \bibinfo{title}{Swissmod a model of the swiss electricity market swissmod-a model of the swiss electricity market}, \bibinfo{howpublished}{\url{http://ssrn.com/abstract=2446807}}, \bibinfo{year}{2014}.
%Type = Article
\bibitem[{Dujardin et~al.(2021)Dujardin, Kahl, and Lehning}]{Dujardin2021}
\bibinfo{author}{J.~Dujardin}, \bibinfo{author}{A.~Kahl}, \bibinfo{author}{M.~Lehning},
\newblock \bibinfo{title}{Synergistic optimization of renewable energy installations through evolution strategy},
\newblock \bibinfo{journal}{Environmental Research Letters} \bibinfo{volume}{16} (\bibinfo{year}{2021}). \DOIprefix\doi{10.1088/1748-9326/abfc75}.
%Type = Article
\bibitem[{Bartlett et~al.(2018)Bartlett, Dujardin, Kahl, Kruyt, Manso, and Lehning}]{Bartlett2018}
\bibinfo{author}{S.~Bartlett}, \bibinfo{author}{J.~Dujardin}, \bibinfo{author}{A.~Kahl}, \bibinfo{author}{B.~Kruyt}, \bibinfo{author}{P.~Manso}, \bibinfo{author}{M.~Lehning},
\newblock \bibinfo{title}{Charting the course: A possible route to a fully renewable swiss power system},
\newblock \bibinfo{journal}{Energy} \bibinfo{volume}{163} (\bibinfo{year}{2018}) \bibinfo{pages}{942--955}. \DOIprefix\doi{10.1016/j.energy.2018.08.018}.
%Type = Techreport
\bibitem[{{ENTSO-g and ENTSO-e}(2022)}]{Entsoe2022}
\bibinfo{author}{{ENTSO-g and ENTSO-e}}, \bibinfo{title}{TYNDP 2022 Scenario Report, Version. April 2022}, \bibinfo{type}{Technical Report}, \bibinfo{year}{2022}.
%Type = Misc
\bibitem[{{Federal Office of Energy}(2024{\natexlab{a}})}]{stromgesetz}
\bibinfo{author}{{Federal Office of Energy}}, \bibinfo{title}{Federal act on a secure electricity supply from renewable energy sources}, \bibinfo{year}{2024}{\natexlab{a}}. \URLprefix \url{https://www.admin.ch/gov/en/start/documentation/votes/20240609/federal-act-on-a-secure-electricity-supply-from-renewable-energy-sources.html}, \bibinfo{note}{(Accessed: 05.08.2024)}.
%Type = Misc
\bibitem[{{Federal Office of Energy}(2024{\natexlab{b}})}]{electricityCHEU}
\bibinfo{author}{{Federal Office of Energy}}, \bibinfo{title}{Electricity agreement {Switzerland - EU}}, \bibinfo{year}{2024}{\natexlab{b}}. \URLprefix \url{https://www.bfe.admin.ch/bfe/en/home/supply/electricity-supply/electricity-agreement-switzerland-eu.html}, \bibinfo{note}{(Accessed: 05.08.2024)}.
%Type = Misc
\bibitem[{{Federal Office of Energy}()}]{federalactsecurityofsupply}
\bibinfo{author}{{Federal Office of Energy}}, \bibinfo{title}{Federal act on a secure electricity supply}, \bibinfo{year}{.} \URLprefix \url{https://www.bfe.admin.ch/bfe/en/home/supply/electricity-supply/federal-act-renewable-electricity-supply.html/}, \bibinfo{note}{(Accessed: 05.08.2024)}.
%Type = Misc
\bibitem[{{Federal Office of Energy}(2024)}]{energieperspektieven2050}
\bibinfo{author}{{Federal Office of Energy}}, \bibinfo{title}{Energieperspektiven 2050+}, \bibinfo{year}{2024}. \URLprefix \url{https://www.bfe.admin.ch/bfe/de/home/politik/energieperspektiven-2050-plus.html}, \bibinfo{note}{(Accessed: 05.08.2024)}.
%Type = Article
\bibitem[{{Van Den Bergh} and Delarue(2015)}]{voll_Kenneth}
\bibinfo{author}{K.~{Van Den Bergh}}, \bibinfo{author}{E.~Delarue},
\newblock \bibinfo{title}{Cycling of conventional power plants: Technical limits and actual costs},
\newblock \bibinfo{journal}{Energy Conversion and Management} \bibinfo{volume}{97} (\bibinfo{year}{2015}) \bibinfo{pages}{70--77}. \DOIprefix\doi{10.1016/j.enconman.2015.03.026}.
%Type = Article
\bibitem[{Löffler et~al.(2019)Löffler, Burandt, Hainsch, and Oei}]{Löffler2019}
\bibinfo{author}{K.~Löffler}, \bibinfo{author}{T.~Burandt}, \bibinfo{author}{K.~Hainsch}, \bibinfo{author}{P.~Y. Oei},
\newblock \bibinfo{title}{Modeling the low-carbon transition of the {European} energy system - a quantitative assessment of the stranded assets problem},
\newblock \bibinfo{journal}{Energy Strategy Reviews} \bibinfo{volume}{26} (\bibinfo{year}{2019}). \DOIprefix\doi{10.1016/j.esr.2019.100422}.
%Type = Misc
\bibitem[{Lazard(2024)}]{Lazard2023}
\bibinfo{author}{Lazard}, \bibinfo{title}{{LCOE}}, \bibinfo{year}{2024}. \URLprefix \url{https://www.lazard.com/media/typdgxmm/lazards-lcoeplus-april-2023.pdf}, \bibinfo{note}{(Accessed: 13.08.2024)}.
%Type = Misc
\bibitem[{FEM(2024)}]{FEM_input}
\bibinfo{title}{The future electricity market model- {FEM}: Model description}, \bibinfo{howpublished}{\url{zhaw.ch/storage/sml/institute-zentren/cee/upload/FEM_model_description.pdf}}, \bibinfo{year}{2024}.

\end{thebibliography}
\end{document}

\begin{frontmatter}
%%%%%%%%%%%%%%%%%%%%%%%%%%%%%%%%%%%%%%%%%%%%%%%%%%%%%%%%%%%%%%%%%%%%%%%%%%%%%%%%%%
% TITLE
%%%%%%%%%%%%%%%%%%%%%%%%%%%%%%%%%%%%%%%%%%%%%%%%%%%%%%%%%%%%%%%%%%%%%%%%%%%%%%%%%%
\title{Policy-relevance of a Model Inter-comparison: Switzerland in the European Energy Transition}

%% Example of authorship
\author[RRE]{Ambra Van Liedekerke}
\author[RRE]{Blazhe Gjorgiev\corref{cor1}} \ead{gblazhe@ethz.ch}
\author[ESC]{Jonas Savelsberg}
\author[UNIGE]{Xin Wen}
\author[EPFL]{Jerome Dujardin}
%\author[EPFL]{Albin Cintas}
\author[ZHAW]{Ali Darudi}
\author[UNIGE]{Jan-Philipp Sasse}
\author[UNIGE]{Evelina Trutnevyte}
\author[EPFL,SLF]{Michael Lehning}
\author[RRE]{Giovanni Sansavini\corref{cor1}} \ead{sansavig@ethz.ch}

\affiliation[RRE]{organization={Reliability and Risk Engineering Laboratory, Institute of Energy and Process Engineering, Department of Mechanical and Process Engineering, ETH Zurich},
            city={Zurich},
            country={Switzerland}}

\affiliation[ESC]{organization={Energy Science Center, ETH Zurich},
            city={Zurich},
            country={Switzerland}}

\affiliation[UNIGE]{organization={Renewable Energy Systems, Institute for Environmental Sciences, Section of Earth and Environmental Sciences, University of Geneva},
            city={Geneva},
            country={Switzerland}}

\affiliation[EPFL]{organization={School of Architecture, Civil and Environmental Engineering, Swiss Federal Institute of Technology in Lausanne (EPFL)},
            city={Lausanne},
            country={Switzerland}}

\affiliation[SLF]{organization={WSL Institute for Snow and Avalanche Research, SLF Davos},
            city={Davos},
            country={Switzerland}}

\affiliation[ZHAW]{organization={ZHAW},
            city={Zurich},
            country={Switzerland}}

\cortext[cor1]{Corresponding author.}

%%%%%%%%%%%%%%%%%%%%%%%%%%%%%%%%%%%%%%%%%%%%%%%%%%%%%%%%%%%%%%%%%%%%%%%%%%%%%%%%%%%%%%%%%%%%%%%%

\end{frontmatter}

%%% OPERATIONS
% Capacity and generation - all scenarios - GA
\begin{figure*}[]
  % \centering
  
  \begin{minipage}[b]{0.45\textwidth}
    \includegraphics[height=0.8\textheight]{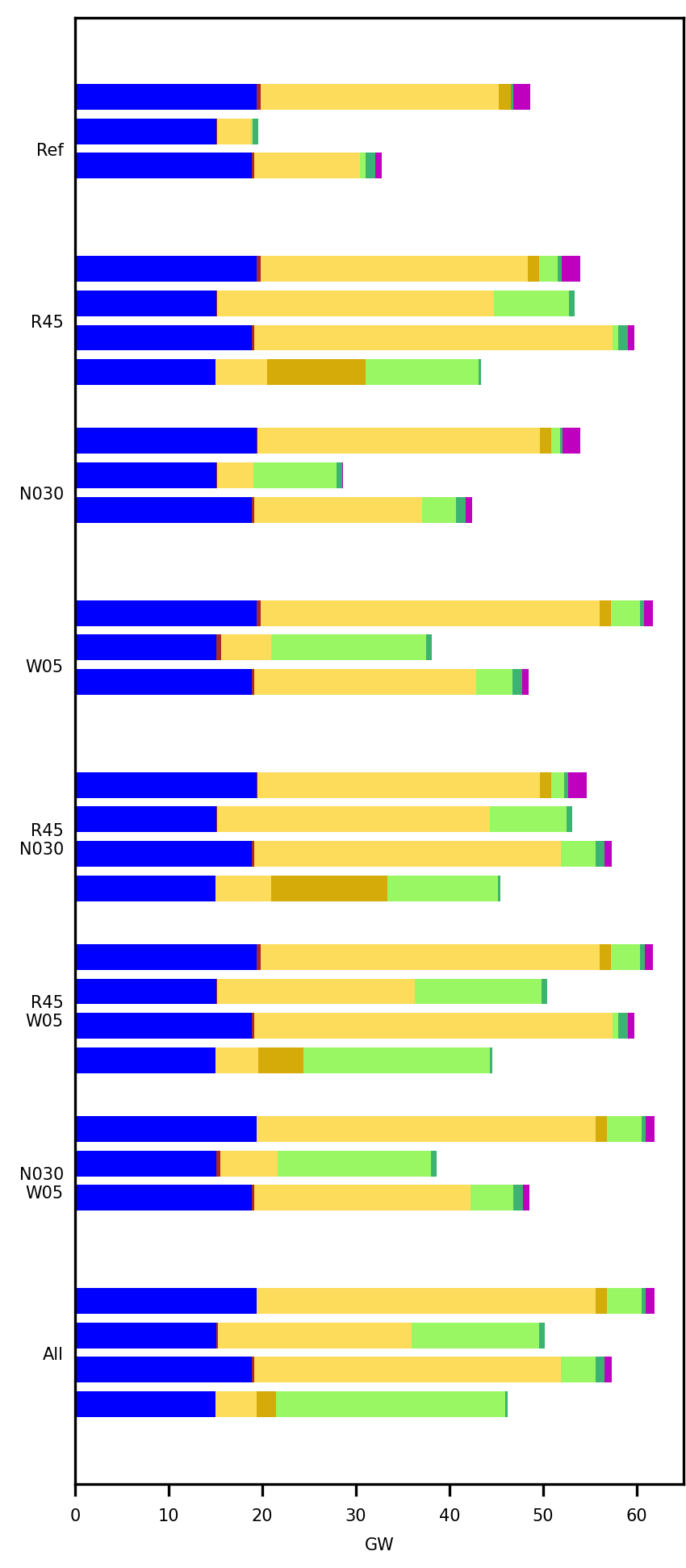}
  \end{minipage}
  \hspace{0.2cm}
  \begin{minipage}[b]{0.45\textwidth}
    \includegraphics[height=0.8\textheight]{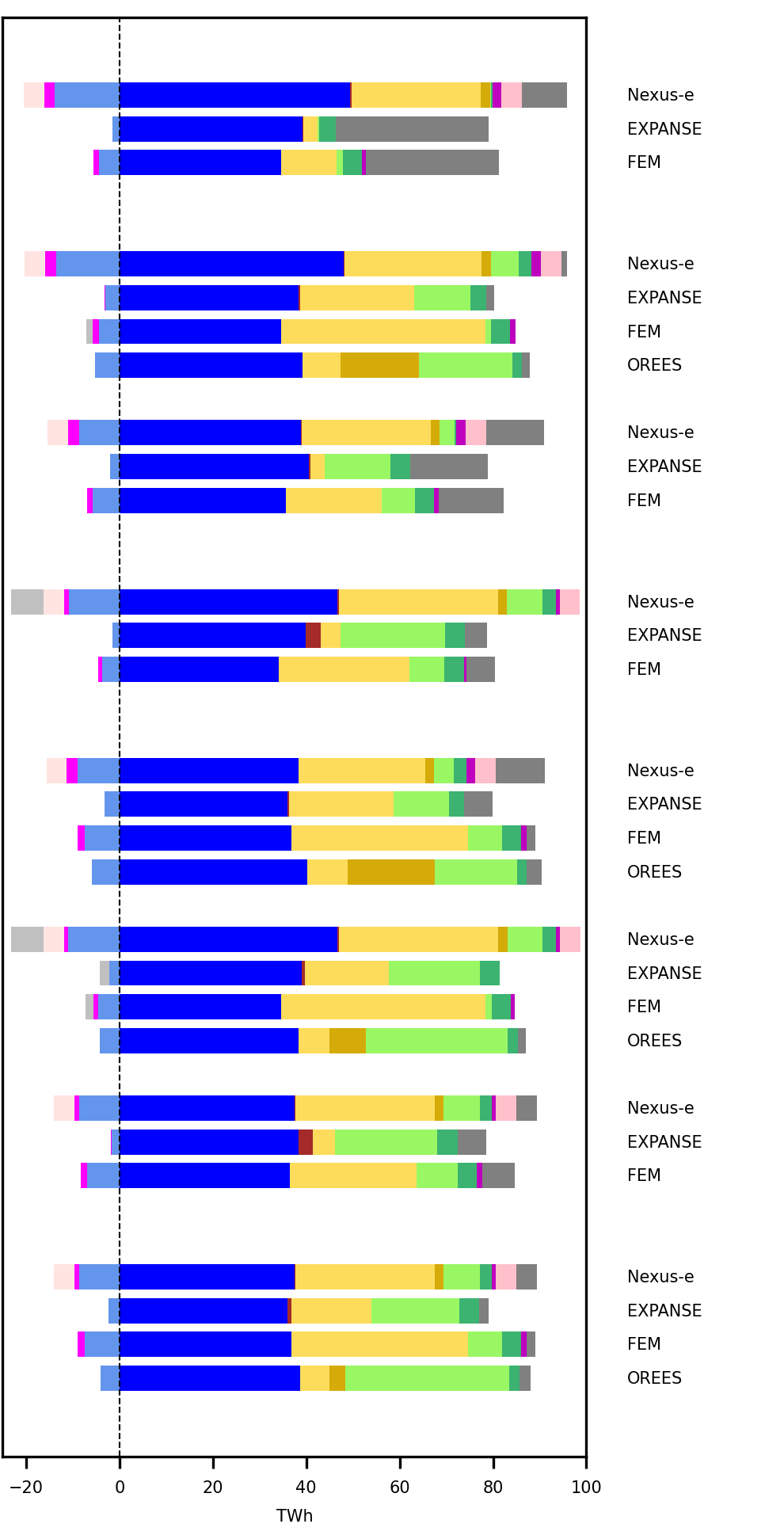}
  \end{minipage}
    
  \begin{minipage}[b]{1\textwidth}
    \includegraphics[width=1\linewidth]{Figs/Generation_legend.png}
    \vspace{-0.5cm}
  \end{minipage}

  \caption{Installed capacity and annual generation by technology type for all scenarios, all four models, considering the GA European development.}
  \label{fig: generationAllGA}
\end{figure*}

% Capacity and generation - all scenarios - DE
\begin{figure*}[]
  % \centering
  
  \begin{minipage}[b]{0.45\textwidth}
    \includegraphics[height=0.8\textheight]{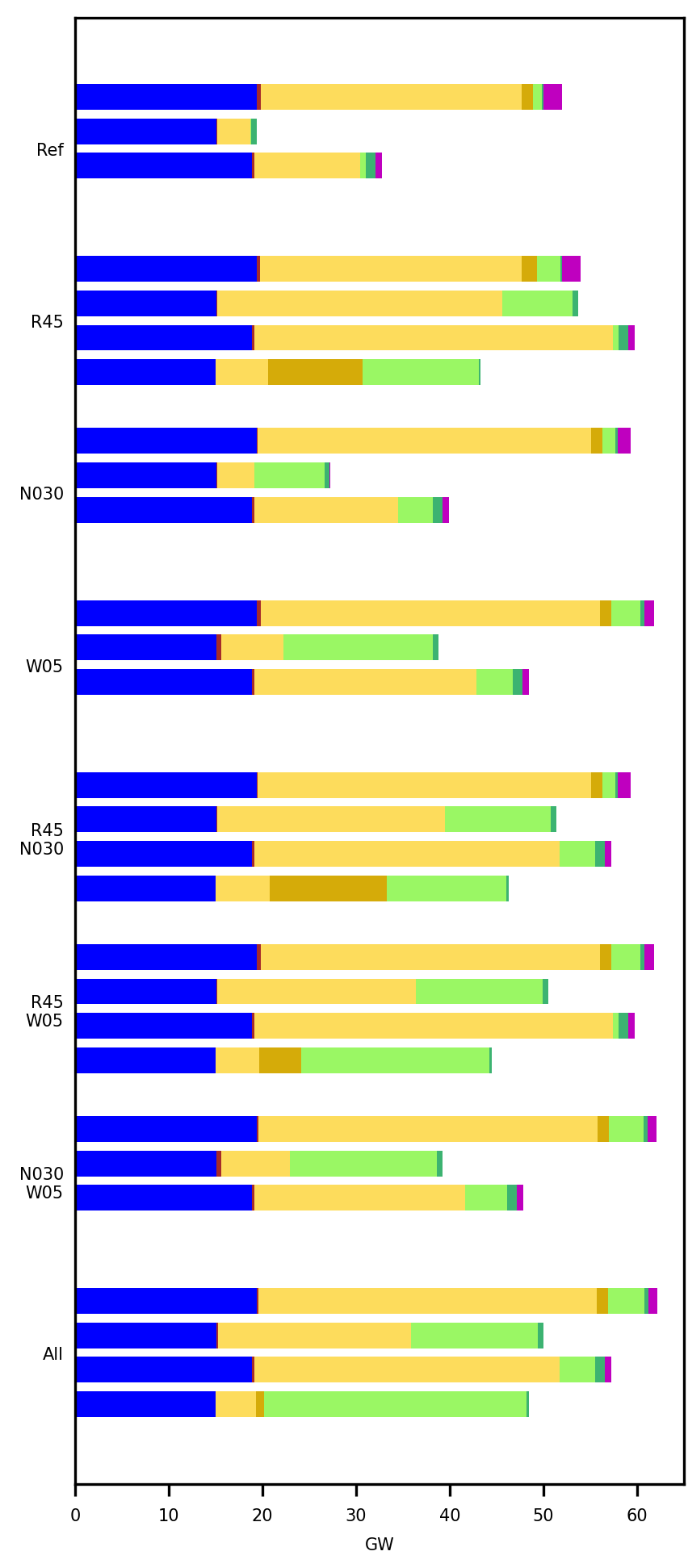}
  \end{minipage}
  \hspace{0.2cm}
  \begin{minipage}[b]{0.45\textwidth}
    \includegraphics[height=0.8\textheight]{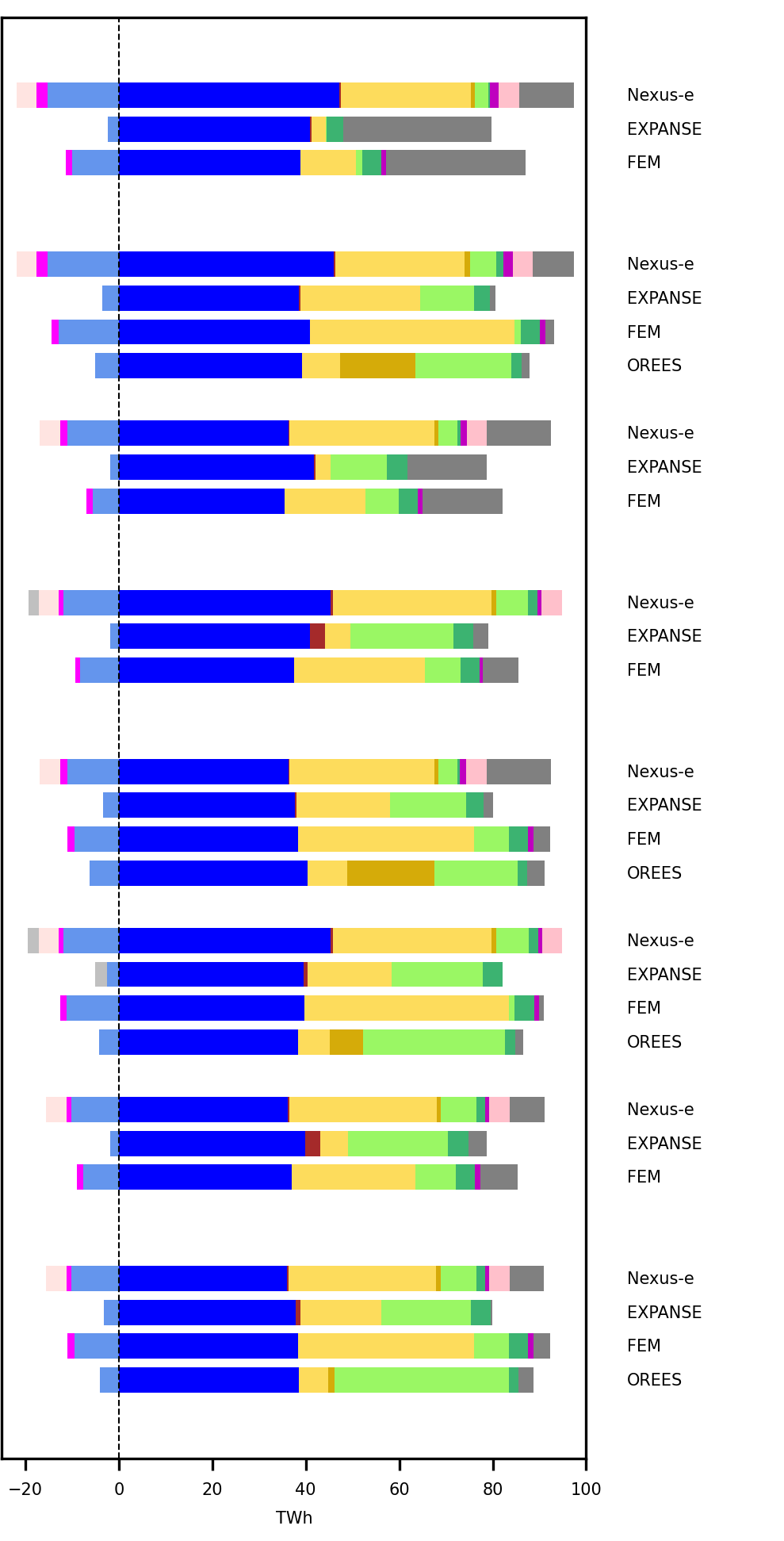}
  \end{minipage}
    
  \begin{minipage}[b]{1\textwidth}
    \includegraphics[width=1\linewidth]{Figs/Generation_legend.png}
    \vspace{-0.5cm}
  \end{minipage}

  \caption{Installed capacity and annual generation by technology type for all scenarios, all four models, considering the DE European development.}
  \label{fig: generationAllDE}
\end{figure*}

% Exchange - GA
\begin{figure*}[]
    \centering
    \includegraphics[width=1\linewidth]{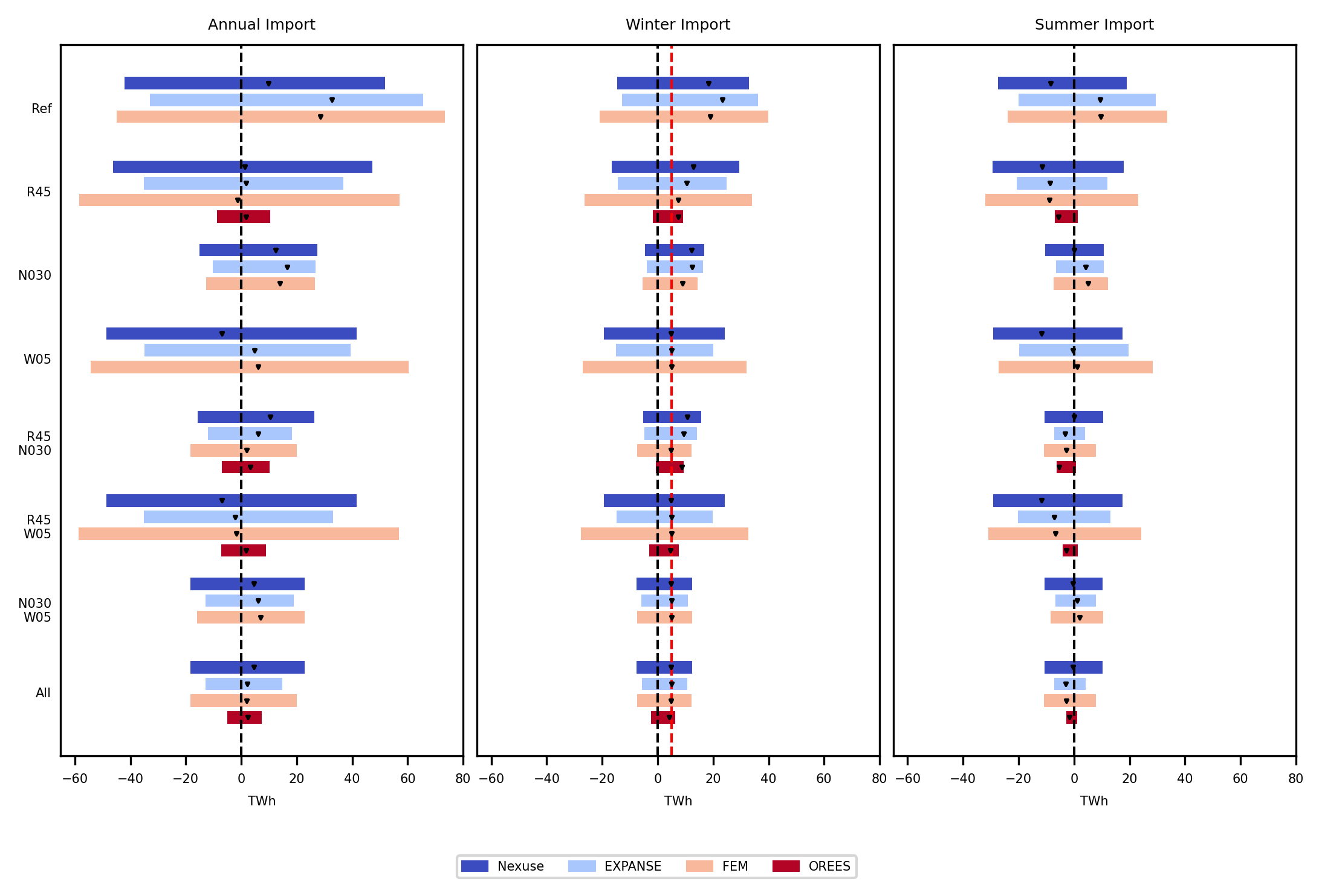}
    \caption{Imports, exports, and net imports over the full year, the winter half, and the summer half. The positive axis denotes imports; the negative axis denotes exports; the reversed triangle shows the net imports (imports - exports).}
    \label{fig:exchange}
\end{figure*}

% Exchange - GA
\begin{figure*}[]
    \centering
    \includegraphics[width=1\linewidth]{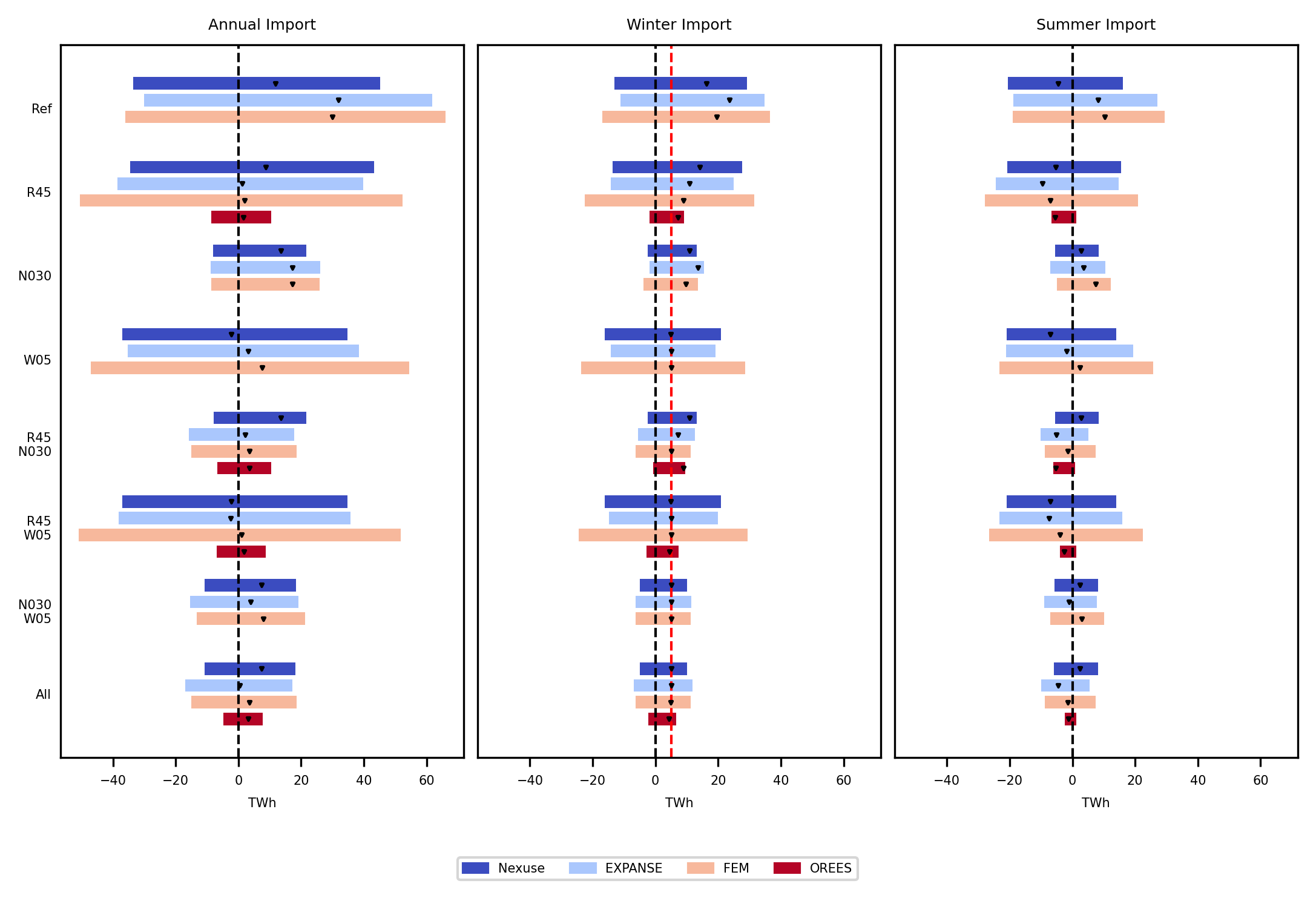}
    \caption{Imports, exports, and net imports over the full year, the winter half, and the summer half. The positive axis denotes imports; the negative axis denotes exports; the reversed triangle shows the net imports (imports - exports).}
    \label{fig:exchange}
\end{figure*}

%%% COSTS
% All costs - percent variation - GA
\begin{figure*}[]
  % \centering
  \begin{minipage}[b]{0.33\textwidth}
    \includegraphics[height=0.48\textheight, keepaspectratio]{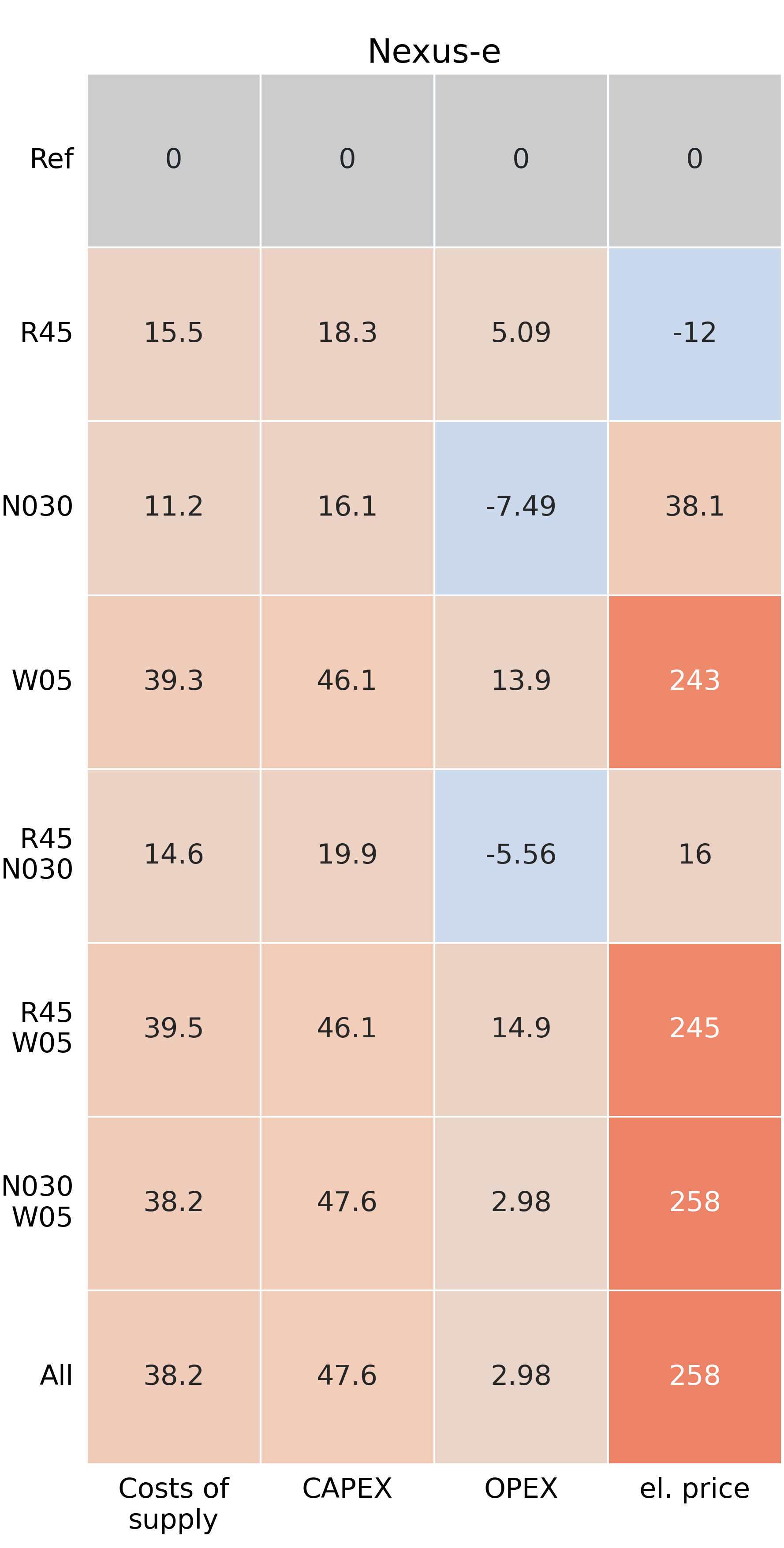}
  \end{minipage}
  \begin{minipage}[b]{0.33\textwidth}
    \includegraphics[height=0.48\textheight, keepaspectratio]{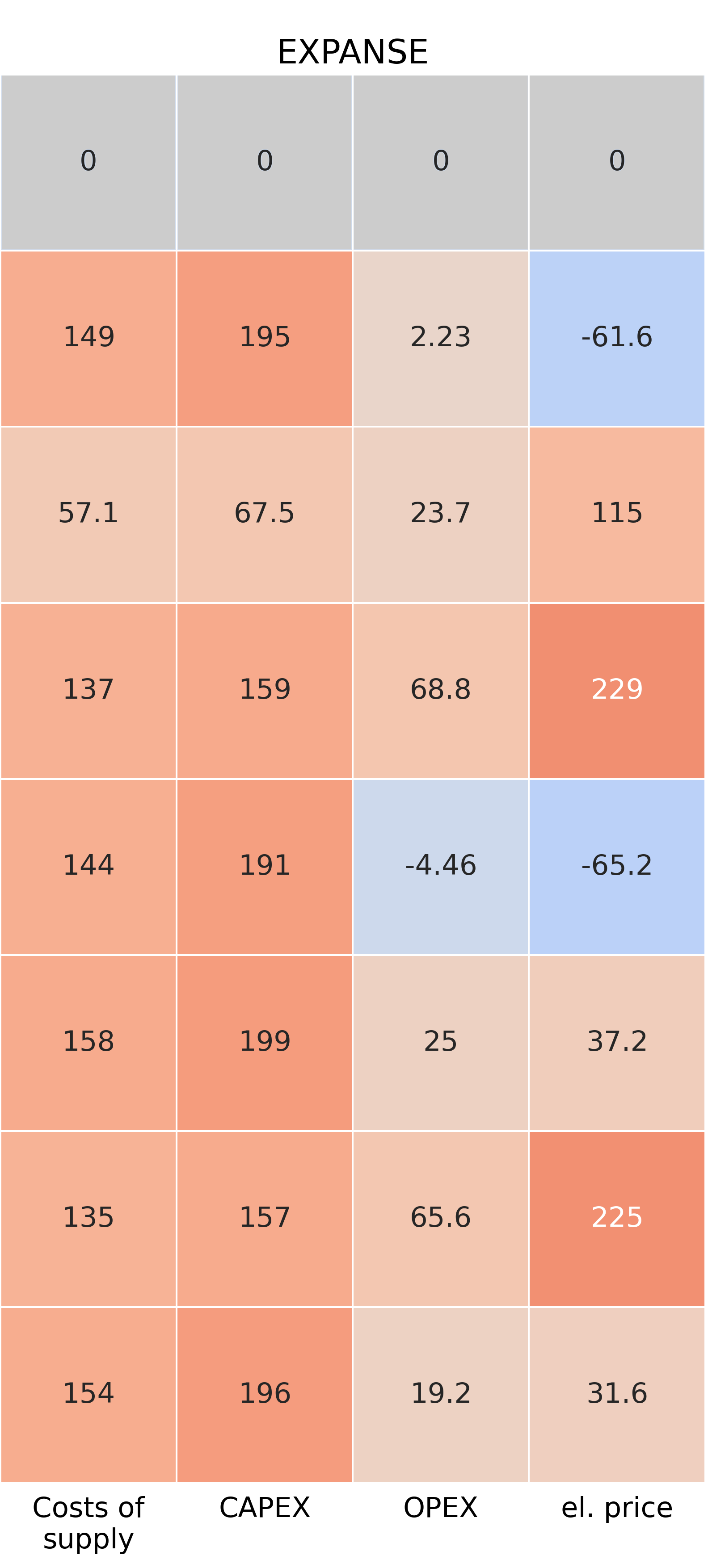}
  \end{minipage}
  \hspace{-0.75cm}
  \begin{minipage}[b]{0.33\textwidth}
    \includegraphics[height=0.48\textheight, keepaspectratio]{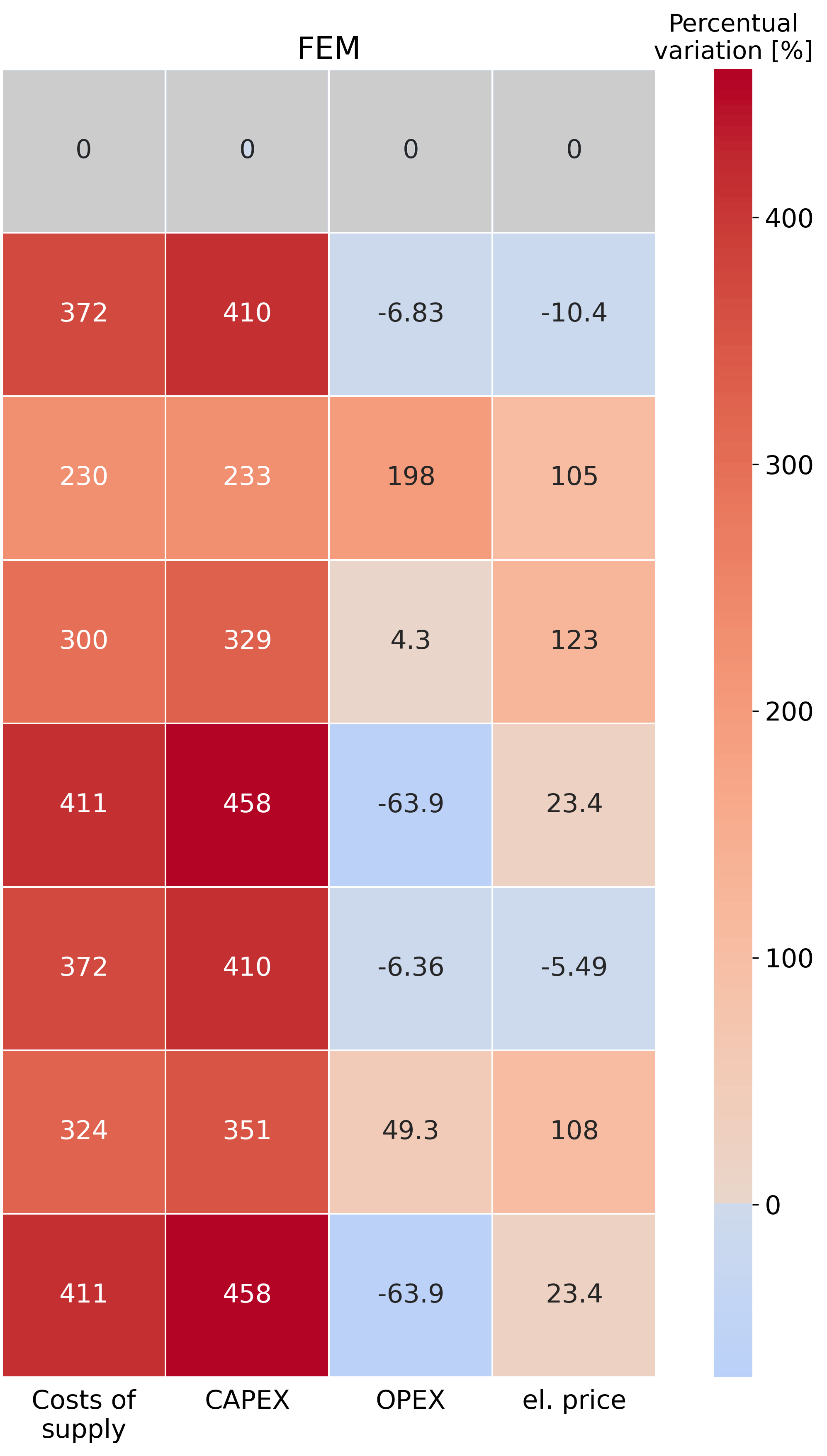}
  \end{minipage}
  
  \caption{Percentual variation of annualized Cost of supply, annualized investment costs (CAPEX), yearly operational costs (OPEX), and average yearly electricity price (el. price) for all scenarios, compared to the Ref. Results for the GA European dimension.}
  \label{fig:costs_all_GA}
\end{figure*}

% All costs - percent variation - DE
\begin{figure*}[]
  % \centering
  \begin{minipage}[b]{0.33\textwidth}
    \includegraphics[height=0.48\textheight, keepaspectratio]{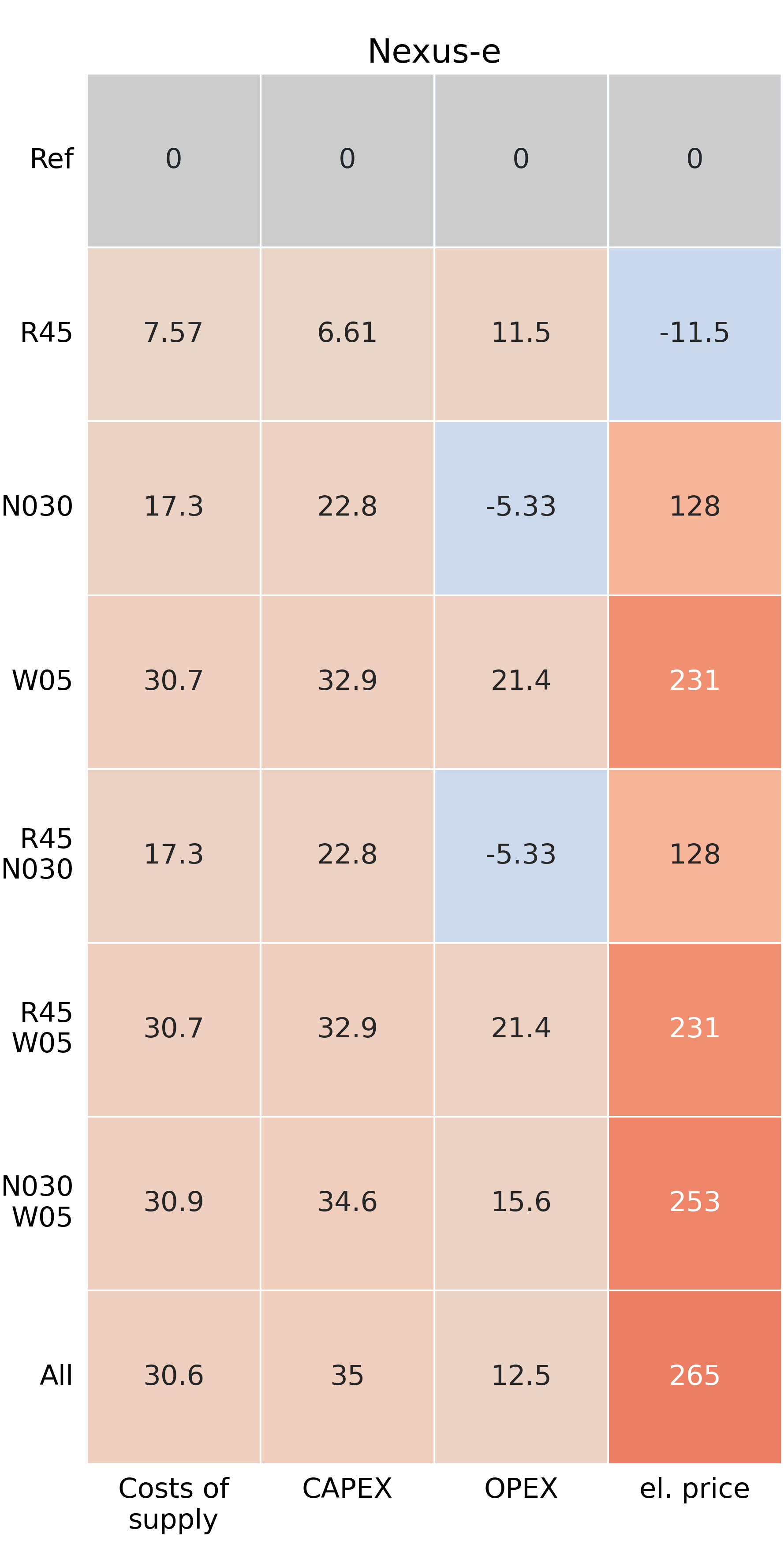}
  \end{minipage}
  \begin{minipage}[b]{0.33\textwidth}
    \includegraphics[height=0.48\textheight, keepaspectratio]{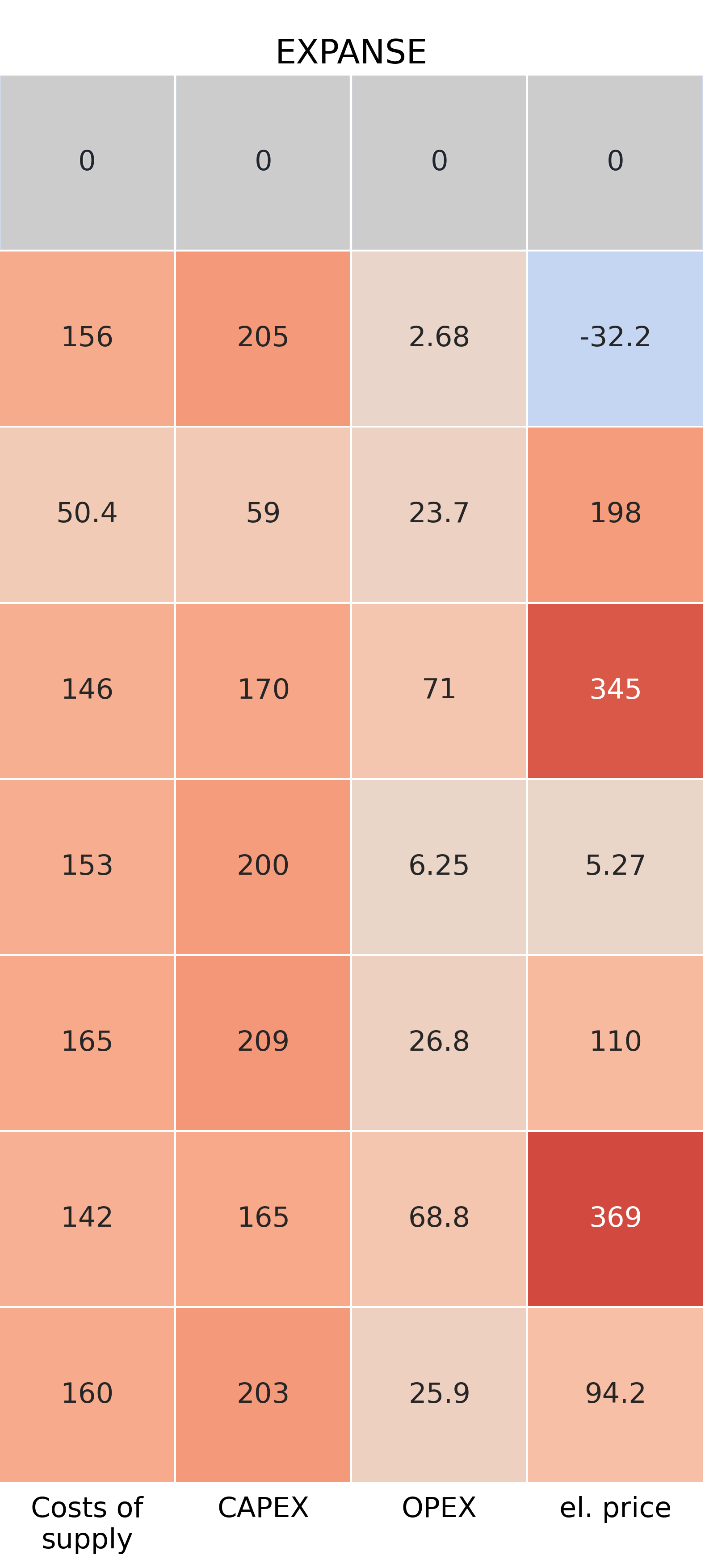}
  \end{minipage}
  \hspace{-0.75cm}
  \begin{minipage}[b]{0.3\textwidth}
    \includegraphics[height=0.48\textheight, keepaspectratio]{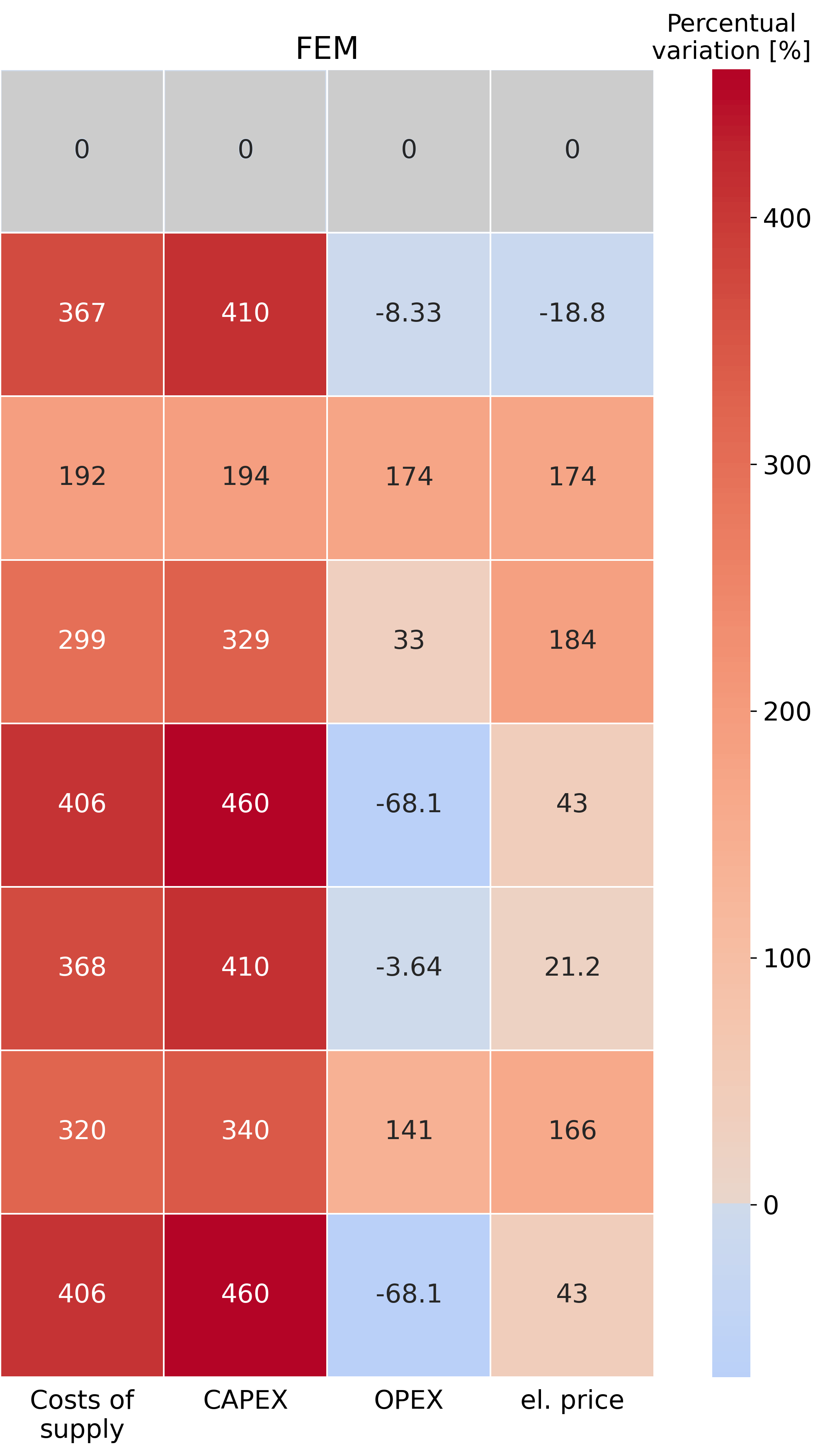}
  \end{minipage}
  
  \caption{Percentual variation of annualized Cost of supply, annualized investment costs (CAPEX), yearly operational costs (OPEX), and average yearly electricity price (el. price) for all scenarios, compared to the Ref. Results for the DE European dimension.}
  \label{fig:costs_all_DE}
\end{figure*}